\documentclass{stavanger}
\usepackage[hypcap=false]{caption}
\usepackage[utf8]{inputenc}
\usepackage{amssymb,amsmath,amsfonts,amsthm}
\usepackage{dsfont}
\usepackage{hyperref}
\usepackage{natbib}
\usepackage{tikz}
\usepackage{graphicx}
\usepackage{cleveref}
\usepackage{mathtools}
\usepackage{placeins}
\usepackage{float}
\usepackage{physics}
\usepackage{xcolor}
\usepackage{ulem}

\begin{document}

\begin{frontmatter}

\begin{fmbox}


\dochead{Preprint}{FP}


\title{In-medium static inter-quark potential on high resolution quenched lattices}


\author[addressref={aff1} ,                	  
   ]{Rasmus N. Larsen}
\author[
   addressref={aff1},                   	  
   corref={aff1},                     		  
   email={gaurang.parkar@uis.no}   		  
]{\inits{GP}\fnm{Gaurang} \snm{Parkar}}
\author[addressref={aff1},                   	  
   ]{Alexander Rothkopf}
\author[addressref={aff3},                   	  
   ]{Johannes Heinrich Weber}

\address[id=aff1]{
  \orgname{Faculty of Science and Technology}, 	 
  \street{University of Stavanger},                     		 
  \postcode{4021}                               			 
  \city{Stavanger},                              				 
  \cny{Norway}                                   				 
}

\address[id=aff3]{
  \orgname{Institut f\"ur Physik \& IRIS Adlershof}, 	 
  \street{ Humboldt-Universit\"at zu Berlin},                     		 
  \postcode{D-12489}                               			 
  \city{Berlin},                              				 
  \cny{Germany}                                   				 
}



\end{fmbox}


\begin{abstractbox}

\begin{abstract} 
We re-investigate the interactions between static color sources in a finite temperature gluonic medium using both high resolution isotropic and anisotropic quenched lattice QCD ensembles. The underlying ill-posed inverse problem, related to the extraction of spectral functions, is attacked with a range of different methods, including Bayesian inference, Pad\'e interpolation and model fits. Among the latter we include a tail amended Gaussian ansatz and a HTL-inspired fit ansatz. We reconfirm the presence of a dominant low-lying spectral feature that supports the existence of a potential picture for the in-medium evolution of the static charges at late real times. Using the raw unmodified lattice data, all applicable methods show clear signs of screening of the real-part of the potential. After applying a subtraction procedure featured in a previous study we find however that screening disappears from the extracted potential. Paths towards the resolution of this puzzle are discussed.
\end{abstract}

\end{abstractbox}

\end{frontmatter}

\tableofcontents

\section{Introduction and Motivation}
The bound states of a heavy quark and antiquark, called heavy quarkonium ($c\bar{c}$ charmonium, $b\bar{b}$ bottomonium), are considered excellent probes for the precision study of the properties of the quark-gluon plasma (QGP), created in relativistic heavy-ion collisions. Their structural simplicity offers unparalleled theoretical access, based on first principles quantum chromodynamics and their experimental investigation benefits from clean decay channels involving dilepton pairs (for recent review see \cite{Rothkopf:2019ipj}, for an experimental perspective see e.g. \cite{Andronic:2015wma}).

Over the past decade efforts have been ongoing to simplify the theoretical treatment of quarkonium in a hot medium by exploiting the inherent separation of scales between the rest mass of the constituent quarks $m_Q$ ($m_c\approx1.2$ GeV, $m_b\approx4.6$ GeV) and environmental scales. Take e.g. temperature $T$, which in recent relativistic heavy-ion collision experiments is estimated indirectly to reach up to values of at most $T^{\rm max}_{\rm LHC}\approx 0.7$GeV \cite{Schenke:2011tv}. 

The sought after theoretical simplification is achieved through effective field theories (EFTs), such as non-relativistic QCD (NRQCD) or potential NRQCD (pNRQCD, see \cite{Brambilla:2004jw} for a review). In the latter framework, the dynamics of the quark-antiquark pair is described in terms of color singlet and color octet wavefunctions. In a weakly-coupled medium these wavefunctions either propagate within each of the color sectors by means of a Schr\"odinger-like equation, governed by a potential, or may transition between the color singlet and octet sector by non-potential interactions induced by exchange of low-energy gluons. Since pNRQCD follows from a systematic expansion of the QCD Lagrangian, the potential itself can be expressed as a series of terms, which contain successively higher powers of the heavy quark velocity $v$ (which is a dimensionless proxy for the inverse mass). Formally the potential constitutes a time-independent but spatially non-local Wilson coefficient of the EFT pNRQCD.

The study of the static heavy-quark potential refers to isolating the potential contribution to quarkonium dynamics, which would be felt by an infinitely heavy quark-antiquark pair, i.e. the contribution to the full quarkonium potential at order $v^0$.  Note that even in this limit, the evolution of the corresponding static color charge-anticharge pair will in general consist of potential and non-potential contributions, which must be disentangled. The evolution of such a static pair of color sources in Minkowski-time is formally described by the Wilson loop
\begin{align}
    W_\square(t,r)=\left\langle {\cal T} {\rm exp}\Big[ i\int_\square dz^\mu T^aA^a_\mu(z)\Big]\right \rangle_{\rm QCD}
\end{align}
where the integration contour describes a rectangle of temporal extent $t$ and spatial extent $r$. The expectation value is taken over the gauge fields $A^a$, which take on values in the $SU(3)$ algebra, represented by the Gell-Mann matrices $T^a$. 

In the seminal work of Ref.~\cite{Laine:2006ns}, Laine et al.~evaluated this Wilson loop perturbatively, selecting a specific hierarchy of energy scales (${1}/{r} \sim gT < T$, where $m_D \sim gT$ is the Debye mass), which is captured by the resummed perturbation theory called hard-thermal loops (HTL). The authors showed that in the late real-time limit, a weakly-coupled Wilson loop at finite temperature evolves solely according to a Schr\"odinger-like equation 
\begin{align}
    \lim_{t\to\infty} i\partial\_t W_{\square}(t,r) = \lim_{t\to\infty} \Phi(t,r) W_{\square}(t,r)=  V(r) W_{\square}(t,r)
\end{align}
with a time-independent \textit{complex} potential $V(r)$, which emerges as the time-independent late-time component of an in-general time-dependent complex function $\Phi(t,r)$. It is the non-potential effects of the evolution, which manifest themselves as the time dependence of $\Phi(t,r)$. One may use the late-time limit to formally define the values of the potential as
\begin{align}
    V(r)=\lim_{t\to\infty}\frac{\partial}{\partial t}{\rm log}[W_\square(t,r)]\label{eq:DefPot}
\end{align}
The HTL result is remarkable, as in the limit of late Minkowski time non-potential effects become irrelevant and only $V(r)$ remains. Later studies \cite{Brambilla:2008cx,Beraudo:2007ky} used different perturbative techniques to compute the static potential contributions in different scale hierarchies. 
While a time-independent potential with a non-vanishing imaginary part is a common feature, screening of the real part is not, i.e. both $gT < T < {1}/{r}$ or $gT < {1}/{r} < T$ lead to a thermally modified yet unscreened real part \cite{Brambilla:2008cx}. 
Laine later argued that the relevant hierarchy for quarkonium melting is $gT < {1}/{r} < T$ \cite{Laine:2008cf}.

Note that the potential $V(r)$ defined as such may be of limited relevance to phenomenological modeling, as \cref{eq:DefPot} does not tell us to what extent the non-potential effects remain relevant to the evolution of the static color charge pair at intermediate times. The sole purpose of this paper is to clarify whether the limit of \cref{eq:DefPot} is well-defined non-perturbatively and if so, isolate from first principles the values of the real- and imaginary part of $V(r)$.

Lattice QCD, the study of the strong interactions by means of numerical simulations of Feynman's path integral on a discretized spacetime grid with lattice spacing $a$, has made vital contributions to our understanding of heavy quarkonium both in vacuum \cite{McNeile:2012qf} and in medium \cite{Kim:2018yhk, Petreczky:2021zmz, Bazavov:2023dci}. It remains one of the few ab-initio approaches to QCD that offer fully non-perturbative access to the physics of quarks and gluons. While it preserves gauge symmetry exactly, it breaks the space-time symmetries of the continuum, which are however believed to be restored when the lattice spacing is extrapolated to the continuum limit (for recent progress in the preservation of space-time symmetries in numerical simulations see \cite{Rothkopf:2023tbp}). 

In contrast to continuum QCD referencing gauge fields $A^a_\mu(x)$, lattice QCD is formulated in terms of $SU(3)$ group-valued link variables 
$U_\mu(x)={\rm exp}[iagA^a_\mu(x+a/2\hat{e}_{\mu})T^a]$.
Correspondingly the Wilson loop becomes a closed product of such links $ W_\square(t,r)=\left\langle \prod_{\square} U_\mu(x)\right \rangle_{\rm QCD}$. More importantly, to carry out actual Monte-Carlo simulations and to avoid the notorious sign-problem, the theory is formulated not in physical Minkowski space-time but instead in an artificial Euclidean time. I.e. the imaginary time coordinate $\tau=it$ resides on the negative imaginary axis of the complex time plane. The subtle relation of statistical mechanics and quantum physics in Euclidean time also allows one to simulate the system at a finite temperature simply by introducing periodic boundary conditions in imaginary time. The extent of the Euclidean time direction is directly related to the inverse temperature. In a computer of finite resources, what can be evaluated in a lattice QCD simulation is the Euclidean counterpart of the real-time Wilson loop at a finite lattice spacing and some non-zero temperature (as the temporal extent is always finite). We will discuss in detail how information regarding real-time dynamics can be extracted from the simulated data.

In order to extract the real-time dynamics of the Wilson loop $W_\square(t,r)$, needed to define $V(r)$, from its Euclidean counterpart $W_\square(\tau,r)$, one can exploit the fact that both encode the same positive definite spectral function \cite{Rothkopf:2011db}
\begin{align}
    W_\square(t,r)=\int_{-\infty}^{\infty} d\omega\,\rho_\square(\omega,r)\,e^{i\omega t}~,\quad W_\square(\tau,r)=\int_{-\infty}^\infty d\omega\,\rho_\square(\omega,r) e^{-\omega\tau}.
\end{align}
If the spectral function $\rho_\square$ exhibits a well defined spectral peak at low frequencies (i.e. in distinction to a wide continuum contribution), associated with the lowest lying pole of the corresponding correlation function, it will be this spectral structure which dominates the late real-time dynamics. Depending on the functional form of this peak, a well defined, i.e. time independent, limit in \cref{eq:DefPot} may or may not ensue  (for a more detailed discussion see \cref{sec:DefPot}). The extraction of spectral functions from Euclidean correlation functions constitutes an ill-posed inverse problem, and several methods have been proposed to attack it. In the past, both Bayesian inference and fits based on model spectral functions have been used to attack this inversion. We will discuss and benchmark several of these methods in detail in \cref{sec:methodsmock}.

At zero temperature the large Euclidean time limit and the large Minkowski time limit in \cref{eq:DefPot} coincidentally agree. One finds that  $\lim_{\tau\to\infty}{\rm log}[\frac{d}{d\tau}W_\square(\tau,r)]$ is indeed well defined and the resulting $V(r)$ qualitatively agrees with the intuition behind the Cornell potential \cite{Eichten:1974af}, featuring a Coulomb-like part at small distances and a linearly rising confined part at intermediate distances \cite{Brambilla:2010cs} (For a discussion of string breaking and the technical difficulties in defining hybrid potentials see e.g. \cite{Bali:2000vr,Bulava:2019iut}). The running of the strong coupling constant that leads to deviations from the naive Coulomb behavior assumed in the Cornell potential has also been revealed in detail (for a recent study see e.g. \cite{Bazavov:2019qoo}). 

At finite temperature, Euclidean time is necessarily finite and thus the late $\tau$ limit simply does not exist. In the past several different thermodynamic quantities, such as the color singlet free energy or internal energies have been suggested as ad-hoc substitutes for the potential, but none of them have been shown to arise as the potential in a QCD based computation. (The only exception is the color singlet free energy, which in lowest order HTL perturbation theory agrees with the real-part of the potential, but already at next to leading order, while still close, does not agree with $\textrm{Re}[V]$ identically.) 

In the past, the late time behavior of static color sources has been investigated in both purely gluonic (quenched) lattice QCD \cite{Rothkopf:2011db,Burnier:2016mxc,Bala:2019cqu} and in the presence of dynamical quarks \cite{Burnier:2015tda,Petreczky:2017aiz,Petreczky:2018xuh}.

The studies in the quenched approximation concluded that the real-time limit of \cref{eq:DefPot} is indeed well defined and that a time-independent potential emerges. In addition, they also concluded that the real part of that potential is modified as the temperature increases beyond the critical temperature $T_c=271$MeV, at which a genuine first order phase transition occurs in $SU(3)$ gauge theory. Above this temperature $\textrm{Re}[V]$ was found to become weaker than at $T<T_C$ and that it asymptotes to a constant value. In particular the real part never exceeds its $T=0$ values. This behavior is in agreement with the presence of screening between the color sources. Screening is predicted from HTL perturbation theory, from simulations in effective field theory \cite{Hart:2000ha}
and has been observed in high temperature Abelian gauge theories \cite{Buchholz:1978ek}(where also the definition of the Debye screening mass does not suffer from the technical difficulty of gauge dependence as in QCD). Let us note that the thermodynamic free energies, obtained on the same quenched lattices, also asymptote to constant values at large separation distances of the color anti-color sources. I.e. they show qualitative consistency with the behavior of the real-part part of the extracted potential. And while the determination of the imaginary component of the potential is technically more difficult than that of its real part (see \cref{sec:DefPot}), the studies concluded that an imaginary part is present and found that within the significant uncertainties it agreed with model predictions \cite{Burnier:2016mxc}.

In QCD with dynamical quarks, we highlight two studies from the literature. The first \cite{Burnier:2015tda} used a legacy lattice QCD formulation (asqtad action) 
with a relatively heavy pion mass, the second one \cite{Bala:2021fkm} is based on state-of-the-art lattices using the highly improved staggered quark action~\cite{Follana:2006rc}. 
Moreover, both QCD studies used the Symanzik-improved L\"uscher-Weisz action in contrast to the quenched studies.
Using Bayesian inference for spectral function reconstruction, the older of the two concluded that the limit of \cref{eq:DefPot} is well defined, arising from a dominant low-lying spectral structure in the spectral function. 
However, as the more recent study \cite{Bala:2021fkm} has shown that the positivity condition for applicability of the Bayesian inference is not met when improved gauge actions are used, these older results have uncontrolled systematic errors that were not known at the time of their publication.
The newer study relied on different analysis methods and turned out less conclusive. 
While it always identifies a dominant lowest lying spectral structure using different methods for the spectral function reconstruction, its position and width depend on the analysis method. 
In particular, the model fits using a Gaussian to regularize the Lorentzian would be inconsistent with a complex time-independent potential if they were carelessly taken at face value. 
Yet as the first few cumulants of the respective model fits are largely independent of the exact fit form (within reasonable constraints) \cite{Bazavov:2023dci}, this does not invalidate the identification of a lowest lying spectral feature in line with a complex time-independent potential.

In the older work both the position and width of the lowest lying structure were found to depend on temperature consistent with gradually screened behavior of the real-part of the potential and a rise of its imaginary part with both distance and temperature. The more recent study on the other hand found indications that the position of the lowest lying peak remains unchanged with temperature, while it is the width of this spectral structure, which grows significantly with temperature. Note that the majority of methods used in that study conclude on the temperature independence of $\textrm{Re}[V]$, while one of them also identified a screened behavior. Interestingly the thermodynamic free energies from the lattice ensembles used in both the older and newer study each show screening \cite{Bazavov:2018wmo}. 

In the light of the unscreened weak-coupling results in the relevant hierarchy $gT < {1}/{r} < T$, a qualitative mismatch between the free energies and the real-part of the potential seems plausible, as the imaginary part of the potential could lead to decoherence of the quarkonium wavefunction \cite{Kajimoto:2017rel} and thus to screened free energies, even as the binding of the color sources, mediated by the real-part, were not screened. 
The equation of state on the same HISQ ensembles shows a transition into a QGP-like phase \cite{Bazavov:2017dsy}, where medium-dressed light quarks and gluons are the relevant d.o.f. instead of hadrons. 
While this indicates that the interquark interaction is indeed much weaker in that phase, the equation of state does not get close to the limit of free quarks and gluons within the range of phenomenologically relevant temperatures. 
Thus, one cannot say a priori whether or not the in-medium interaction strength is insufficient for quarkonium binding.

In light of this discrepancy between the quenched QCD and full QCD sector, the present study sets out to re-investigate the interactions of color sources in quenched QCD, based not only on Bayesian inference, but also on those methods that have been deployed in the more recent paper. Our goal is to both reconfirm the previous results in quenched QCD, which were based on the Bayesian Reconstruction (BR) method for spectral function reconstruction and obtain a better understanding of how other methods compare to this established method, in a system, where an apples to apples comparison among methods is possible.

This paper is organised as follows: In \cref{sec:DefPot} we discuss the definition of the potential $V(r)$ from spectral functions. In \cref{sec:methodsmock} we introduce the different methods for spectral function reconstruction deployed in this study and carry out extensive mock data testing to explore their strengths and weaknesses. \Cref{sec:latticeres} showcases our results obtained from high-resolution isotropic and anisotropic quenched lattices. We conclude with a discussion in \cref{sec:conclusions}.

\section{Wilson Loop Potential and Spectral Structure}\label{sec:pot}
\label{sec:DefPot}

Let us retrace in more detail, how the potential defined via 
\begin{align}
    \lim_{t\to\infty} i\partial_t W_{\square}(t,r) = \lim_{t\to\infty} \Phi(t,r) W_{\square}(t,r)=  V(r) W_{\square}(t,r)\label{eq:limitPot}
\end{align}
can be related to the spectral structure encoded in the Wilson loop spectral functions in a fully non-perturbative fashion, as was established in a series of prior publications \cite{Burnier:2012az,Rothkopf:2011db,Rothkopf:2009pk}. We then discuss under which conditions we may deduce the existence of the potential $V(r)$ from an inspection of spectral structure.

By definition, the Wilson loop (in both continuum and on the lattice) exhibits a simple behavior under time reversal. As a complex quantity, it turns into its complex conjugate 
\begin{align}
    W_\square(-t,r)=W^*_\square(t,r)
\end{align}
This relation arises from the fact that $W_\square$ is defined from an exponentiation of Hermitean matrices multiplied with the imaginary unit and thus is valid non-perturbatively.

At first, let us assume that the limit in \cref{eq:DefPot} is well defined, i.e. that the potential $V(r)$ actually exists. We may then define a characteristic time scale $t_{\rm relax}$ beyond which the function $\Phi(t,r)$ relaxes to the time independent $V(r)$. Below we will confirm that $t_{\rm relax}$ indeed is connected to the region in time over which non-potential effects remain relevant for the evolution of the Wilson loop.

We may therefore \textit{without loss of generality} write $\Phi(t,r)=V(r)+\phi(t,r)$ with $\phi(t>t_{\rm relax},r)=0$\footnote{If we instead had an arbitrary relation  $\Phi(t,r)=F[V(r); t,r]$ with $F[V(r); t>t_{\rm relax},r]=V(r)$ we may always define the function $\phi(t,r)$ as $\Phi(t,r)=V(r)-(F[V(r);t,r]-V(r))\equiv V(r)-\phi(t,r)$.}. 
Since, as an SU(3) group element, the Wilson loop can only take on finite values, the imaginary part of of the potential $\textrm{Im}[V](r)<0$ must be negative. Again we note that no reference to weak coupling has been made so far.

Let us proceed to consider the fully general equation of motion of the Wilson loop
\begin{align}
    i\partial_t W_\square(t,r)=\Phi(t,r)W_\square(t,r),
\end{align}
whose solution may be written in the following form, taking into account the properties of the Wilson loop established above
\begin{align}
W_\square(t,r)={\rm exp}\Big[ -i\Big(& {\rm Re}[V](t)t + {\rm Re}[\sigma](t,r)\Big) -|{\rm Im}[V](r)|t+{\rm Im}[\sigma](t,r) \Big]\label{Eq:WLPotMod}.
\end{align}
We have introduced the quantity $\sigma(t,r)=\int_0^t\phi(t,r)dt$ making reference to the function $\phi(t,r)$ and which thus asymptotes to $\sigma_\infty(r)=\sigma(|t|>t_{\rm relax},r)=\int_0^\infty\phi(t,r)dt$.

Next we wish to connect $V(r)$ to the spectral function associated with the above solution of the Wilson loop equation of motion. To this end we turn to the spectral decomposition of the Wilson loop, which has been established non-perturbatively by taking the infinite mass limit of the point split heavy-meson correlation function \cite{Rothkopf:2009pk} in 
\begin{align}
    W_\square(t,r)=\int_{-\infty}^{\infty} d\omega \; \rho_\square(\omega,r)\,e^{i\omega t}\label{eq:specdecreal}~.
\end{align}
Combined with the definition of the potential \cref{eq:limitPot} we thus may relate the potential directly to the spectral function via
\begin{align}
V(r)=\lim_{t\to\infty}\frac{ \int_{-\infty}^\infty d\omega\; \omega\; e^{-i\omega t} \;\rho_\square(\omega,r)}{\int_{-\infty}^\infty d\omega \; e^{-i\omega t} \; \rho_\square(\omega,r)} \label{eq:realpot}.
\end{align}

As the spectral decomposition amounts to a simple Fourier transform, it can be readily inverted to yield
\begin{align}
\nonumber&\rho_\square(r,\omega)=\\
&\frac{1}{2\pi}\int_{-\infty}^\infty dt\; {\rm exp}\Big[ i \Big(\omega-{\rm Re}[V](r)\Big) t 
 -i{\rm Re}[\sigma](r,|t|)\mathrm{sign}(t) -|{\rm Im}[V](r)||t|+ {\rm Im}[\sigma](r,|t|)\Big]
\end{align}

Under the assumption that the function $\sigma(t,r)$ asymptotes to a constant for times $t>t_{\rm relax}$, we may decompose the integral into two parts, one over times smaller then $|t|<t_{\rm relax}$ and one over the whole time range
 \begin{align}
&\rho_\square(r,\omega)=\label{eq:decompspecwl}\\
 \nonumber&\frac{1}{2\pi}e^{{\rm Im}[\sigma_\infty](r)}\int_{-\infty}^\infty  dt\; {\rm exp}\Big[ i \Big(\omega-{\rm Re}[V](r)\Big)t -|{\rm Im}[V](r)||t|  -i{\rm Re}[\sigma_\infty](r)\mathrm{sign}(t) \Big]\\
\nonumber &+\frac{1}{2\pi}\int_{-t_{\rm relax}}^{t_{\rm relax}}  dt\; {\rm exp}\Big[ i \Big(\omega-{\rm Re}[V](r)\Big) t  -|{\rm Im}[V](r)||t|\Big] \times \\
 \nonumber &\qquad \qquad \qquad \Bigg(e^{ -i{\rm Re}[\sigma](r,|t|)\mathrm{sign}(t) + {\rm Im}[\sigma](r,|t|)}-e^{-i{\rm Re}[\sigma_\infty](r)\mathrm{sign}(t) + {\rm Im}[\sigma_\infty](r)}\Bigg).
\end{align}

This decomposition now reveals how the potential emerges in the spectral structure. The integral involving integration over the whole time regime can be carried out analytically and becomes a skewed Lorentzian peak located at the position of the real-part of the potential, its width being related to imaginary part. Around the maximum of that peak $(\omega-\textrm{Re}[V])t_{\rm relax}\ll1$ we can expand the remaining integral over the finite time regime in a Taylor series leading to a background for the potential peak
\begin{align}
\nonumber&\rho_\square^{\rm pot}(r,\omega)=\\
\nonumber&\frac{1}{\pi}e^{{\rm Im} \sigma_\infty](r)}\frac{|{\rm Im}[V](r)|{\rm cos}[{\rm Re}[\sigma_\infty](r)]-({\rm Re}[V](r)-\omega){\rm sin}[{\rm Re}[\sigma_\infty](r)]}{ {\rm Im}[V](r)^2+({\rm Re}[V](r)-\omega)^2}+\\
&c_0(r)+c_1(r)t_{Q\bar Q}({\rm Re}[V](r)-\omega)+c_2(r)t_{Q\bar Q}^2({\rm Re}[V](r)-\omega)^2+\cdots \label{eq:PotFitFunc}
\end{align}
If one reinserts $\rho_\square^{\rm pot}$ into the relation \cref{eq:realpot} one finds that only the pole in the skewed Lorentzian matters and indeed a time independent complex potential $V(r)={\rm Re}[V](r)-i{\rm Im}[V]$ ensues.

In practice we attempt to extract a spectral function from lattice QCD simulations, inspect it for a dominant low-lying structure and aim to understand, whether or not it leads to a well defined time-independent potential $V(r)$. This requires to reverse the above argumentative chain.

Let us start with shape of such a dominant spectral peak. One may ask what would happen if a distribution without the extended tails of a skewed Lorentzian arises? Taking e.g. a Gaussian $\rho_{\rm G}(\omega,r)=A(r){\rm exp}[-(\omega-m(r))^2/g^2(r)]$ and inserting into \cref{eq:realpot} one finds that it does not lead to a well defined potential, as a time dependent and even divergent imaginary part ensues $V_{G}(r,t)=m(r)-ig(r)^2t$. This result shows that tail structures play an important role in establishing a time independent potential. 

In order to get insight into the tail structures, we can use the HTL spectral functions as a concrete and non-trivial example. While the real-time evolution of the Wilson loop does not place strict restrictions on the tail structure of the potential peak (i.e. the integral of \cref{eq:specdecreal} is well defined for a skewed Lorentzian), the Euclidean counterpart of the spectral decomposition
\begin{align}
    W_\square(\tau,r)=\int_{-\infty}^\infty d\omega\, \rho_\square(\omega,r)\,e^{-\omega\tau}\label{eq:euclWL}
\end{align}
requires the spectral function to be exponentially suppressed at negative frequencies, incompatible with a naive Lorentzian peak. As shown in Fig.\ref{fig:wspecs_HTL} the HTL Wilson loop spectral function computed from the real-time correlator, indeed shows exactly this exponential suppression, which cuts off the tails of the dominant skewed Lorentzian potential peak. Note that this cut-off spectrum is compatible with our non-perturbatively derived decomposition in \cref{eq:decompspecwl} and tells us that it is the non-potential effects which affect the low frequency spectral structure. It is important to note that this modification of the Lorentzian peak at negative frequencies \textit{does not} introduce an additional pole at small frequencies that would interfere with the late time limit. I.e. in HTL the Lorentzian potential peak is exponentially cut off but remains the relevant structure for the asymptotically late-time evolution of the charge-anticharge pair. 

The example of the HTL spectral function is encouraging. It supports the conclusion that if one finds a dominant lowest lying spectral peak in a lattice spectral function, with a skewed Lorentzian behavior around its maximum, this peak is likely to encode the asymptotic late real-time behavior of the real-time correlator, even if the spectral function is exponentially suppressed at negative frequencies. 

Accurately determining the existence of a low lying, well defined spectral peak and reconstructing its shape with high fidelity from simulated Euclidean lattice data is therefore of paramount importance to clarify the existence of a time-independent potential. We will discuss and benchmark relevant methods, recently deployed for this task in the subsequent section.

\section{Extraction Methods and Mock Data Tests}
\label{sec:methodsmock}

\begin{figure}[t]
    \includegraphics[scale=.4]{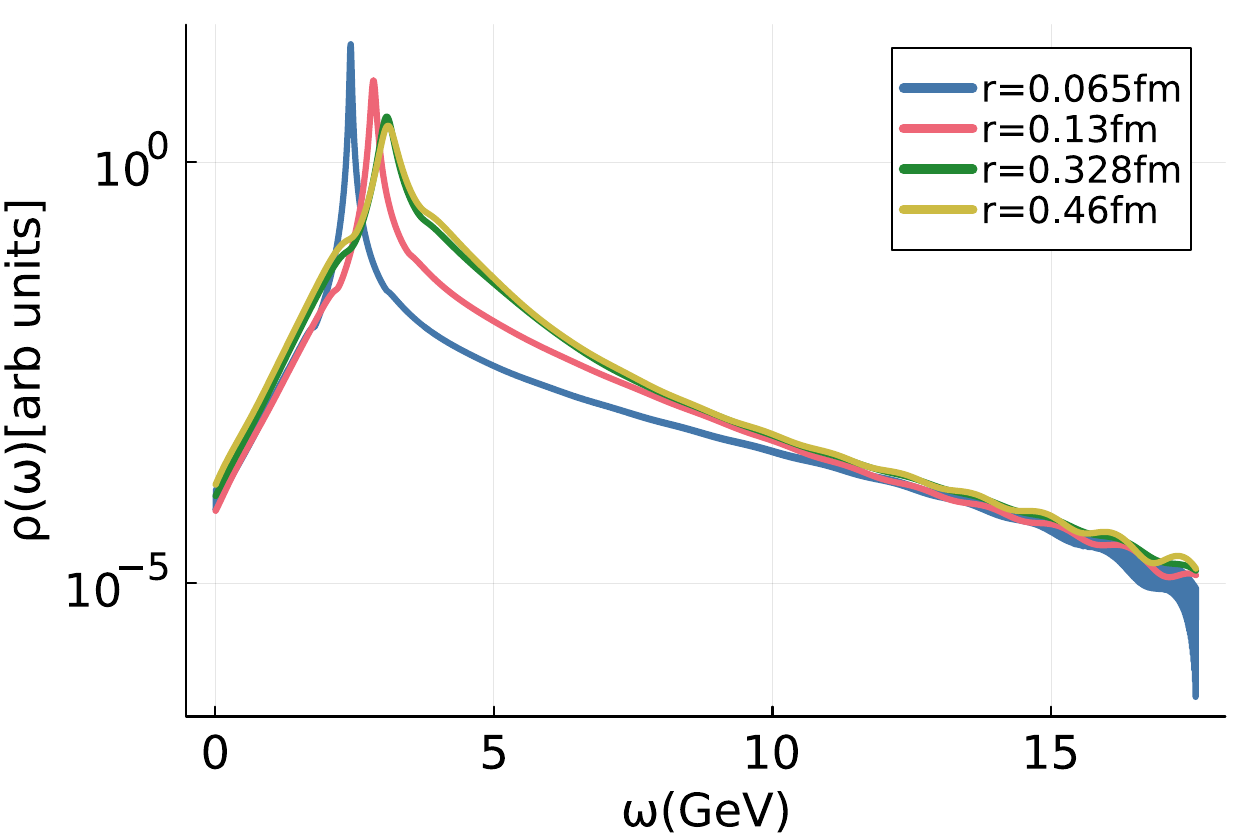}
    \includegraphics[scale=.4]{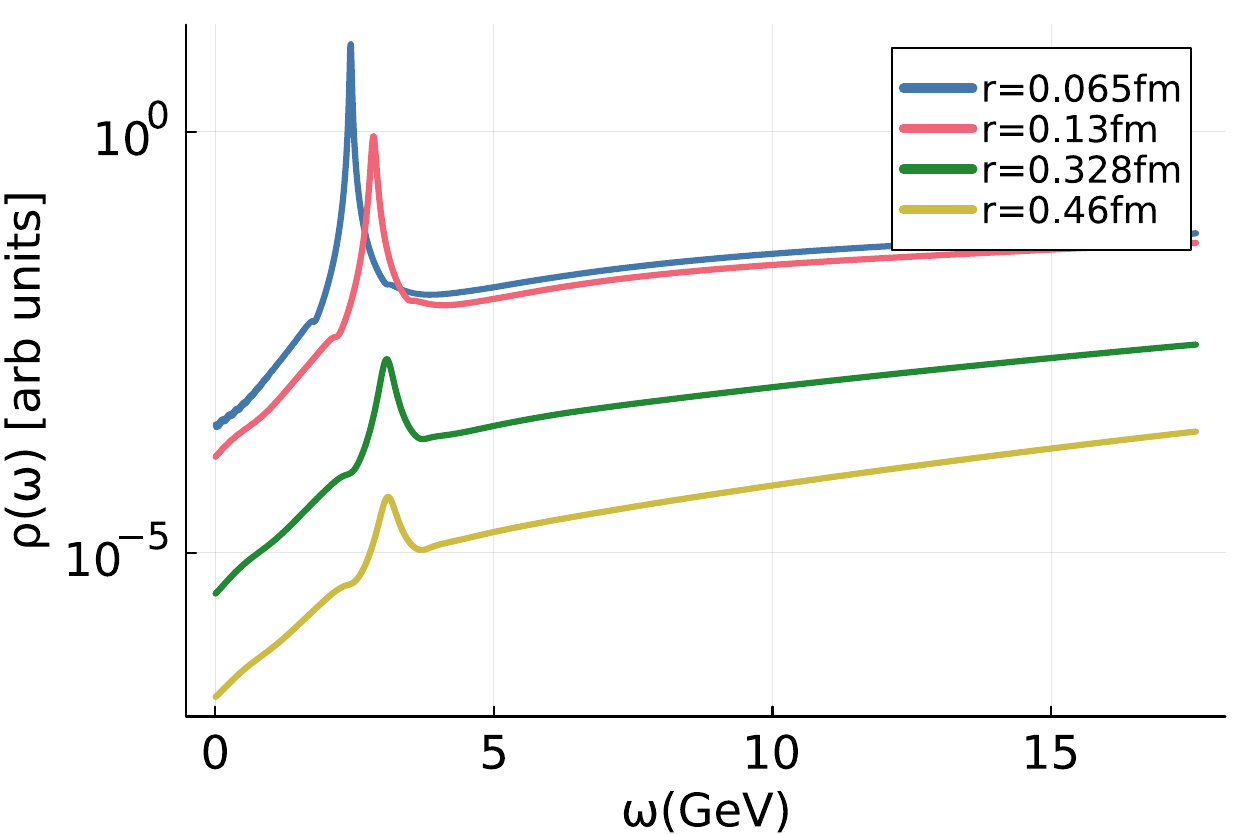}
    \caption{Semi-analytically computed spectral function of the (top) Coulomb-gauge Wilson line correlator and (bottom) Wilson loop in HTL perturbation theory at $T=2.33T_C$. The four curves denote spectra at different spatial separation distances. Note that the peak encoding the potential has the same position and width while different amplitudes for Wilson loop and lines. The Wilson loop spectrum contains a sizable UV continuum induced by cusp divergences.}
    \label{fig:wspecs_HTL}
\end{figure}

In this section, we will compare the quantitative performance of different extraction methods for spectral information from Euclidean correlators. As non-trivial testing ground we deploy the analytically computed Wilson loop and Coulomb-gauge Wilson line correlators from hard-thermal-loop perturbation theory \cite{Burnier:2013fca} at $T=2.33T_C$, whose spectral functions are known. We introduce a momentum cutoff to mimic the presence of a finite grid and evaluate the Euclidean data at the same number of discrete imaginary time steps accessible on the lattice. 
We however, recognise that this may not fully mimic the effects of the lattice cutoff on the spectrum, namely, that there is a larger density of modes near the lattice cutoff that could be attributed to scattering states. Nonetheless, using the HTL data for benchmarking our methods does provide us with non-trivial structure in the Wilson loop spectral function at high frequency coming from the large background extending far into the UV along with cusp divergences that introduce kinks which would present a challenge for our reconstruction methods.
This ideal data is salted with Gaussian noise with constant relative errors similar (or larger) than those encountered in actual simulations, providing us with a realistic mock input data set. This mock data is provided to four different approaches previously deployed in the study of the in-medium static quark potential in the literature: Bayesian inference (BR method) \cite{Burnier:2013nla}, Pad\'e interpolation \cite{Schlessinger:1968}, the HTL-inspired fit \cite{Bala:2019cqu,Bala:2019boe} and generalized Gaussian fits \cite{Larsen:2019bwy}.
In the top panel of \cref{fig:wspecs_HTL} we show four spectral functions $\rho_{||}^{\rm HTL}(\omega,r)$ of the Coulomb-gauge Wilson line correlator at different spatial distances $r\in[0.065\ldots0.46]$fm. A dominant skew-Lorentzian peak encoding the potential is located on top of a weak shoulder structure, which decays exponentially at small and algebraically at large frequencies. In case of the Wilson loop spectral function $\rho_\square^{\rm HTL}(\omega,r)$ shown in the bottom panel, we find peaks at the same position and with the same width as in $\rho_{||}^{\rm HTL}$ but with amplitudes significantly suppressed with increasing distance. In addition the peak is embedded into a substantial background, which extends far into the UV region at high frequencies at a significant fraction of the peak amplitude \cite{Burnier:2013fca}. I.e. the different correlators encode the same potential but very different non-potential effects.

\begin{figure}[ht]
    \includegraphics[scale=0.4]{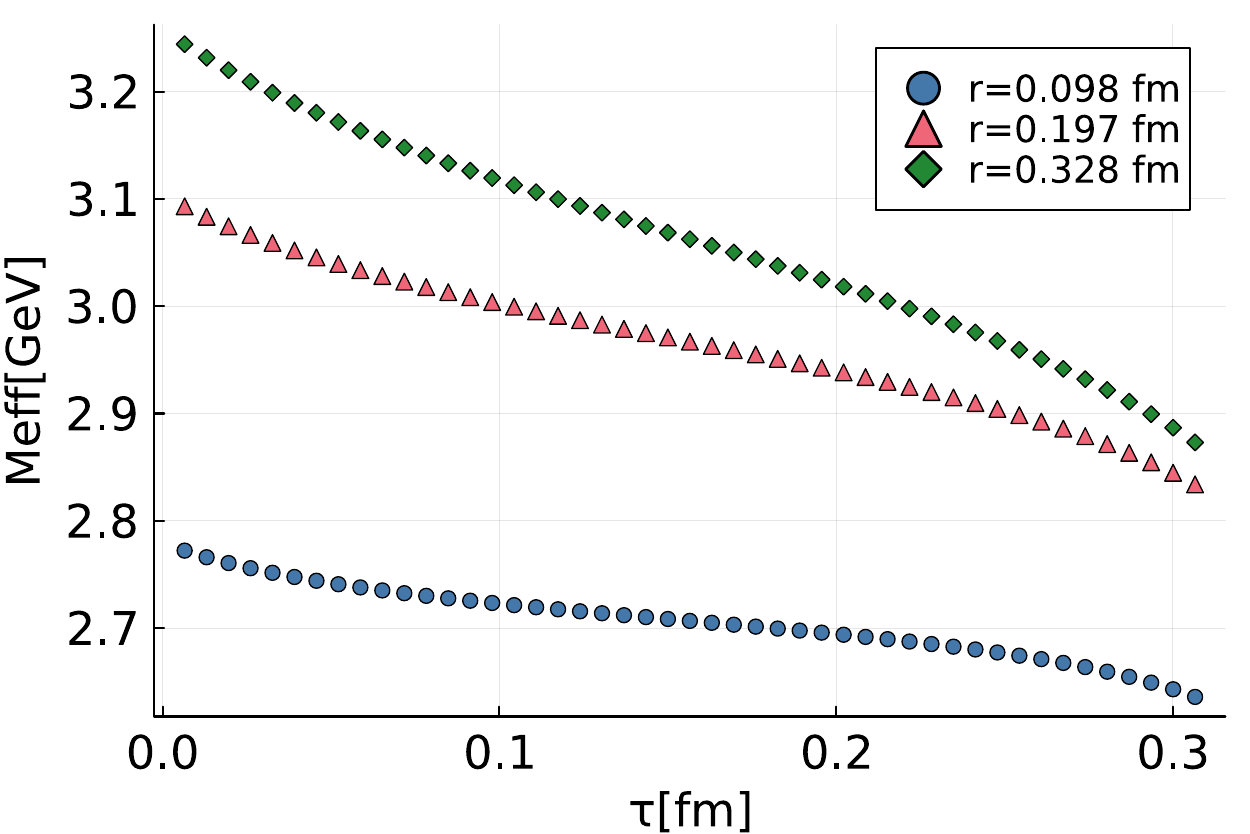}
    \includegraphics[scale=0.4]{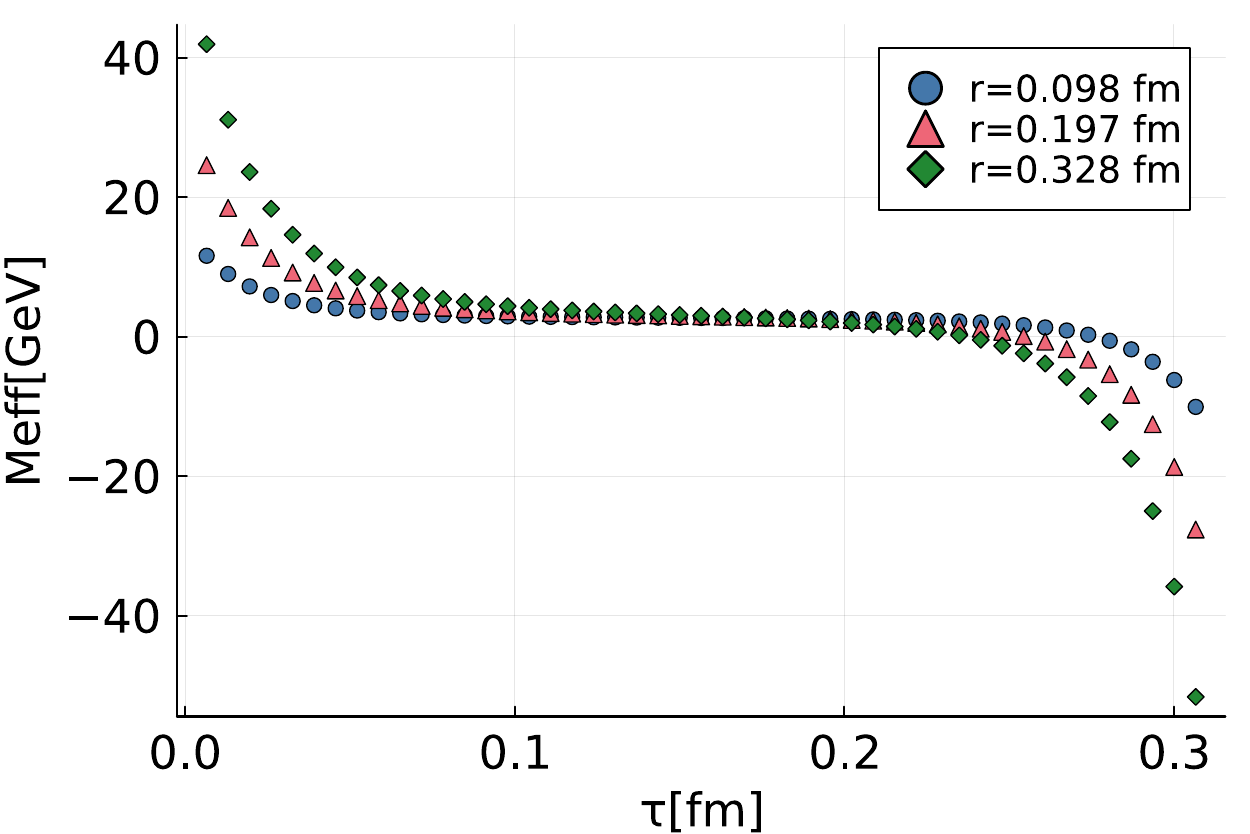}
\caption{Effective masses from the (top) Coulomb-gauge Wilson line correlator and (bottom) Wilson loop at $T=2.33T_C$ in HTL perturbation theory.}
    \label{fig:meffs_HTL}
\end{figure}

To obtain a first insight into the spectral structure from euclidean correlators we inspect their effective masses, evaluated at discrete imaginary times $\tau_i$ with spacing $a$, which are defined as 
\begin{align}
m_{\rm eff}(\tau_i,r)=\frac{1}{a}{\rm log}\Big[\frac{W(\tau_i+a,r)}{W(\tau_i,r)}\Big]
\end{align}
Their values from the HTL Wilson lines and Wilson loop at three spatial distances are given in the top and bottom plot of \cref{fig:meffs_HTL} respectively. One sees that the dominant spectral peak in $\rho_{||}$ in the absence of a strong UV contribution leads to an $m_{\rm eff}$ that is almost linear around the middle of the imaginary time interval $\tau=1/(2T)$ with some curvature at early and late $\tau$. Contrast this to the Wilson loops, where the substantial UV contribution leads to a strong curvature at small $\tau$, which overshadows the linear behavior at intermediate imaginary times, before another region of strong curvature from the low frequency tail of the spectrum induces a downward trend at $\tau\approx1/T$. 

\begin{figure}[ht]
    \includegraphics[scale=0.4]{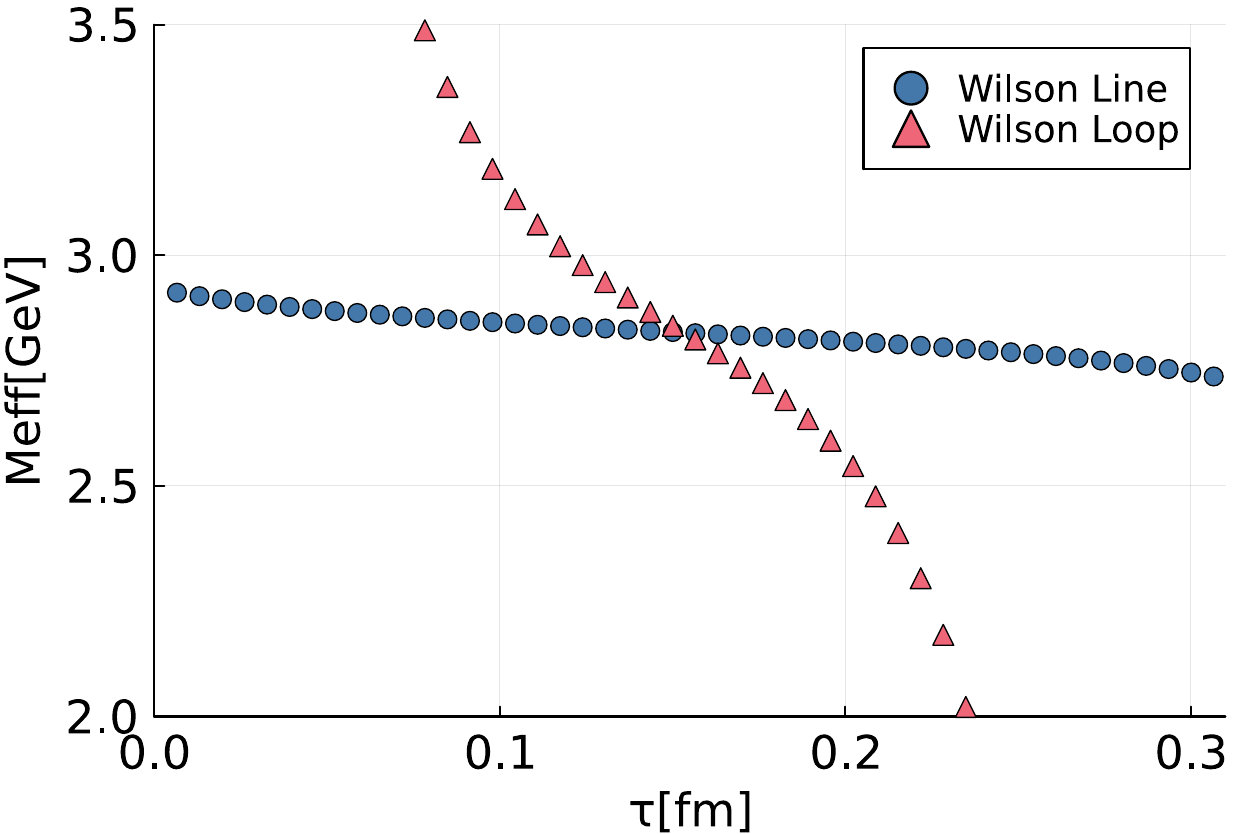}
    \caption{A comparison of HTL effective masses of the Wilson loop (triangle) and Coulomb-gauge Wilson lines correlator at $T=2.33T_C$ evaluated at a separation distance of $r=0.131$fm. Both encode the same spectral peak position and width related to the common value of the complex static quark potential. Their behaviors constitute limiting test cases for the effective masses observed in actual lattice QCD simulations.}
    \label{fig:cmpHTLmeff}
\end{figure}

In \cref{fig:cmpHTLmeff} we show a direct comparison of the effective masses of the Wilson loop and Coulomb-gauge Wilson lines at a single $r=0.131$fm to illustrate that their behavior resembles limiting cases for the behavior of effective masses observed in actual lattice simulations. 
Nonetheless, it is not a priori clear whether or not the effects of the modes near the lattice cutoff can be mimicked appropriately by the UV part of the HTL Wilson loop spectrum.

\subsection{An ill-posed inverse problem}

In order to extract the spectral function from Euclidean correlators, we need to invert the discretized version of \cref{eq:euclWL}
\begin{align}
W(\tau_i,r)=\sum_{l=1}^{N_\omega}\Delta\omega_l {\rm exp}\big[-\omega_l\tau_i\big]\rho(\omega_l,r). \label{eq:specdecdiscr}
\end{align}
where a relatively small number of $N_\tau \sim \order{10-100}$ data-points inform us about the spectral function discretized along $N_\omega\gg N_\tau$ frequency bins with resolution $\Delta\omega_l$.

This task of estimating a finely resolved $\rho$ from sparse $W_i=W(\tau_i)$ and noisy $\Delta W>0$ data-sets is a classic ill-posed inverse problem. Since the data and spectral function are related by an integral convolution over an exponential the task is akin to inverting a Laplace transform. There exist an infinite number of $\rho$'s that reproduce $W$ within its statistical uncertainties. In addition, if we were to naively invert \cref{eq:specdecdiscr}, due to the exponential form of the integral kernel, the outcome would be highly sensitive to the uncertainty in the input data.

As discussed in more detail in \cite{Rothkopf:2022ctl} a regularization of the inverse problem is needed to give it meaning. One may utilize ab-initio knowledge about the spectral function, such as that it can only take on positive values. 
For the lattice correlators with improved gauge action, the positivity does not hold, which can be seen explicitly for small $r$ and small $\tau$ \cite{Bala:2021fkm}. This reduces the available functional space of $\rho$'s to fit the data but does not yet give a unique result. Thus additional assumptions, such a smoothness or certain analyticity properties are deployed in the literature. 
Yet again, smoothness may be violated in the lattice correlators at small $\tau$ close to the inverse lattice cutoff.

Another option is to model the spectral functions based on insight obtained from simple models and a careful inspection of the correlator data. If the spectral function can be parameterized by a set of few parameters, these can be uniquely determined via a maximum likelihood approach conditioned on the modeling hypotheses.

In each case the dependence of the result of the spectral function extraction must be tested on its dependence on the data uncertainty (statistical error) and the model hypotheses (systematic errors), to establish the full uncertainty budget. 

\subsection{Method I: Bayesian Inference}

Bayesian inference of spectral functions builds upon Bayes theorem 
\begin{align}
    P[\rho|D,I]\propto P[D|\rho,I]P[\rho|I] = {\rm exp}[-L+\alpha S_{\rm BR}],
\end{align}
to introduce prior domain knowledge on the spectral function to regularize the inverse problem. The posterior $P[\rho|D,I]$ describes the probability of a test function $\rho$ to be the correct spectral function given correlator data $D_i=W_\Box(\tau_i,r)$ and prior information $I$. It is expressed as product of the Gaussian likelihood $P[D|\rho,I]$, which describes the distribution of the sampled data and connects the estimated mean of the correlator to the spectral function via \cref{eq:euclWL},
\begin{equation}
L=\frac{1}{2}\sum_{i,j=1}^{N_\tau}(D_i-W[\rho]_i)C_{ij}^{-1}(D_j-W[\rho]_j).
\end{equation}
where $C_{ij}$ denotes the covariance matrix w.r.t. the mean and $ W[\rho]_i=\int_{-\infty}^\infty d\omega\, \rho(\omega,r)\,e^{-\omega\tau}$~. It is the prior probability $P[\rho|I]$ that acts as the regulator. In this study we choose the BR prior, related to the gamma-distribution
\begin{align}
S_{\rm BR}=\int d\omega \big( 1- \frac{\rho(\omega)}{m(\omega)} + {\rm log}\big[ \frac{\rho(\omega)}{m(\omega)} \big]\big),
\end{align}
which enforces the positivity and smoothness of the spectral functions, while being the weakest regulator on the market. It therefore allows information encoded in the data to manifest easily in the reconstructed $\rho$, but at the same time is more susceptible to ringing artifacts compared to stronger regulators. The prior is parameterized by a default model $m(\omega_l)$, which encodes its mode and we marginalize apriori over the width of the prior assuming full ignorance (for more details see \cite{Burnier:2013nla}).

The explicit dependence of the posterior on the data and prior assumptions allows to test the uncertainties in each of them by carrying out a resampling (Jackknife) analysis and by repeating the analysis for different choices of the default model $m(\omega)$. For a more detailed discussion of Bayesian spectral function reconstruction see \cite{Burnier:2013nla}.

\begin{figure}[ht]
    \includegraphics[scale=0.4]{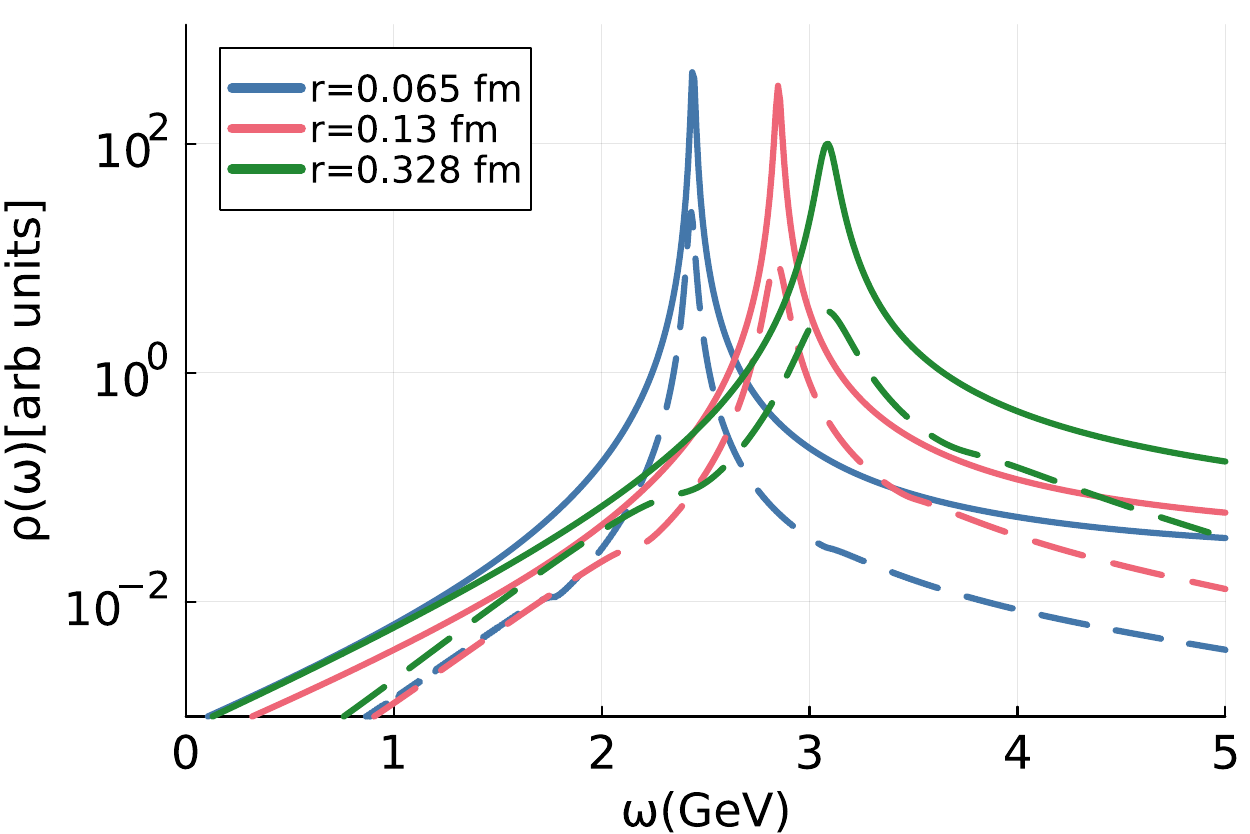} 
    \includegraphics[scale=0.4]{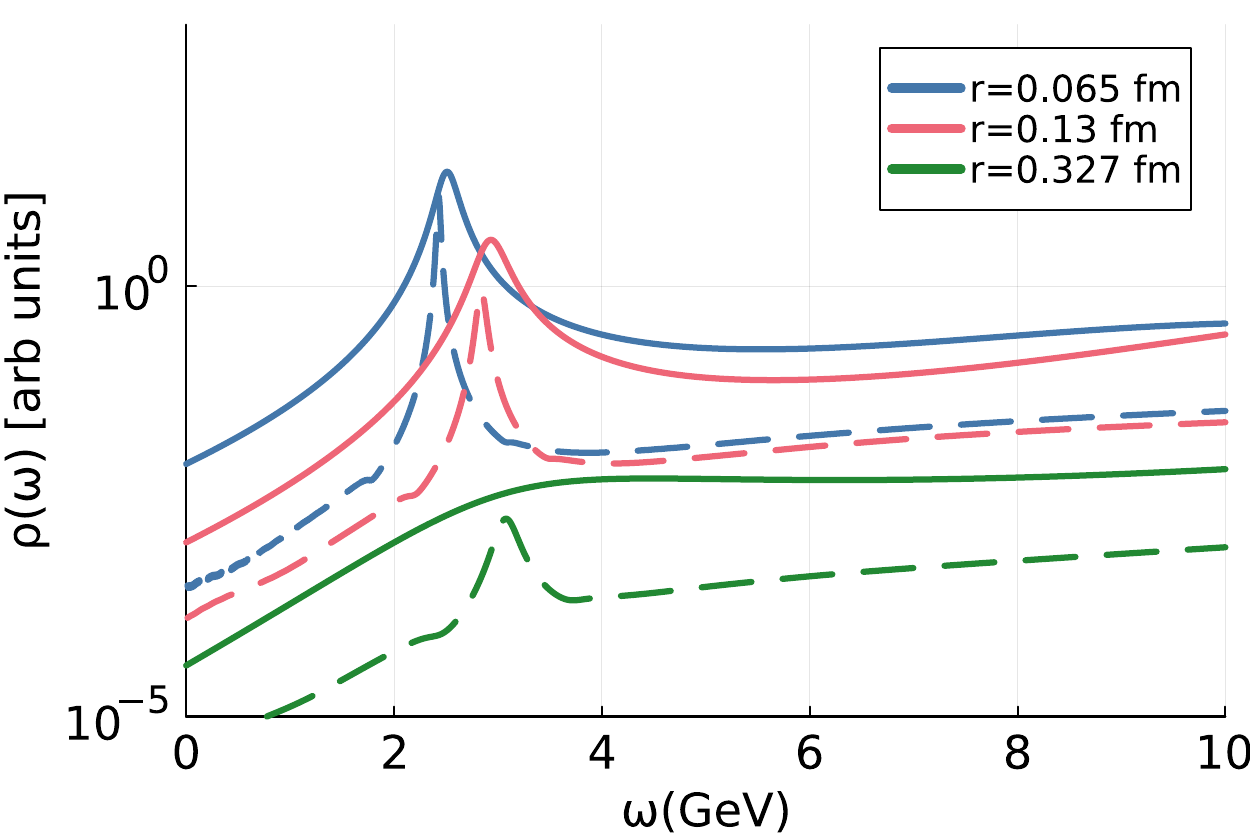}
\caption{Bayesian inference reconstructed (solid line) and semi-analytical (dashed line) spectral functions of the (top) Coulomb-gauge Wilson line correlator $\rho_{||}$ and (bottom) Wilson loop $\rho_\square$ in HTL perturbation based on $22$ datapoints with $\Delta D/D=10^{-2}$. The three curves each denote spectra at different spatial separation distances.}
\label{fig:HTLBRdD2spec}
\end{figure}

We carry out the Bayesian spectral function reconstruction on the raw correlator data discretized along $N_\tau=24$ datapoints representative for the $N_\tau$ accessible on the lattice. We discard the first and last point, as those are also discarded in lattice data due to the presence of divergent contributions. We assign constant relative errors $\Delta D/D=\kappa$ with $\kappa=10^{-2}$ or $\kappa=10^{-3}$ and will show in the next section that $\kappa=10^{-3}$ is actually achieved in our lattice data.

\begin{figure}[!ht]
    \includegraphics[scale=0.4]{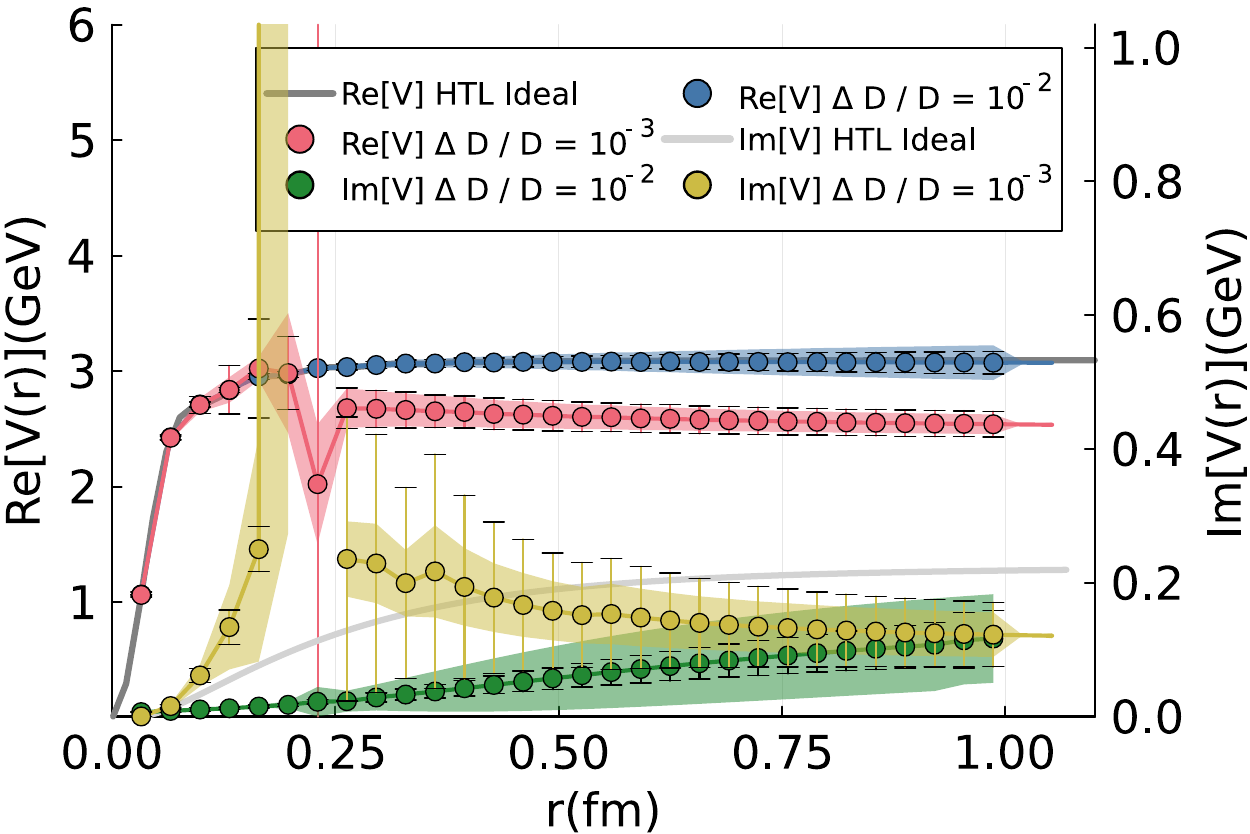}
    \includegraphics[scale=0.4]{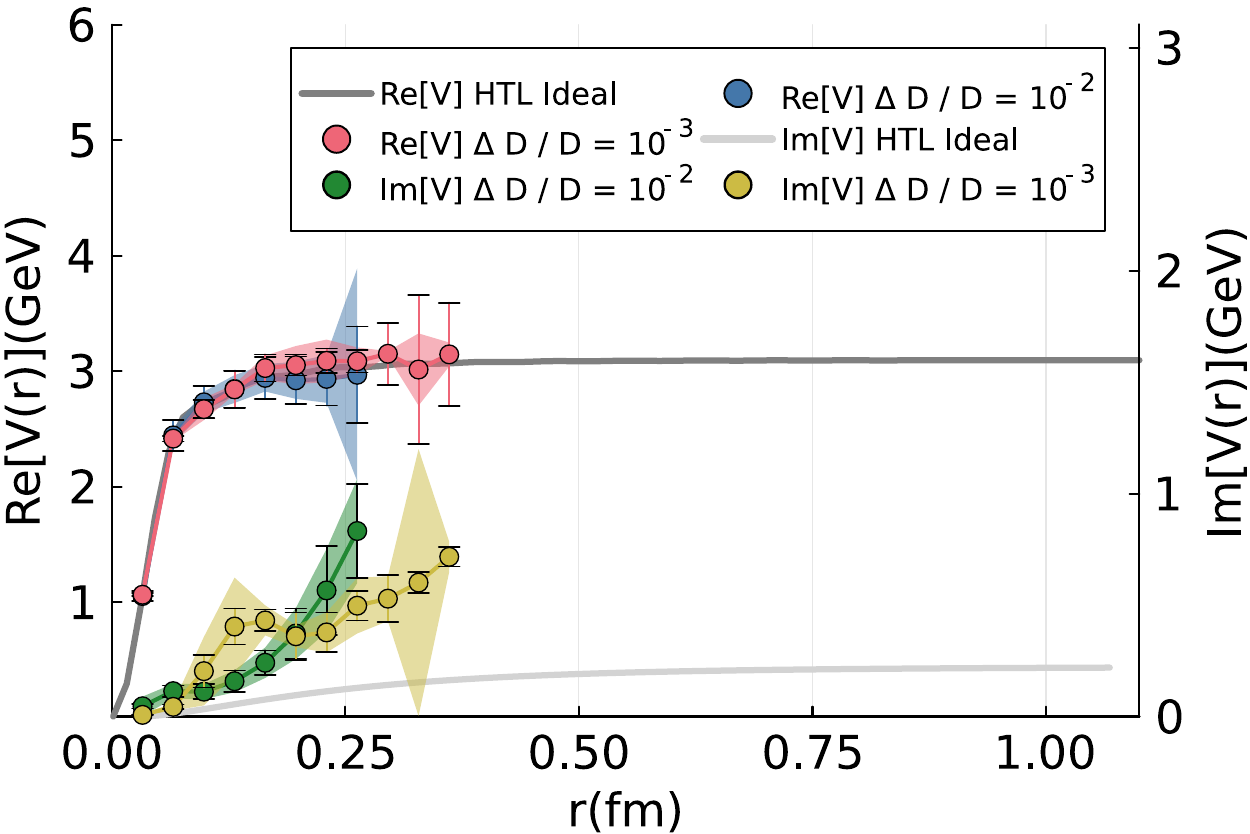}
    \caption{Bayesian inference reconstructed (colored symbols) and analytic (solid line) values for the real- and imaginary part of the complex static quark potential from the (top) Coulomb-gauge Wilson line correlator $V_{||}(r)$ and (bottom) Wilson loop $V(r)$ in HTL perturbation based on $22$ datapoints with $\Delta D/D=10^{-2}$ and $\Delta D/D=10^{-3}$. Note the onset of ringing in the top panel (red, yellow) manifests in a  jump of $\textrm{Re}[V]$.} 
    \label{fig:HTLBRdD3pot}
\end{figure}

The reconstruction with $\Delta D/D=10^{-2}$ from $W_{||}$ manages to locate the peak position already excellently (see \cref{fig:HTLBRdD2spec}), since a single peak dominates the spectrum. For $W_\square$ the position of the peak is less accurately determined (see \cref{fig:HTLBRdD3pot}). Note that the true position of the peak from the Bayesian reconstruction is found to \textit{be approached from above}. It is the structures that exists above the Lorentzian peak which pull the reconstructed peak to higher frequencies. This is an important finding, as it tells us that even in the presence of large non-potential effects the true peak position is likely lying at or slightly below the reconstructed value.

Interestingly the width of the potential peak is underestimated when reconstructed from $W_{||}$ and overestimated when using $W_\square$. Thus the Wilson loop and line results bracket the correct imaginary part.

\begin{figure}[!ht]
    \includegraphics[scale=0.4]{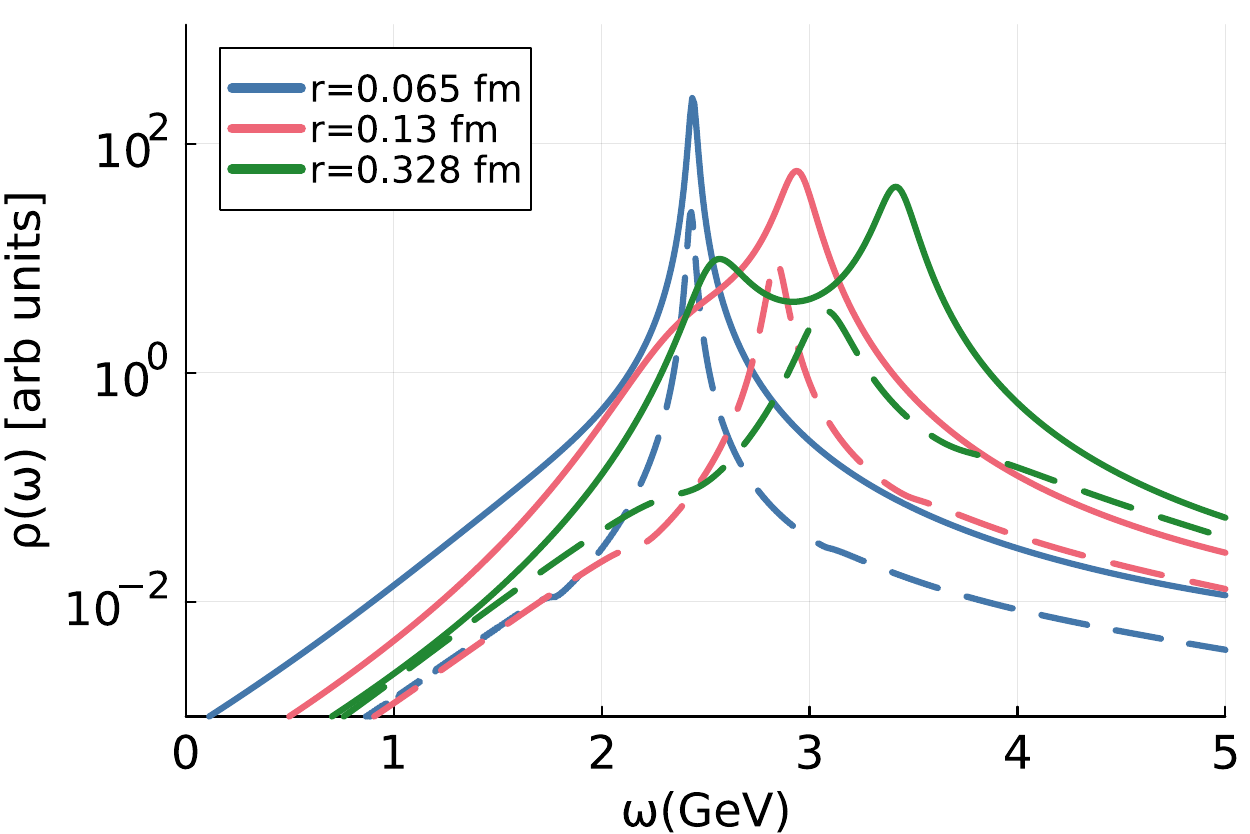}
    \includegraphics[scale=0.4]{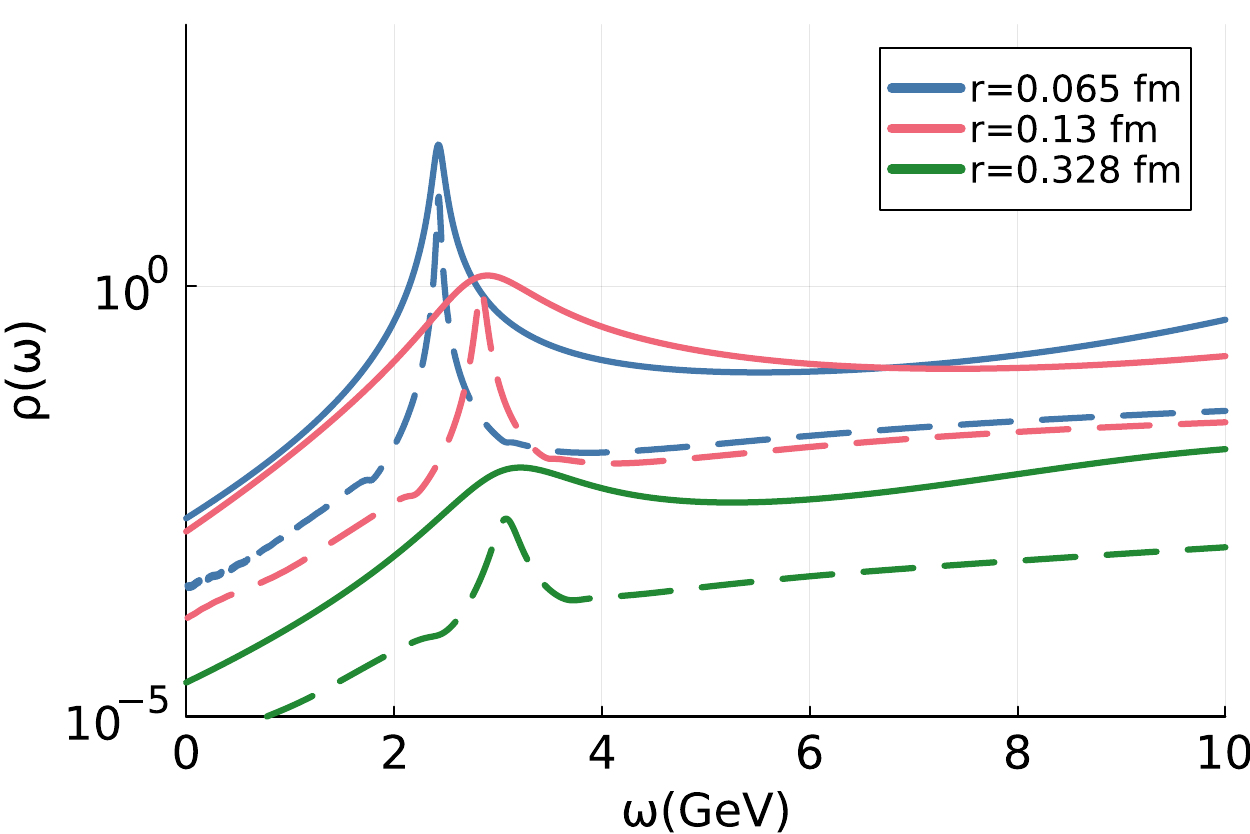}
    \caption{Bayesian inference reconstructed (solid line) and semi-analytical (dashed line) spectral functions of the (top) Coulomb-gauge Wilson line correlator $\rho_{||}$ and (bottom) Wilson loop $\rho_\square$ in HTL perturbation based on $22$ datapoints with $\Delta D/D=10^{-3}$. The three curves each denote spectra at different spatial separation distances.}\label{fig:HTLBRdD3spec}
\end{figure}

Let us take a look at the reconstruction results when the errors on the input data is reduced to $\Delta D/D=10^{-3}$ in \cref{fig:HTLBRdD3spec}. The Wilson loop reconstructions in the bottom panel show a significant improvement compared to the $\Delta D/D=10^{-2}$ case. In the top panel we can also see an improvement in the reconstruction from $W_{||}$ at small distances $r$ but find the occurrence of ringing at $r=0.327$fm. The true single peak structure is split up into two structures, one lying below, one above the true peak. Reducing the errors on the input data without at the same time increasing the number of datapoints thus may lead the reconstruction method to favor such a multipeak structure. If we were to take the position of the lowest structure as indicative of $\textrm{Re}[V]$ it would induce a significant jump between $r=0.13fm$ and $r=0.327$fm (see \cref{fig:HTLBRdD3pot}). 

We learn from the mock analysis that for the Bayesian reconstruction method when studying lattice data, we need to test for ringing artifacts. To this end we will repeat the reconstructions at different levels of $\Delta D/D$ and check whether at some point single peaks split into multiple structures. Let us note at this point that these crosschecks show that the lattice data analysis presented in the next section does not suffer from ringing.

\begin{figure}[!ht]
    \includegraphics[scale=0.4]{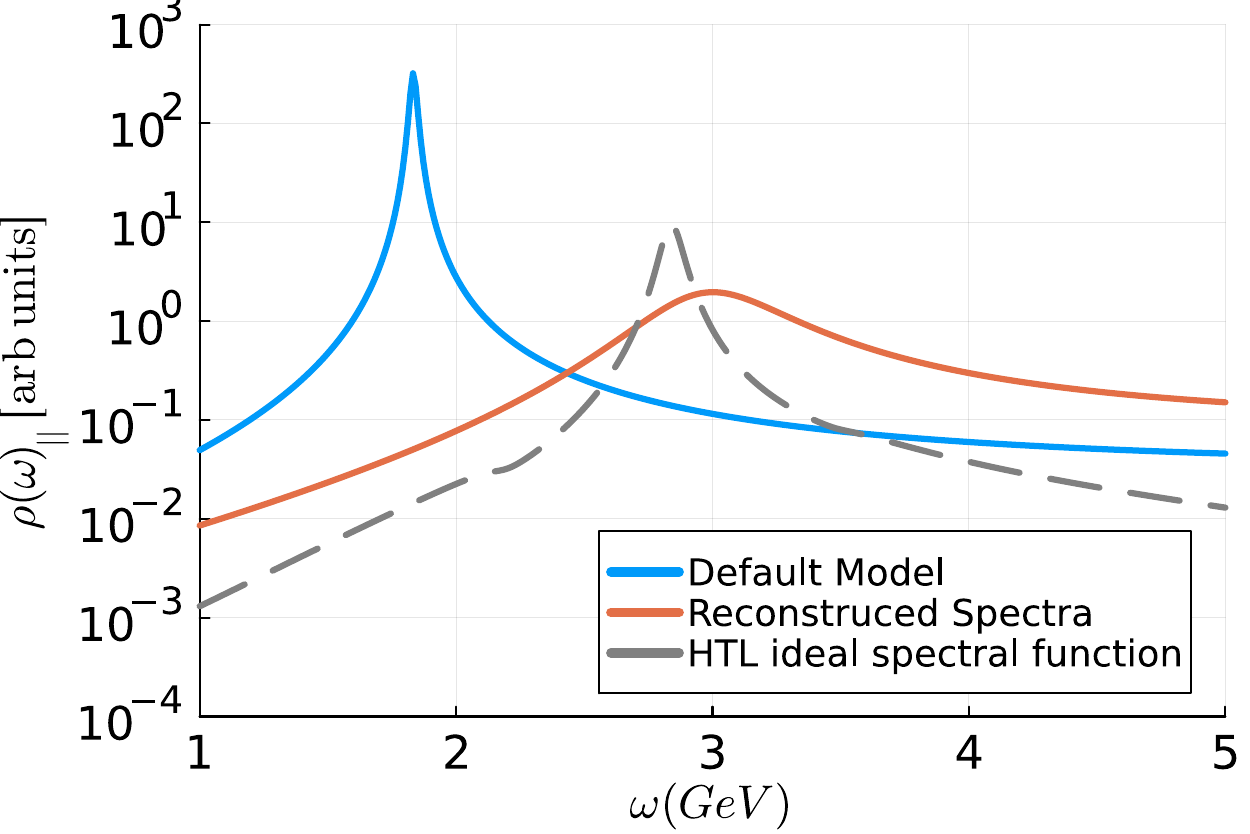}
    \includegraphics[scale=0.4]{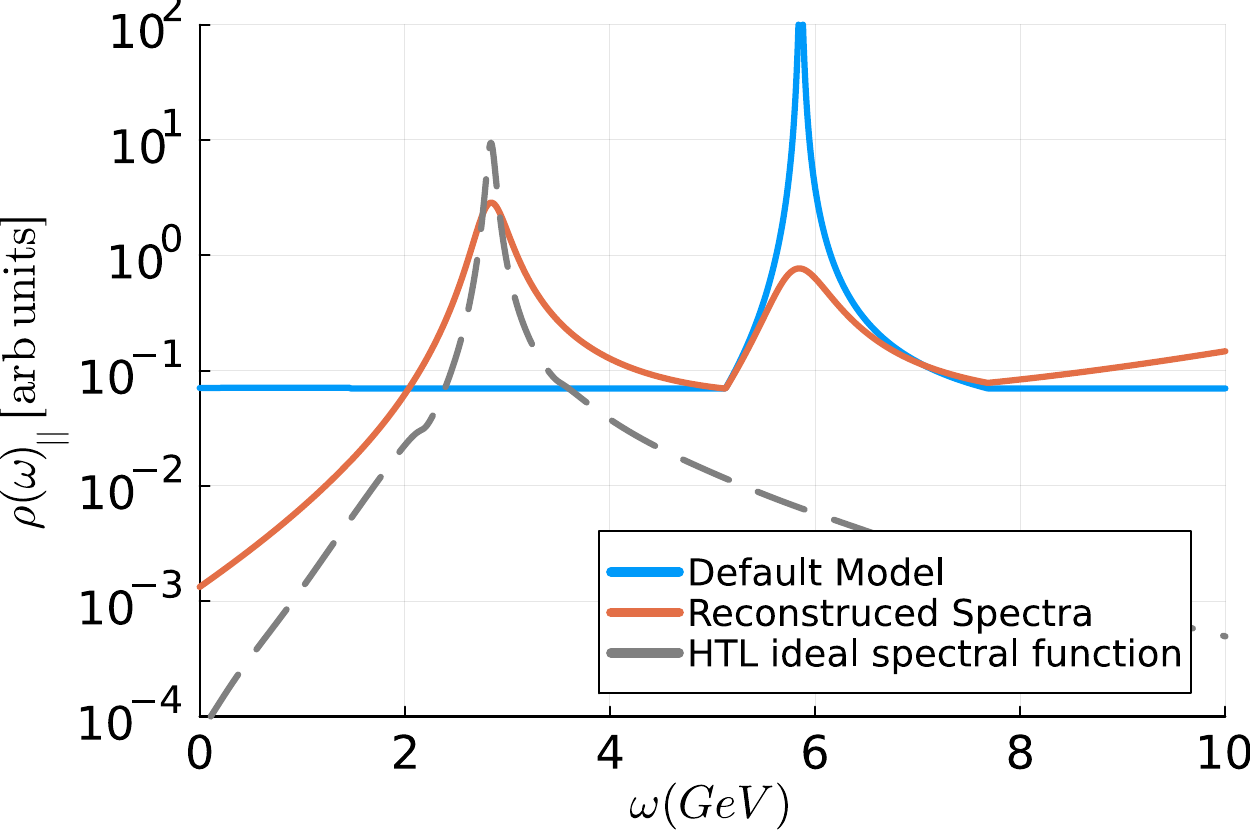}
    \caption{Bayesian inference reconstructed (solid line) and semi-analytical (dashed line) spectral functions of the (top) Coulomb-gauge Wilson line correlator $\rho_{||}$ 
    in HTL perturbation theory based on $22$ datapoints with $\Delta D/D=10^{-2}$ at a separation distance of 0.13 fm. The reconstruction has been carried out with an artificial adversarial peak inserted in the default model (blue). }
    \label{fig:HTLBRdD3specmtst}
\end{figure}

One may ask how susceptible the reconstruction is to adversarial attacks, such a by supplying false prior information in the form of peak structures in the default model. In \cref{fig:HTLBRdD3specmtst} we show the robustness of the Bayesian reconstruction against a set of artificial peaks located both above and below the true peak position. At $\Delta D/D=10^{-3}$ and $N_\tau=24$ the presence of adversarial peak located above the true position leaves the position of the reconstructed dominant peak relatively unaffected, however when the adversarial peak is located below the true position we see a shift in the reconstructed peak towards a slightly higher value. We see that incorrect prior only moves the peak to a higher $\omega$ and \textbf{not} lower $\omega$.

After a qualitative inspection of spectral reconstructions, let us plot the corresponding extracted values of $\textrm{Re}[V]$ and $\textrm{Im}[V]$ from fitting the dominant peak with \cref{eq:PotFitFunc} in \cref{fig:HTLBRdD3pot}. It is reassuring that even in the presence of very strong non-potential effects in $\rho_\square$ the Bayesian reconstruction appears able to quantitatively accurately reproduce the real part of the potential with the number of datapoints and errorlevels accessible to us in this study.

\subsection{Method II: Pade interpolation}
\label{sec:pade}
The second method we inspect is a complex generalization of the resonances via Pad\'e approach \cite{Tripolt:2016cya}, based on the Schlessinger Pad\'e interpolation \cite{Schlessinger:1968}.

After transforming the Euclidean correlator evaluated at discrete imaginary times $\tau_i$ to Matsubara frequencies via discrete Fourier transform
\begin{align}
W(\tilde \omega_n,r)=\sum_{j=0}^{N_\tau-1}e^{ia \tilde \omega_n j } W(\tau_i,r), \quad \tilde \omega_n=2 \pi n/aN_{\tau}. \label{eq:specdec}
\end{align}
we assign these datapoints to improved lattice frequencies
\begin{align}
\tilde \omega_n \rightarrow \omega_n=2 {\rm sin}\big(\frac{\pi n}{ N_\tau}\big)/a
\end{align}
in a similar spirit as the Sommer improvement  \cite{Luscher:1996sc} 
in coordinate space reduces lattice artifacts in distances.

One subsequently constructs a rational interpolation of the complex valued Matsubara data, corresponding to a Pad\'e approximation, where the order of the polynomial in the numerator is either of the same order or one order higher than that of the denominator. The interpolation formula for $N_\tau$ datapoints in the form of a continued fraction, according to Schlessinger, reads
\begin{align}
\nonumber&C_{N_\tau}(i\omega,r)=\frac{W(r,\omega_0)}{1+} \frac{a_0(r)[\omega-\omega_0]}{1+}\frac{a_1(r)[\omega-\omega_1]}{1+}\\
 &\ldots\frac{a_{N_\tau-2}(r)[\omega-\omega_{N_\tau-2}]}{1+}a_{N_\tau-1}(r)[\omega-\omega_{N_\tau-1}].
 \label{Eq:ContFrac}
\end{align}
and its coefficients can be computed recursively via 
\begin{align}
a_l(& r)(\omega_{l+1}-\omega_l)=-\Big\{ 1+\\
\nonumber &\frac{a_{l-1}(r)[\omega_{l+1}-\omega_{l-1}]}{1+}\frac{a_{l-2}(r)[\omega_{l+1}-\omega_{l-2}]}{1+}\cdots \\ \nonumber &\cdots\frac{a_{0}(r)[\omega_{l+1}-\omega_0]}{1-[W(r,\omega_0)/W(r,\omega_{l+1})]}\Big\}.
\end{align}

\begin{figure}[!ht]
    \includegraphics[scale=.4]{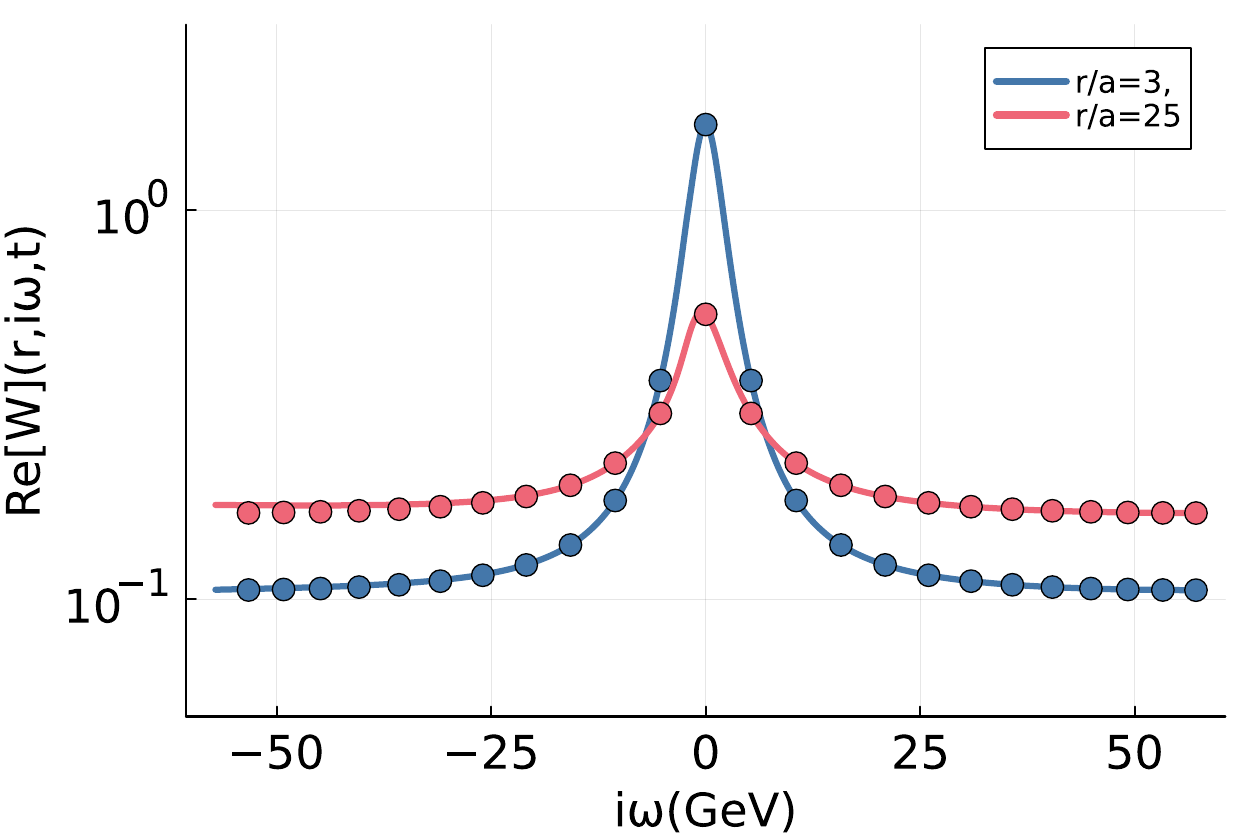}
    \includegraphics[scale=.4]{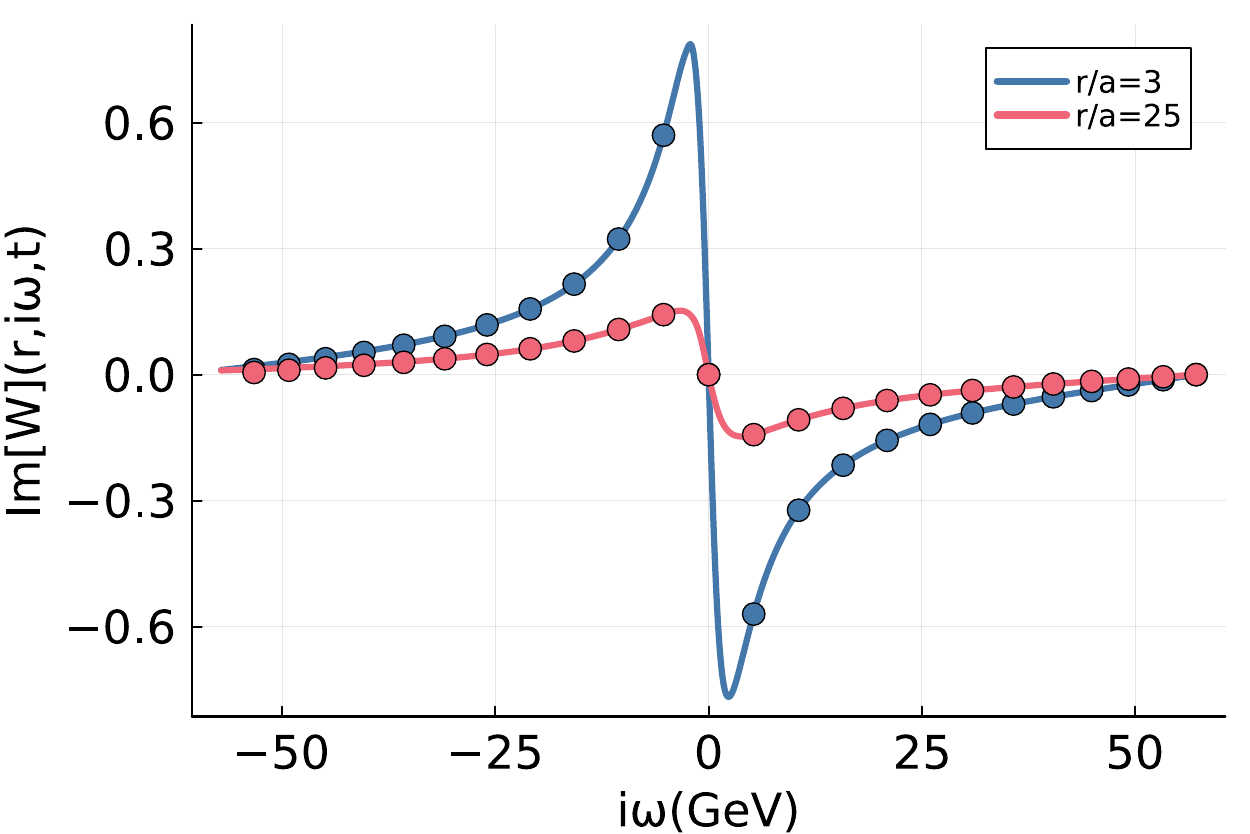}
    \caption{Matsubara frequency Wilson line correlators at two different separation distances for the $N_\tau=24$ ($T=3.11T_c$) anisotropic lattices. The top part shows the real part and the bottom part shows the imaginary parts. The dots show the Fourier transformed lattice data and the lines show the interpolated function.}
    \label{fig:HTLMatscorr}
\end{figure}

In \cref{fig:HTLMatscorr} we plot the real (top panel) and imaginary part (bottom panel) of the HTL Coulomb gauge Wilson line correlator along improved Matsubara frequencies (filled symbols) compared to the rational interpolation via \cref{Eq:ContFrac} (solid lines).

One can use \cref{Eq:ContFrac} to carry out an analytic continuation $\rho_r(\omega,T)=-\frac{1}{\pi}{\rm Im}[ C_{N_\tau}(r,\omega,T) ]$ and subsequently fit the lowest lying structure via \cref{eq:PotFitFunc}. Alternatively one can inspect the poles of the rational function \cref{Eq:ContFrac} directly and read off the dominant peak structure, as it belongs to the pole that is located most closely to the real frequency axis. If a well separated and pronounced pole emerges, its real- and imaginary part correspond to the real- and imaginary part of the potential. Having checked that the results obtained from both options agree within uncertainties, we choose to inspect poles, as it is computationally cheap.

In contrast to Bayesian spectral reconstruction, the result of the Pad\'e method does not necessarily reproduce the Euclidean input data \cite{Cyrol:2018xeq}, i.e. it is known to violate the spectral decomposition. On the other hand, as we show below, it is able to reproduce relevant structures of the HTL spectral functions, in particular the position of the dominant peak. Since no regularization is present in the Pad\'e interpolation formula, one also interpolates the error on the input data, making the reconstruction susceptible to statistical uncertainties. We therefore benchmark the reconstruction here only with $\Delta D/D=10^{-3}$, which corresponds to the uncertainty level present in our lattice data.

\begin{figure}[!ht]
    \includegraphics[scale=0.4]{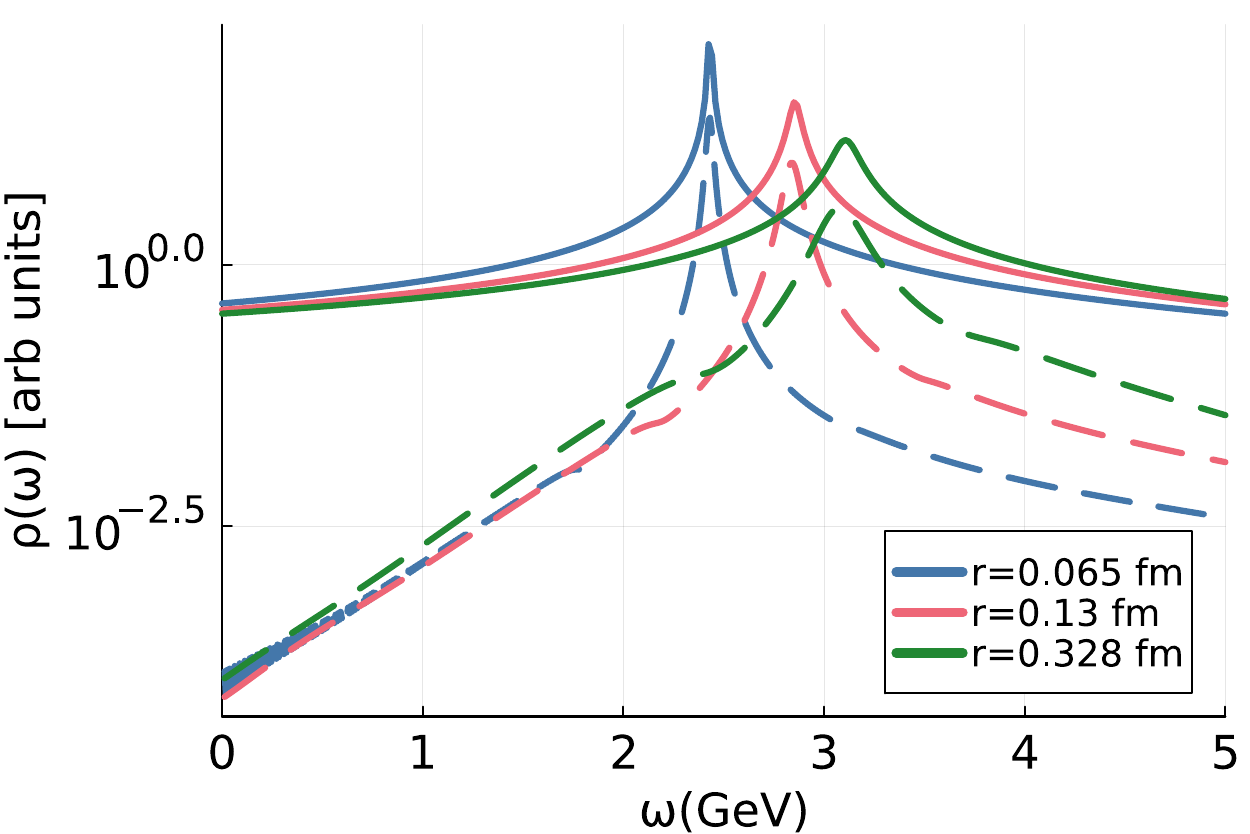}
    \includegraphics[scale=0.4]{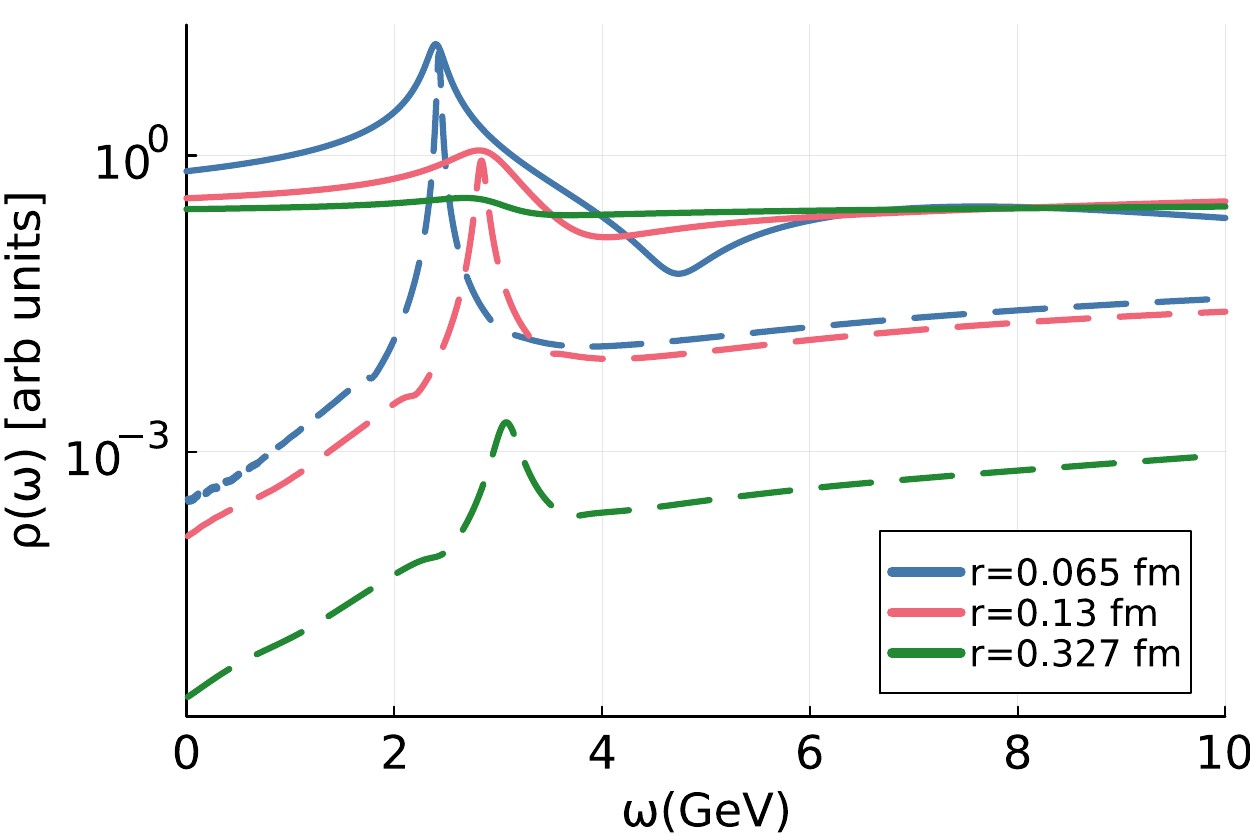}
    \caption{Pad\'e interpolation reconstructed (solid line) and semi-analytical (dashed line) spectral functions of the (top) Coulomb-gauge Wilson line correlator $\rho_{||}$ and (bottom) Wilson loop $\rho_\square$ in HTL perturbation based on $N_\tau=24$ datapoints with $\Delta D/D=10^{-3}$. The three curves each denote spectra at different spatial separation distances.}\label{fig:HTLPadedD3spec}
\end{figure}

As shown in the top panel of \cref{fig:HTLPadedD3spec}, the Pad\'e approach when applied to discrete HTL Wilson line data is able to locate the dominant peak structure and place the reconstructed peak accurately. 

However the peaks are too sharp close to the maximum and strongly broaden later, indicating that the true width is underestimated.
Note that the Pad\'e reconstruction fails to describe the falloff of the spectral function both at frequencies below and above the peak in an accurate fashion. As expected, this entails that the resulting $\rho$, reinserted into the spectral decomposition does not reproduce the Euclidean input data.

In case of the Wilson loop, the performance of the Pad\'e is more limited. With the available $N_\tau=24$ datapoints at $\Delta D/D=10^{-3}$, it manages to identify the peak up to intermediate distances $r<0.25$ fm  after which no discernible structure (nor an individual pole close to the real frequency) can be found.

\begin{figure}[!ht]
    \includegraphics[scale=.4]{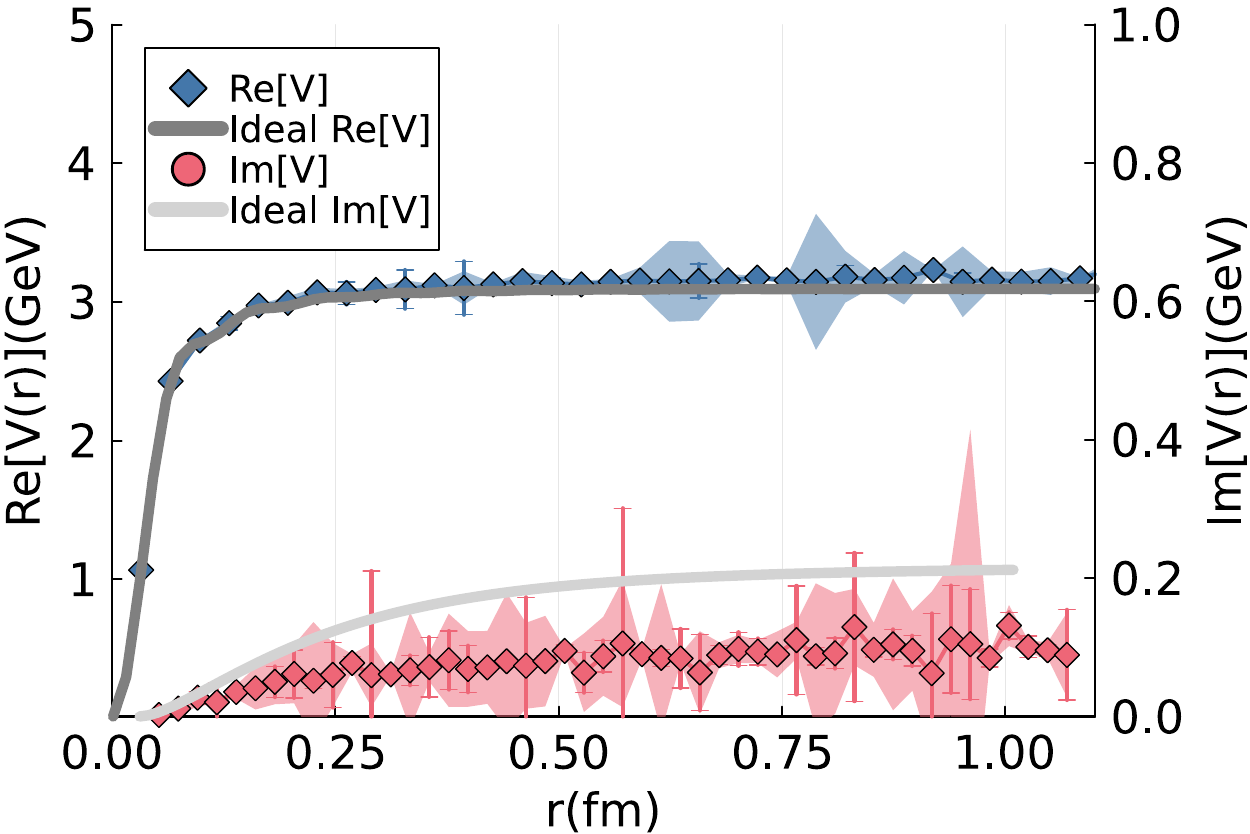}
    \includegraphics[scale=.4]{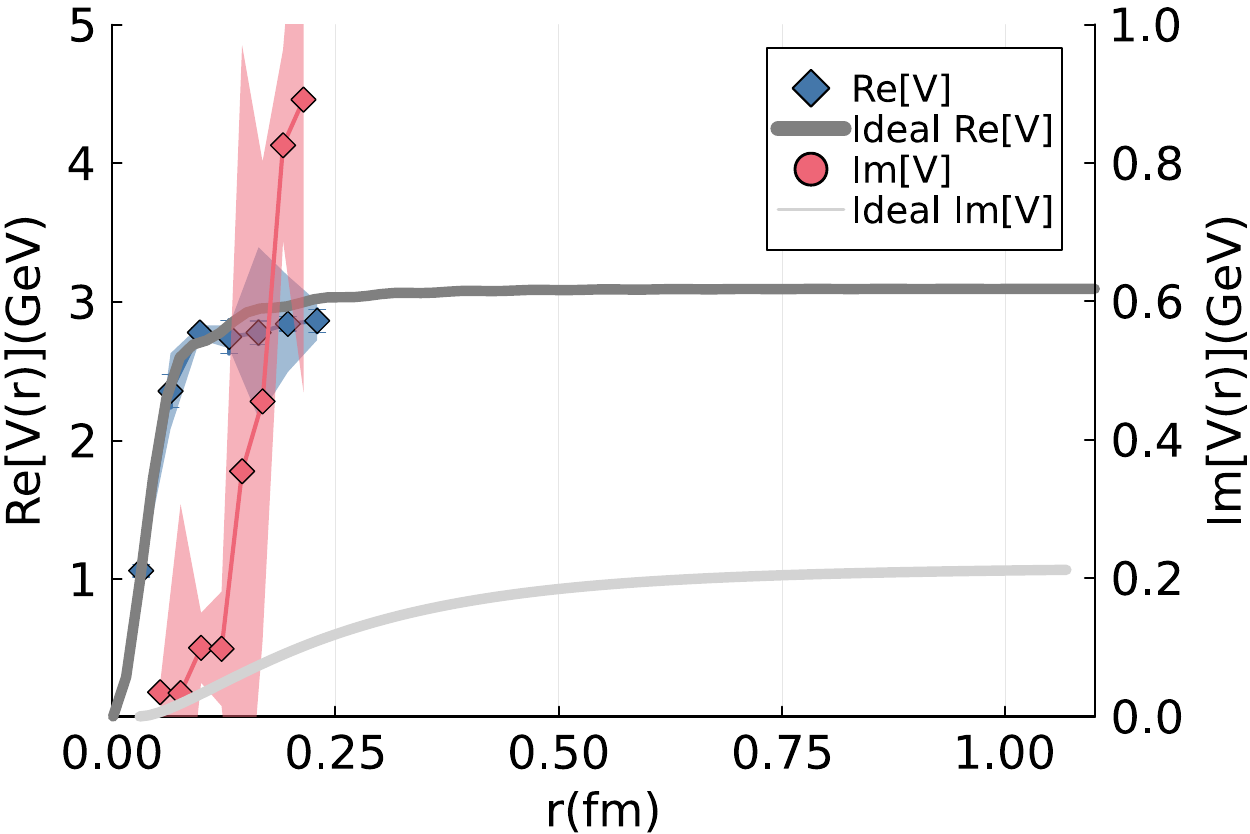}
    \caption{Pad\'e reconstructed (colored symbols) and analytic (solid line) values for the real- and imaginary part of the complex static quark potential from the (top) Coulomb-gauge Wilson line correlator $V_{||}(r)$ and (bottom) Wilson loop $V(r)$ in HTL perturbation based on $N_\tau=24$ datapoints with $\Delta D/D=10^{-3}$.}
    \label{fig:HTLPadedD3pot}
\end{figure}

Turning to a quantitative assessment, we plot in the top panel of \cref{fig:HTLPadedD3pot} the real- and imaginary part of the static potential extracted from discrete HTL $W_{||}$ along $N_\tau=24$ with $\Delta D/D=10^{-3}$. Even though the reconstructed spectral function does not reproduce the input data, we find that a robust estimation of the position of the dominant pole, i.e. of ${\rm Re}[V]$ is possible with realistic data, up to large distances of $r=1$fm. As expected from visual inspection of the reconstructed $\rho$ the imaginary part is systematically underestimated by around a factor of two.

In case of the more challenging Wilson loop data $W_\square$, the Pad\'e manages to locate the dominant peak up to $r\approx 0.225$fm. Within this region, its position is correctly reproduced within the statistical uncertainties, which are larger than for $W_{||}$. The imaginary part is overestimated but suffers from strong variance even with the favorable $\Delta D/D=10^{-3}$, a testament to the missing regularization of such a direct projection approach.

\subsection{Method III: HTL-inspired fits}
\label{sec:bala_datta}

In this approach (see \cite{Bala:2019cqu} and \cite{Bala:2019boe} for more details), one constructs a set of observables, which, by taking inspiration from HTL perturbation theory, are designed to reveal more clearly the physics of the real- and imaginary part of the potential, as is possible from effective masses alone, which were designed for the study of a purely real potential at $T=0$.

The basic idea is to consider an expansion of the correlator around $\tau=\beta/2$, instead of the $\tau=0$ expansion underlying $m_{\rm eff}$. This is achieved by combining the correlator into an antisymmetric and a symmetric quantity
\begin{align}
    &A(\tau)={\rm log}\Big[\sqrt{\frac{W(\tau,r)}{W(\beta-\tau,r)}}\Big]\\
    &P(\tau)={\rm log}\Big[\sqrt{ W(\tau,r)W(\beta-\tau,r) }\Big].
\end{align}

In leading order HTL perturbation theory, the quantity $A^{\rm HTL}(\tau)=(\beta/2-\tau){\rm Re}[V]$ is directly related to the real-part of the potential, while in case that the spectral peak is shaped as a Gaussian, one would obtain $A^{\rm Gaussian}(\tau)=(\beta/2-\tau)({\rm Re}[V]+{\rm Im}^2[V])$, where both ${\rm Re}[V]$ and ${\rm Im}[V]$ contribute to the value of $A$. In any case, a behavior proportional to $(\beta/2-\tau)$ is indicative of the presence of a dominant peak structure. It is interesting to note that a exponentially cut-off skewed Lorentzian and the Gaussian both lead to an exactly linear behavior in this quantity.

\begin{figure}[!ht]
    \includegraphics[scale=0.4]{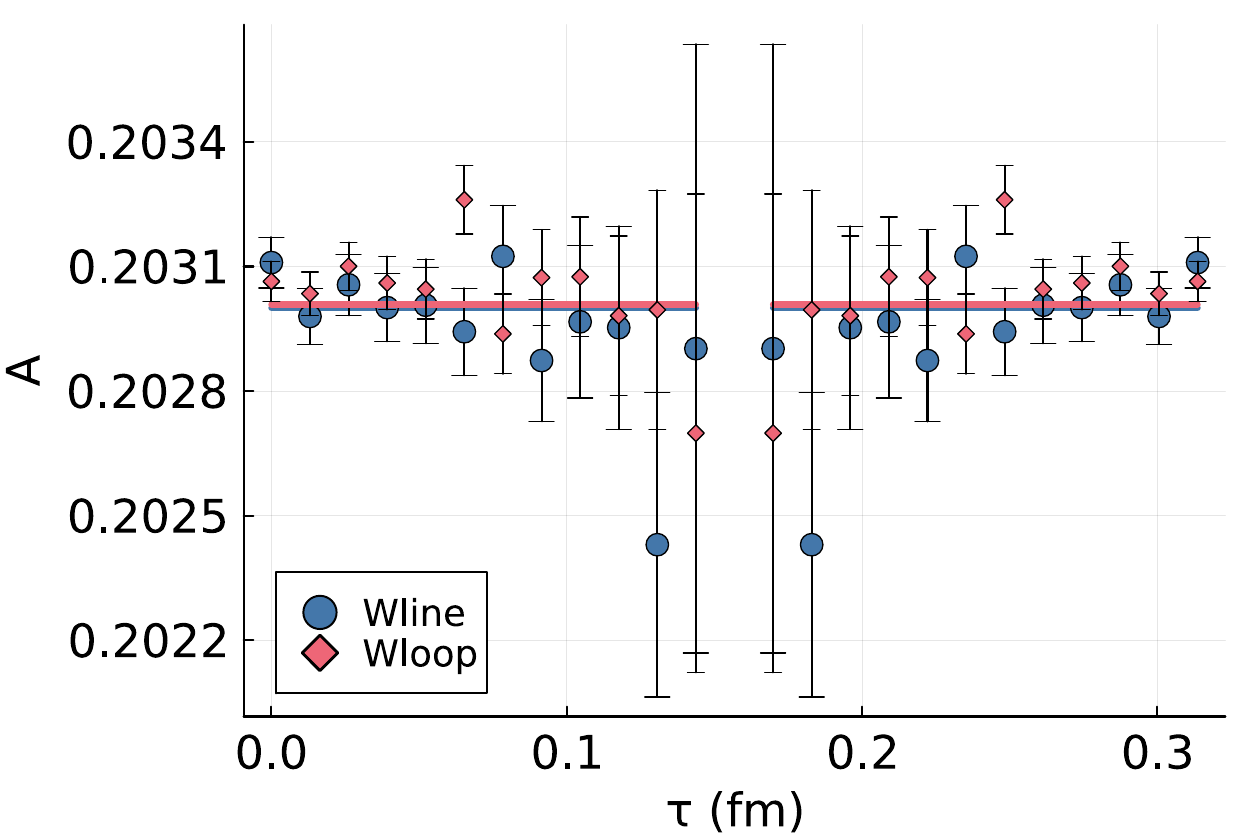}
    \caption{Comparison of the quantity $A$ from HTL Coulomb-gauge Wilson lines (blue) and Wilson loops (red) for $N_\tau=24$ and $\Delta D/D=10^{-3}$ at a separation distance of $r=0.0329$ fm. The HTL-inspired fit with $(\beta/2-\tau)$ is shown as solid line.}
    \label{fig:AsfromHTL}
\end{figure}
The symmetric component $P(\tau)$ in HTL perturbation theory can be shown to encode both the physics of the imaginary part of the potential, as well as non-potential effects. Its derivative to first order gives
\begin{align}
    \frac{d}{d \tau} P^{\rm HTL}(\tau)= \frac{1}{\pi}  {\rm Im}[V](r){\rm log}\big[ \frac{\tau}{\beta-\tau}]+\ldots
\end{align}
where the symbols $\ldots$ indicates a series of terms that includes non-potential effects. 

One of the benefits of the way these observables are constructed is that they appear to lessen the influence of the UV continuum and are dominated by the low-lying peak structures. Let us plot the quantity $A$ for both the HTL Coulomb-gauge Wilson line (red) and Wilson loops (green) in \cref{fig:AsfromHTL}. Note that while the effective masses for the Wilson loop in \cref{fig:cmpHTLmeff} show a significantly different slope around $\tau=\beta/2$, which overshadows the contribution from the potential peak visible in the Wilson line effective masses, here the quantity $A$ shows very similar slopes and only close to the fringes of the imaginary time interval exhibits curvature.

\begin{figure}[!ht]
    \includegraphics[scale=0.4]{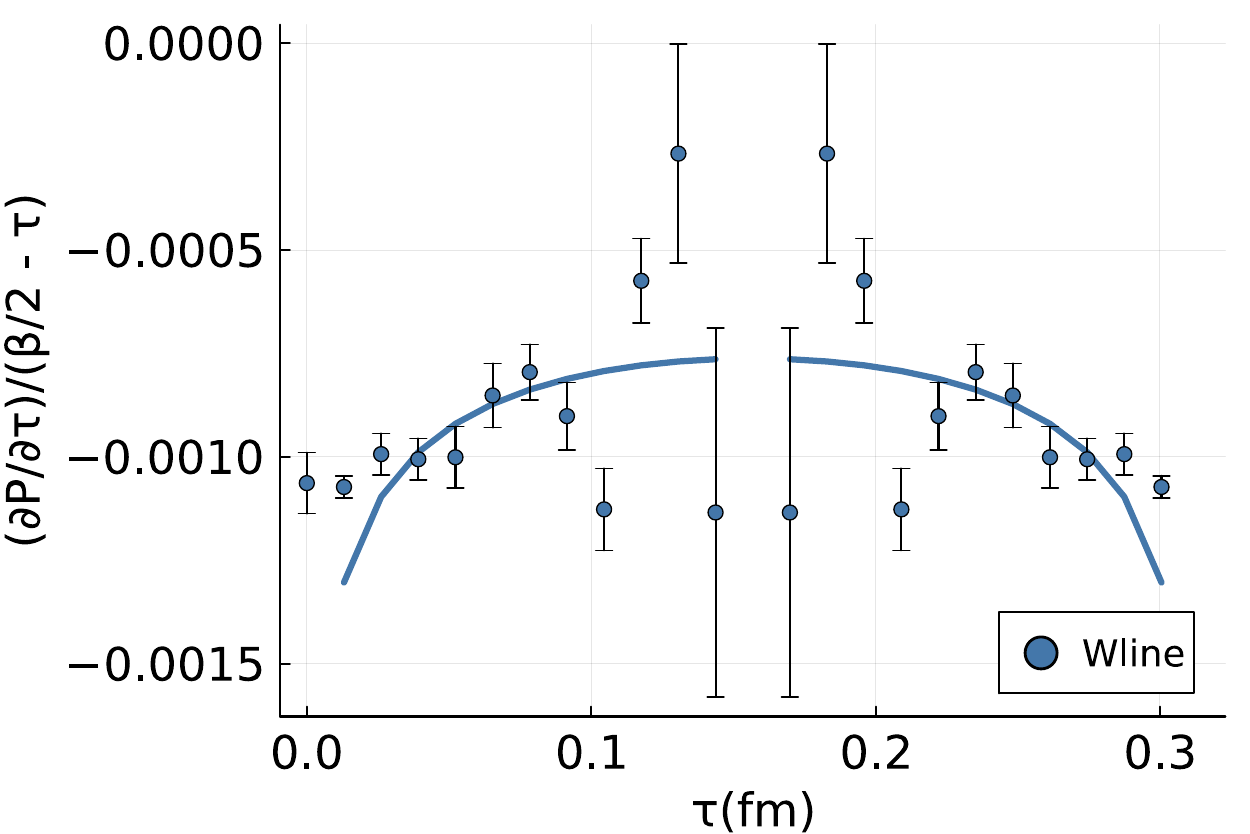}
    \includegraphics[scale=0.4]{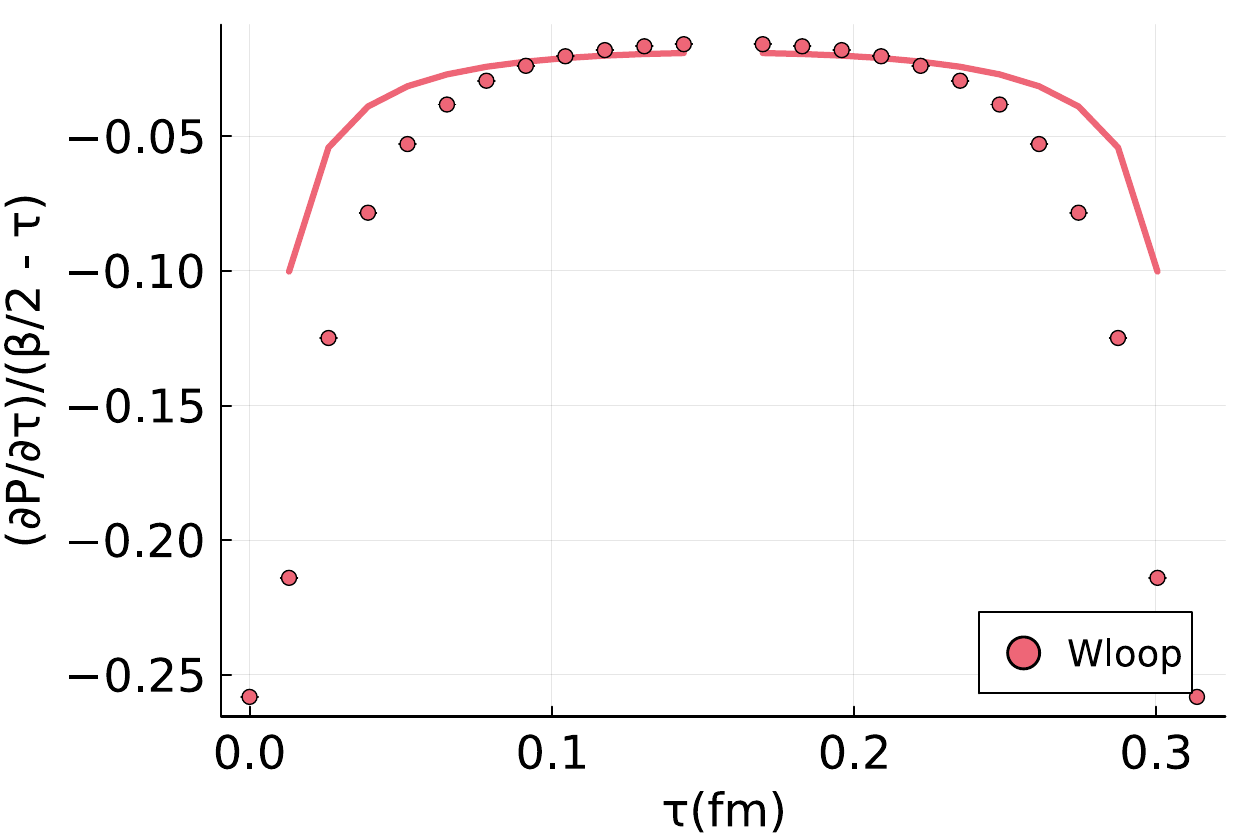}
    \caption{Comparison of the quantity $dP/d\tau$ from HTL Coulomb-gauge Wilson lines (blue, top) and Wilson loops (red, bottom) for $N_\tau=24$ and $\Delta D/D=10^{-3}$ at a separation distance of $r=0.329$ fm. The HTL-inspired fit with ${\rm cot }[\pi\tau/(\beta)]$ is shown as solid line.
    }
    \label{fig:dPdtausfromHTL}
\end{figure}
Benchmarking the extraction of the imaginary part from the derivative of $P$ in HTL with realistic data turns out to be more difficult. Since in HTL the imaginary  part is relatively small, the signal to noise for $\Delta D/D=10^{-3}$ is bad. As shown in the next section, when we investigate actual lattice data, we find a much larger curvature of $dP/d\tau$, so that even with a $\Delta D/D=10^{-3}$ an estimate for the imaginary part can be obtained. For demonstration purposes we thus show in \cref{fig:dPdtausfromHTL} the values from HTL for $\Delta D/D=10^{-3}$, together with the lowest order HTL-inspired fit around $\tau=\beta/2$ proportional to ${\rm log}[\tau/(\beta-\tau)]$. We can see that the fit Ansatz now is only compatible with the data around $\tau = \beta/2$. As \cref{fig:dPdtausfromHTL} suggests, we would need to consider higher order non-potential terms to fit the data to a larger $\tau$ range. To fit the data over a considerably larger $\tau$ range we instead use a fit form:
\begin{equation}
    W(\tau,r) = e^{-\rm{Re}[V](r)(\tau - \beta/2) - \frac{\beta}{\pi}\rm{Im}[V](r)\log \sin \frac{\pi \tau}{\beta} + G_{\mathrm{NP}}(\tau,r,\beta)} W(\beta,r)
\end{equation}
With the higher order terms representing non-potential effects being,
\begin{align}
    &G_{NP}(\tau,r,\beta) = \sum_l c_l \int _{\beta/2}^\tau \tilde G _l (\tau) \\
    &\tilde G_l(\tau) = \frac{(2l!)}{\beta^{2l +1}} \Big( \zeta (2l + 1, 1-\frac{\tau}{\beta}) - ( \zeta (2l + 1, \frac{\tau}{\beta})\Big)  \nonumber
\end{align}
Here $\zeta$  is the generalized zeta function $\zeta(s,x) = \sum_{n=1} ^\infty \frac{1}{(x+n)^s}$. Instead of fitting to the correlator we fit to effective masses (as suggested in \cite{Bala:2021fkm}) which is done by taking the log of derivative on both sides.
\begin{align}
&m_{\text{eff}}(r,\tau)a=\log\left(\frac{W(r,n_\tau,N_{\tau})}{W(r,n_\tau+1,N_\tau)}\right)\nonumber\\
		&=\mathrm{Re[V]}(r,T)\,a-\frac{\mathrm{\textrm{Im}[V]}(r,T)a N_\tau }{\pi }\,\log\left[\frac{\sin(\pi n_\tau/N_\tau)}{\sin(\pi  (n_\tau+1)/N_\tau)}\right] + \sum_l c_l \tilde{G}_l(\tau)
\label{eq:BDfit}
\end{align}

\begin{figure}[!ht]
    \includegraphics[scale=.4]{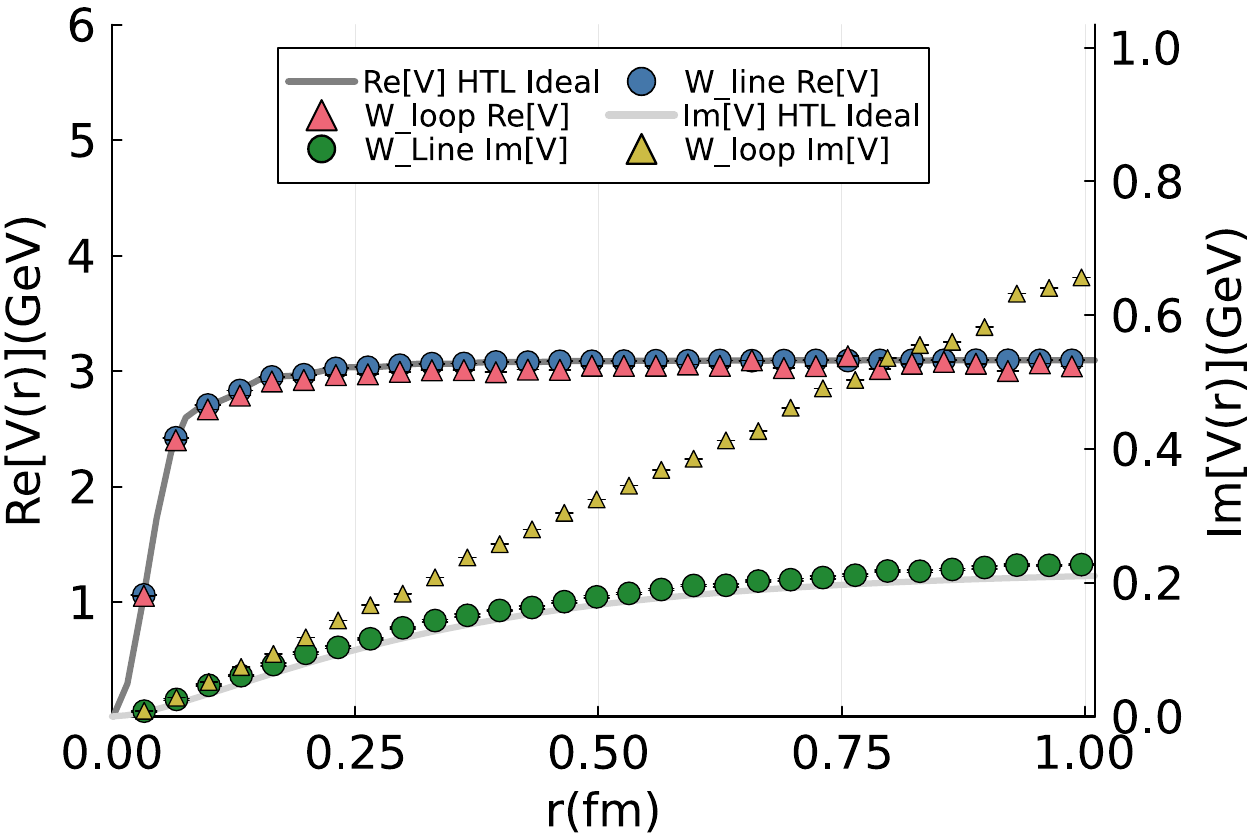}
    \caption{HTL-inspired fit reconstructed (colored symbols) and analytic (solid line) values for the real- and imaginary part of the complex static quark potential from the  Coulomb-gauge Wilson line correlator $V_{||}(r)$ and  Wilson loop $V(r)$ in HTL perturbation based on $N_\tau=24$ datapoints with $\Delta D/D=10^{-3}$.
    }
    \label{fig:HTLBaladD3pot}
\end{figure}

In our fits we found that the first two terms with parameters $c_1$and $c_2$ were sufficient to fit the data to a large $\tau$ region (see \cref{sec:quality}) with the exceptions being a few small $\tau$ and a few points around $\tau = \beta$ (1-3 depending on the separation distance). We will discuss the quality of the fits on lattice data in \cref{sec:meff_lat}. Using the fit form as described in \cref{eq:BDfit}, we extract ${\rm Re}[V]$ and ${\rm Im}[V]$ from the mock data with results given in \cref{fig:HTLBaladD3pot}.

We find that for the Wilson line correlators we are able to recover both the correct real and imaginary parts using data with errors $\Delta D/D = 10^{-3}$ up to a high separation distance of 1 fm. For the Wilson loop data we were able to extract the correct real part, but however we were only able to recover the correct imaginary part up to $r \sim 0.15$fm. 

\subsection{Method IV: Moment analysis \& extended Gaussian fits}
\label{sec:meff_sub}
In a prior study of the static potential on lattices in full QCD a remarkable behavior in UV subtracted effective masses was observed. At small $\tau$ behavior of the effective masses at $T=0$ and at finite temperature took on virtually the same values within statistical uncertainties. Such behavior had not been observed in quenched QCD studies before.

It led the authors of \cite{Larsen:2019bwy} to suggest the following subtraction procedure. They first identified the lowest lying delta-peak-like contribution to the spectral function of Coulomb-gauge Wilson line correlators at $T=0$ and subtracted that component out from the correlator. Under the assumption that the remaining spectral contributions are separated well enough from that lowest peak and that they remain temperature independent, they proceeded to subtract these $T=0$ UV contributions from the correlators at finite temperature. They observed that the subtracted $T>0$ correlators all showed effective masses, which exhibit a straight line behavior for $\tau<\beta/2$. Such a behavior could be reproduced with a dominant low-lying Gaussian or cut-Lorentzian peak in the spectral function. Lastly, the non-potential effects in the low-frequency part could be parameterized with a single delta-function \cite{Bazavov:2023dci}.
Adding two delta peaks to the Gaussian, one above and one below, the authors constructed a model spectrum which they fit to the subtracted lattice correlators. 
They found that the position of the peak did not change with temperature and instead only its width grew.
We cannot test this approach directly with HTL perturbation theory, as no zero temperature
subtraction can be carried out.

In the subsequent section we will carry out a subtraction analysis that reveals an almost linear behavior of the effective masses after subtraction and we will carry out the extended Gaussian fits to estimate the real- and imaginary part of the potential.

\subsection{Mock data test summary}

We find from extensive testing, based on non-trivial mock data that for the realistic scenario of $N_\tau=24$ and $\Delta D/D=10^{-2}$ the approaches based on Bayesian inference and Pad\'e approximation are able to extract reliably the position of the dominant peak encoded in the HTL spectrum. The width of the peak, when extracted from Coulomb-gauge Wilson lines is systematically underestimated, while from Wilson loops tends to be overestimated. A reliable reconstruction of the position of the peak thus would require that the uncertainties in our data $\Delta D/D =10^{-2}$.

The HTL inspired method we find is able to distinguish between between the HTL and single dominant Gaussian peak hypothesis. Under the (in this case correct) assumption that the data encodes an HTL-like spectral function we can reliably extract both ${\rm Re}[V]$ and ${\rm Im}[V]$ using either Wilson lines or loops.

\section{Potential from high resolution quenched lattices}
\label{sec:latticeres}

In this part of our study, we deploy the methods benchmarked in the previous section to the extraction of spectral information of Coulomb gauge Wilson line correlators computed on high-resolution isotropic and anisotropic quenched lattices. Both sets of ensembles are based on the naive Wilson action, the former utilizes a parameter set $(\beta=7.196)$, which has been used in the study of relativistic bottomonium physics at finite temperature in \cite{Burnier:2017bod} with a transition temperature of 313 MeV and lattice spacing $a=0.0176$ fm with total volume of $96^3 \times N_\tau$ with $N_\tau =16,24,48$, the latter $(\beta=7,\xi=3.5)$ in previous studies of quarkonium melting in \cite{Burnier:2016mxc} with transition temperature  $T_c=270$ MeV, spacial lattice spacing $a_s=0.039$ fm and temporal lattice spacing $a_\tau=a_s/4$ with total volume of $64^3 \times N_\tau$ with $N_\tau =24,32,40,56,96$. 
The anisotropic lattices were generated using the openQCD code by the fastsum collaboration which has been extended to anisotropic lattices \cite{glesaaen_2018_2216356} by utilizing the Fram supercomputer maintained by Sigma2 in Norway. The configurations were gauge fixed to Coloumb gauge with a tolerance of $\Delta_{GF}=10^{-12}$ and the Wilson line correlators were measured using the SIMULATeQCD code \cite{HotQCD:2023ghu} on the MARCONI100 computing cluster in Italy. The isotropic lattices were generated at the University of Bielefeld and some of the measurements were also performed there.
A summary of measurements at various temperatures for isotropic lattices are shown in \cref{table:isotropic} and for anisotropic lattices are shown in \cref{table:anisotropic}.  

\begin{table}
\begin{tabular}{||c c c c ||} 
 \hline
 $N_\tau$ & 16 & 24 & 48 \\ [0.5ex] 
 \hline\hline
 T[MeV] & 698 & 466 & 232  \\ 
 \hline
 $T/T_C$ & 2.23 & 1.49 & 0.74 \\
 \hline
 $N_{meas}$ & 8700  & 4500  & 1600 \\
 \hline 
\end{tabular}
\caption{A table of isotropic quenched lattice ensembles at $\beta=7.196$, lattice spacing $a_s=0.0176$ fm with $N_s=96$.}
\label{table:isotropic}
\end{table}

\begin{table}
\begin{tabular}{||c c c c c c||} 
\hline
$N_\tau$ & 24 & 32 & 40 & 56  & 96  \\ [0.5ex] 
\hline\hline
T[MeV] & 839 & 629 & 503  & 359  & 210 \\ 
\hline
$T/T_C$ & 3.11 & 2.33 & 1.86  & 1.33 & 0.78 \\
\hline
$N_{meas}$ & 2500  & 1014  & 735 & 708 & 172 \\
\hline 
\end{tabular}
\caption{A table of quenched anisotropic lattice ensembles at $\beta=7$ and $\xi=3.5$, lattice spacing $a_s=0.039$ fm with $a_\tau = a_s/4$ and $N_s=64$.}
\label{table:anisotropic}
\end{table}

\subsection{Effective Masses}
\label{sec:meff_lat}

\begin{figure}[!ht]
    \includegraphics[scale=.4]{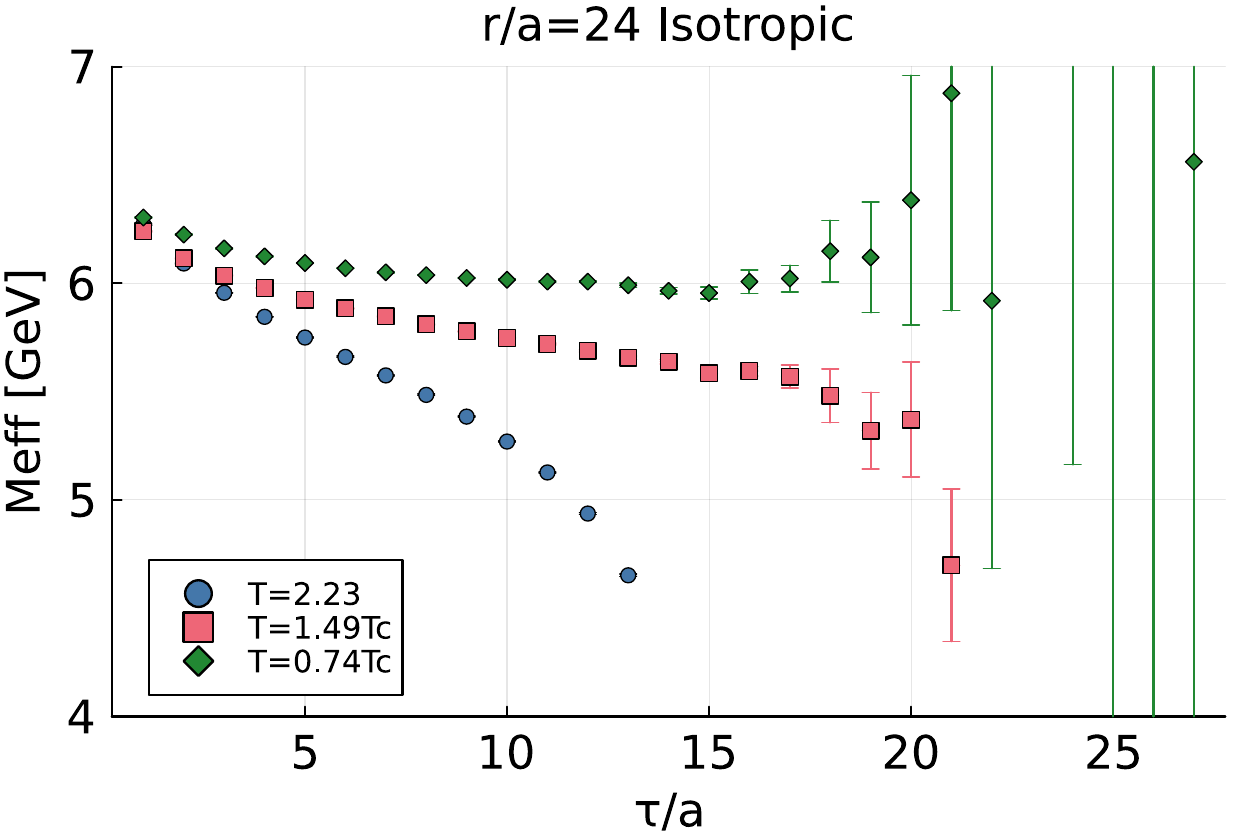}
    \includegraphics[scale=.4]{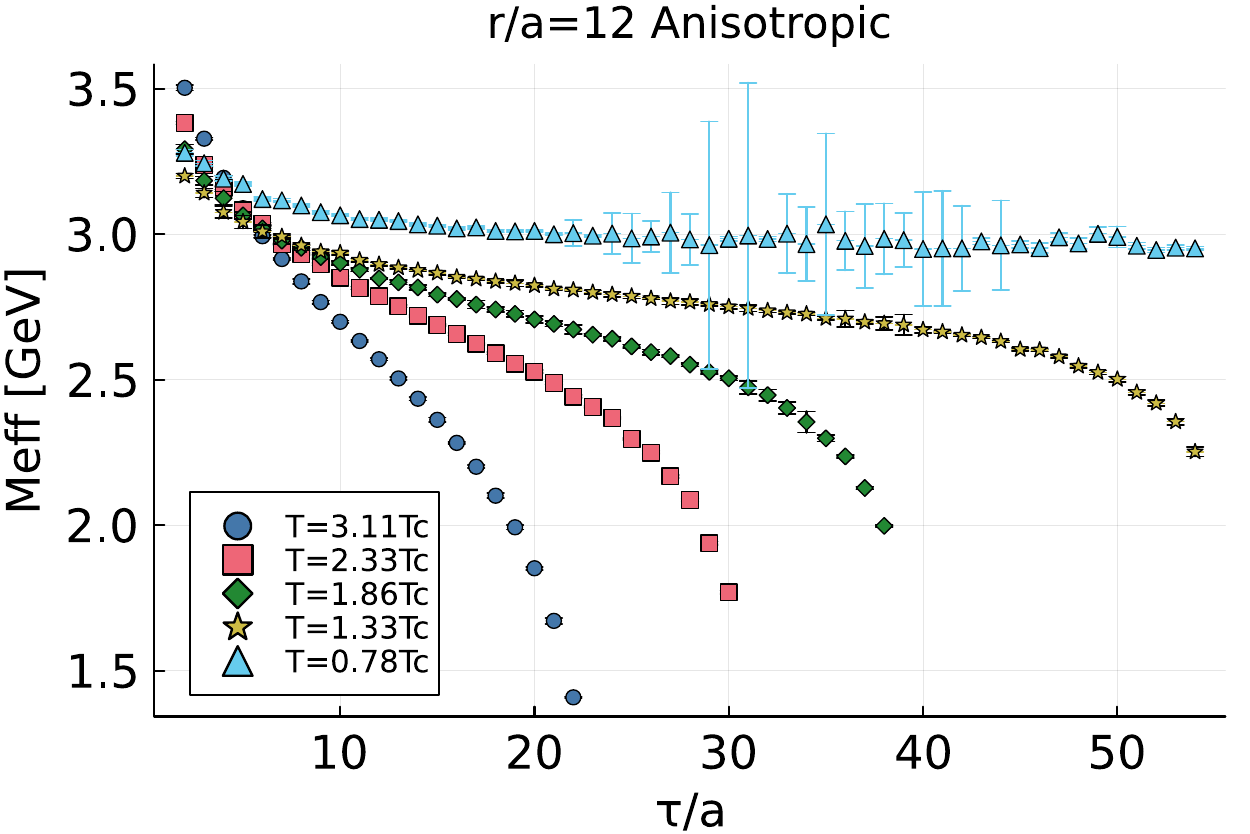}
    \caption{A representative selection of effective masses, obtained from the Coulomb-gauge Wilson line correlator computed on (top) isotropic ($\beta=7.196$) and (bottom) anisotropic ($\beta=7, \xi=4$) quenched lattices.}
    \label{fig:qnchdmeff}
\end{figure}
Following in the footsteps of the mock data test, we start by analysing the effective masses of the Wilson line correlator on the quenched lattices in \cref{fig:qnchdmeff}. The observation we have both for the isotropic and anisotropic lattices is that for $T<T_c$ the effective masses show a tendency to plateau at intermediate $\tau$. The small $\tau$ behaviour in the correlation function represents the large frequency contribution (UV) in the spectral function. The presence of a plateau at intermediate $\tau$ suggests the spectral function comprises of a lowest lying delta peak plus additional UV contribution that comes from the small $\tau$ regime; this is reminiscent of zero temperature behaviour where one expects there to be a well defined lowest lying peak described by a delta function. Such behaviour is to be expected since in the absence of dynamical fermions there exists a first order transition from the confined to the deconfined phase.\\

At $T>T_c$ we see that the effective masses obtain a non-zero slope at intermediate $\tau$, which suggests the presence of a finite width in the spectral function. The structure of the small $\tau$ behaviour is rather interesting. 
On the one hand, for the anisotropic lattices for $T=3.11 T_c$ and $T=2.33 T_c$ there are significant deviations at small $\tau$ when comparing the $T>T_c$ and $T<T_c$ case. These differences are diminished as the temperatures are reduced to around $1.86 T_c$ and below. 
On the other hand, for the isotropic lattices we find agreement at small $\tau$ for both available temperatures ($T=2.23 T_c$ and $T=1.49 T_c$)
 as compared with a recent study of (\texorpdfstring{\boldmath{$2+1$}}{2+1})-flavour QCD \cite{Bazavov:2023dci}. 
However, as all recent QCD studies were based on the improved L\"uscher-Weisz gauge action, and also on gradient flow smearing in the case of \cite{Bazavov:2023dci}, these two factors might contribute to the different observations as well.
Qualitatively similar to the mock tests, we see a bending in the effective masses at large $\tau$.
We understand this bending effect at $\tau \lesssim 1/T$ as being due to the interaction of a forward propagating static quark-antiquark pair and backward propagating states of the medium \cite{Bala:2021fkm}.

 The behaviour in quenched QCD is quantitatively different from the behaviour of effective masses in the mock HTL Wilson line data. There are significantly more UV contributions present in the lattice data which is manifested in the upward bending at small $\tau$. However, the magnitude of UV contribution is less than that the case of Wilson loop HTL mock data. The magnitude of the UV contributions in lattice data lie in between those of the Wilson line and the Wilson loop mock data. From the analysis of effective masses alone we cannot make a statement about the structure of the high frequency behaviour of the spectral function.

\begin{figure}[!ht]
    \includegraphics[scale=.4]{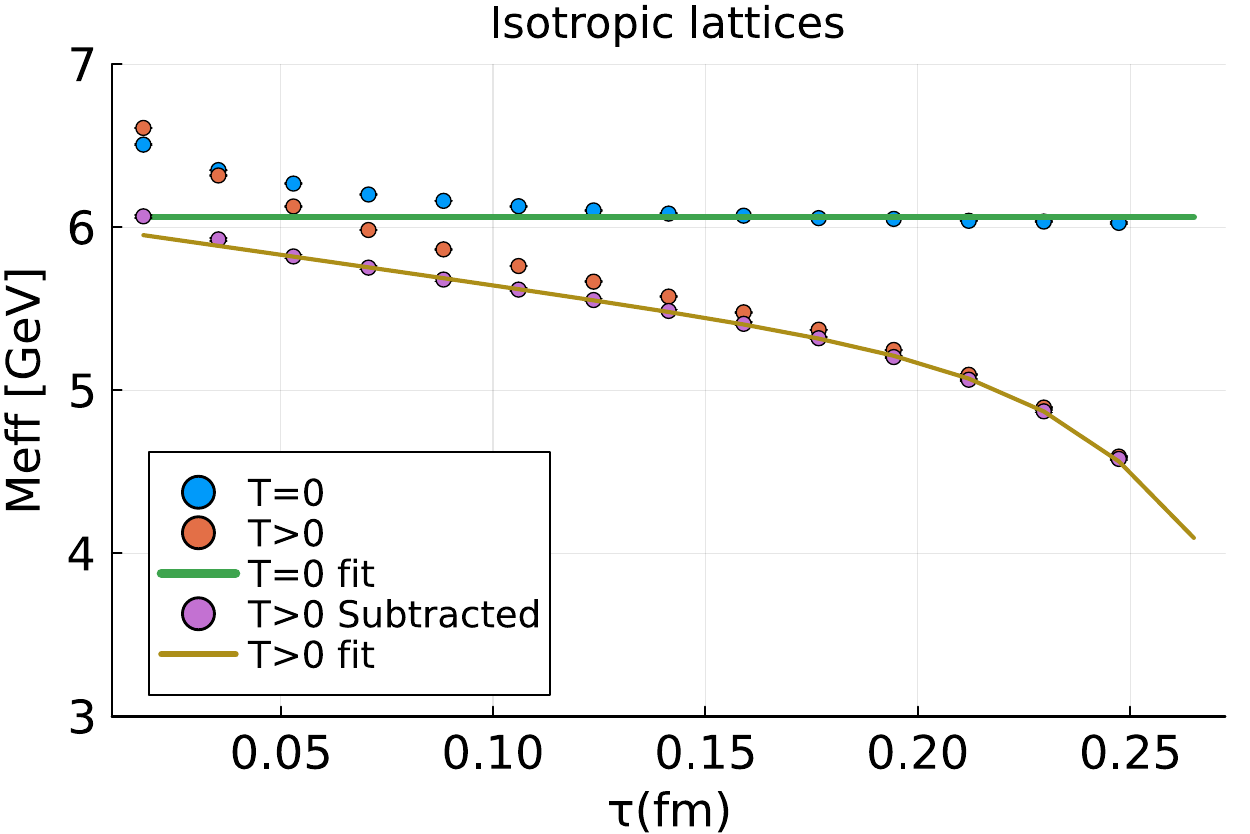}
    \includegraphics[scale=.4]{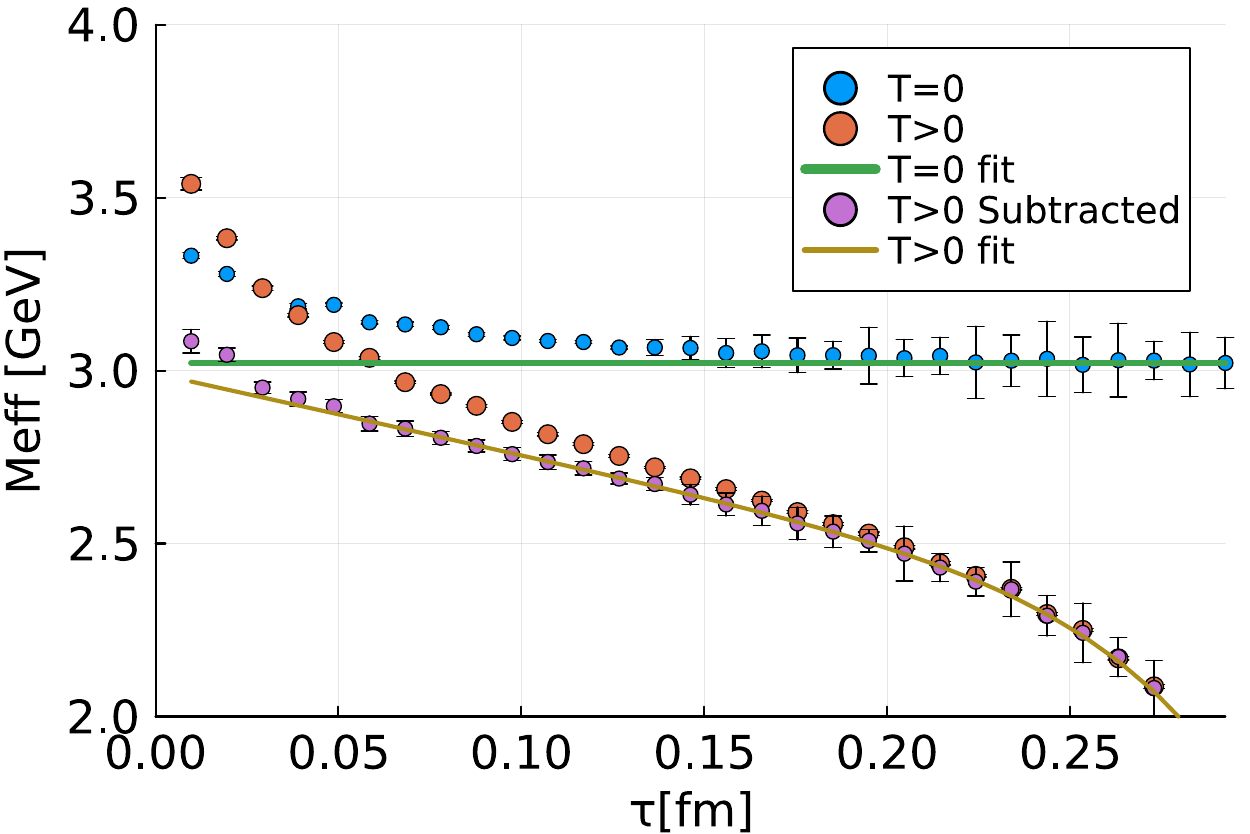}
    \caption{An example of subtracted effective masses for (top) $r=0.441$ fm for the $T=2.23 T_c$ isotropic lattices and (bottom) $r=0.467$ fm at $T=2.33 T_c$  for anisotropic lattices. Here the $N_\tau=48 (0.74 T_c)$ and $N_\tau=96 (T=0.78 T_c)$ lattices are used as the zero temperature reference . }
    \label{fig:Meff_sub}
\end{figure}

Despite different behaviour of $M_{\rm eff}$ on quenched lattices at small $\tau$, we follow the steps as described in \cref{sec:meff_sub} and show the subtracted correlator analysis.  We also note the fact that we might be over-subtracting or under-subtracting the high-frequency part. If we are indeed over-subtracting we might be doing something non-trivial to the spectral function that is relevant for the potential effects. An example of the high-frequency subtracted effective masses are shown at $T=2.23 T_c$ (isotropic) and $T=2.33 T_c$ (anisotropic) in \cref{fig:Meff_sub}. We carry out the subtraction procedure for all temperatures on the anisotropic lattices for $T>T_c$ at all but the highest temperature as the small $\tau$ behaviour is very far from being compatible with the zero temperature behaviour for the $T=3.11 T_c$ anisotropic lattices. The details of these fits are shown in \cref{fig:gauss_fit_good_isotropic} and \cref{fig:gauss_fit_good_anisotropic} in the appendix.

Thus, for each temperature we obtain two different correlators to perform spectral reconstructions on, i.e. the subtracted correlator and the raw correlator. We expect the spectral functions of the raw correlator to show structures at large frequencies owing to the high-frequency effects at small $\tau$ and these effects will be suppressed on the subtracted correlator. Since these high-frequency structures are related to be non-potential effects there is little need in reconstructing them accurately. Thus, the subtraction procedure attempts to get rid of irrelevant information while at the same time keeps the number of points in the $\tau$ direction fixed from the input data, in turn the information content on the small frequency regime should increase thus improving the outcome of spectral reconstruction. However, the subtraction procedure could affect the analysis in other ways. The subtraction is done by dividing the zero temperature raw correlator data into 16 jackknife bins and then performing a weighted fit to a single exponential. Once the fit parameters are obtained, the single exponential is subtracted from each jackknife bin. The remaining part of the correlator is used for subtraction from the finite temperature case. The subtraction will introduce new errors (coming from a mismatch in statistics in zero and finite temperature data) in the data and the spectral reconstruction will be limited by the statistics of the zero temperature data. The statistics on the zero temperature data for anisotropic lattices is limited in our case (we have about 175 uncorrelated measurements). Nevertheless, we expect that the gains made from reducing the severeness of ill-posed problem will outweigh the losses coming from limited statistics in the subtracted data.

\begin{figure}[!ht]
    \centering
    \includegraphics[scale=0.4]{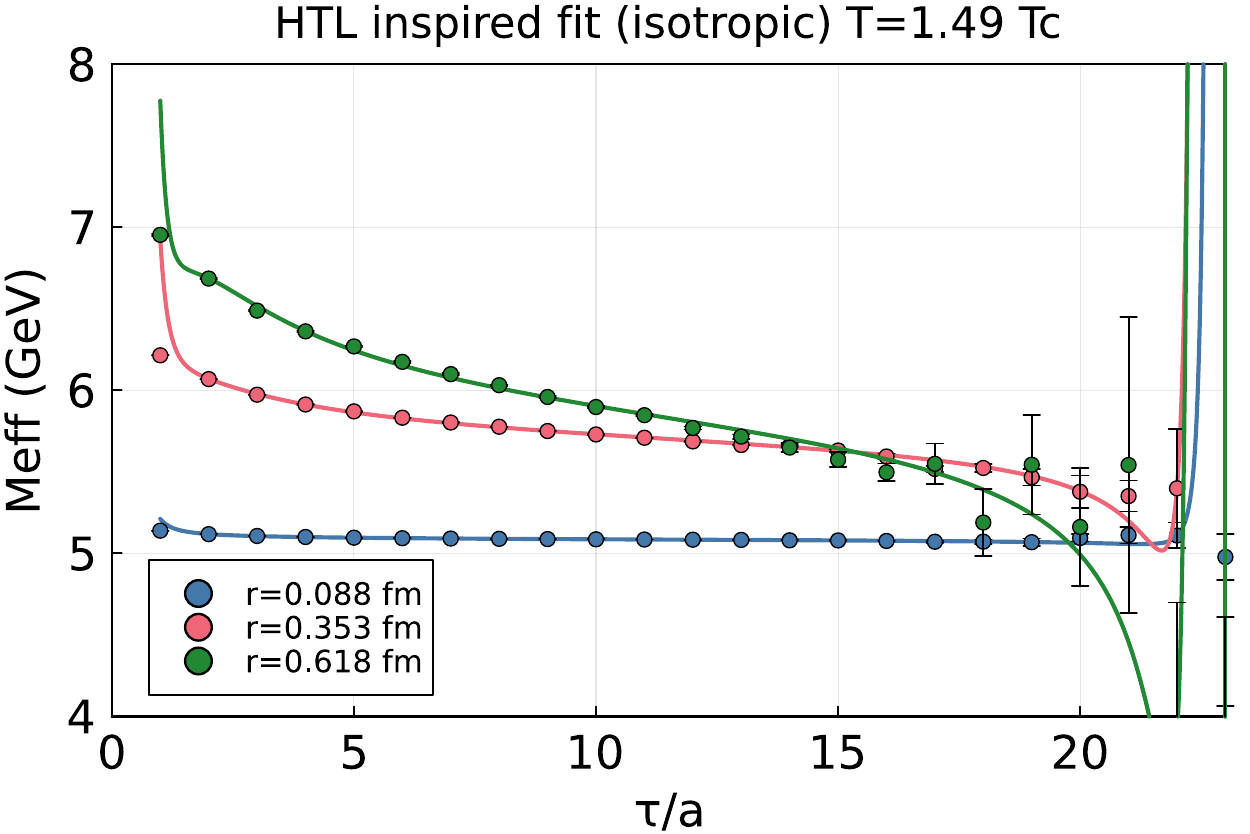}
    \includegraphics[scale=0.4]{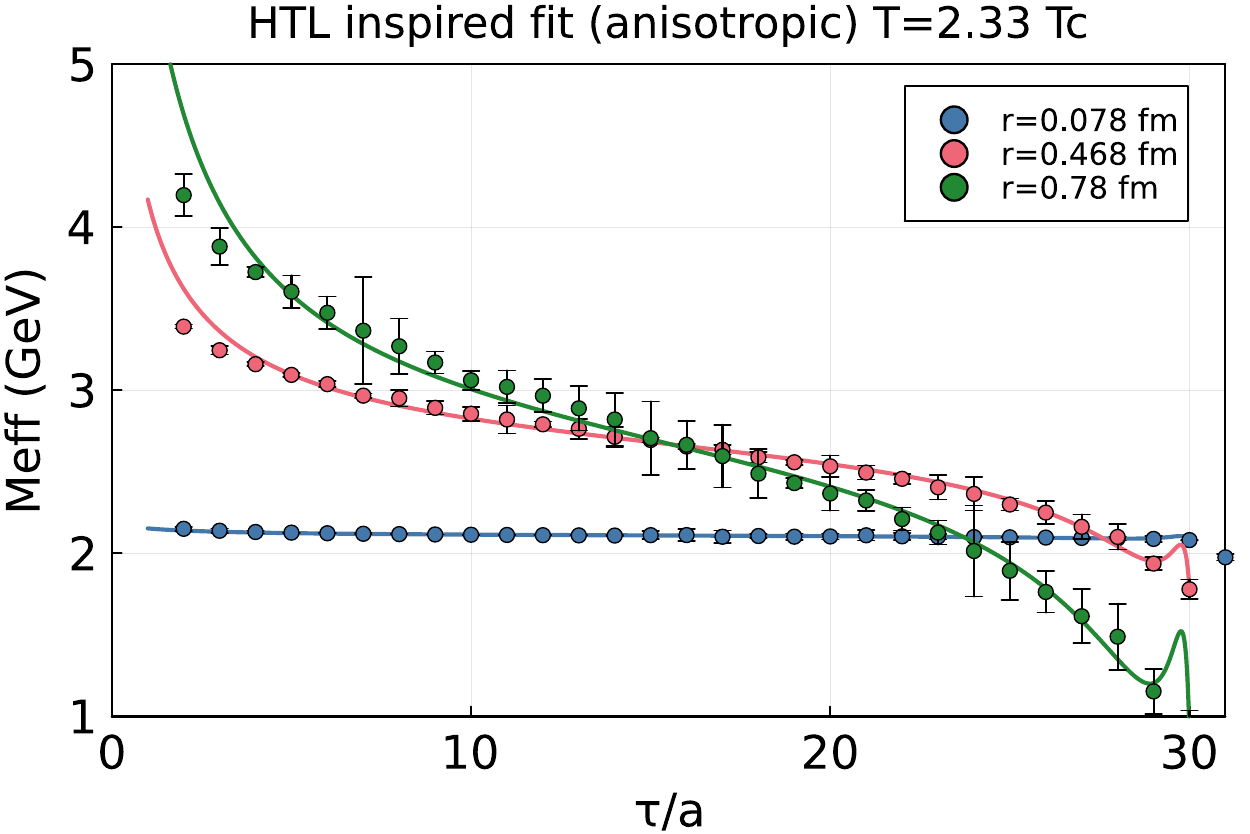}
    \caption{The figure shows HTL-inspired fits on effective mass data on isotropic (top) and anisotropic (bottom) lattices at three different separation distances.}
    \label{fig:HTL_fit}
\end{figure}

Next we show the HTL-inspired fits as outlined in \cref{sec:bala_datta} and use the fit form as described in \cref{eq:BDfit} to obtain the value of the potential. We were able to fit the data quite well upto a large $\tau$ region with two higher order terms $c_1$ and $c_2$. The fit on effective masses are shown in \cref{fig:HTL_fit}. 
We were able to get the fit within the margin of error for all but the smallest and largest $\tau$ data. We have shown the details of the fits in \cref{fig:HTL_fit_good_anisotropic} and \cref{fig:HTL_fit_good_isotropic} in the appendix. We find that the fits are somewhat worse than the Gaussian fits. For eg, considering the case of $T=2.33 Tc$  anisotropic lattices ($N_\tau = 32$), at a separation distance of $r=0.585$fm with first 5 and last 3 points excluded we get an error ($\chi$ as usual in statistics) of 4.16 per degree of freedom (number of $\tau$ points) as compared with the Gaussian fits where this number is 0.172 for the same. The HTL fits perform considerably better when the $\tau$ range is restricted close to $\tau = \beta/2$. The same quantity reduces to about 0.24 when the first 8 and the last 5 points are excluded in the HTL fits.
The somewhat counterintuitive observation that the HTL-inspired fits work better at lower temperatures is understood from the fact that a smaller fraction of the points are affected by discretization artifacts at $\tau \approx 0$ or $\tau \lesssim 1/T$ that is not accounted for at all in the HTL inspired Ansatz.

\subsection{Spectral Reconstructions}
\label{sec:Spec_lat}
In this section we use the Bayesian BR method and the Pad\'e interpolation method to reconstruct the spectral function from the Wilson line correlators.  We use a large frequency range  $\omega \in [-56,112]$ GeV for isotropic lattices and  $\omega \in [-101,202]$ GeV for anisotropic lattices  with $N_\omega = 2000$ in both cases, starting with the most uninformative default model i.e. $m(\omega)=\rm const$.  
\begin{figure}[!ht]
    \centering
    \includegraphics[scale=0.4]{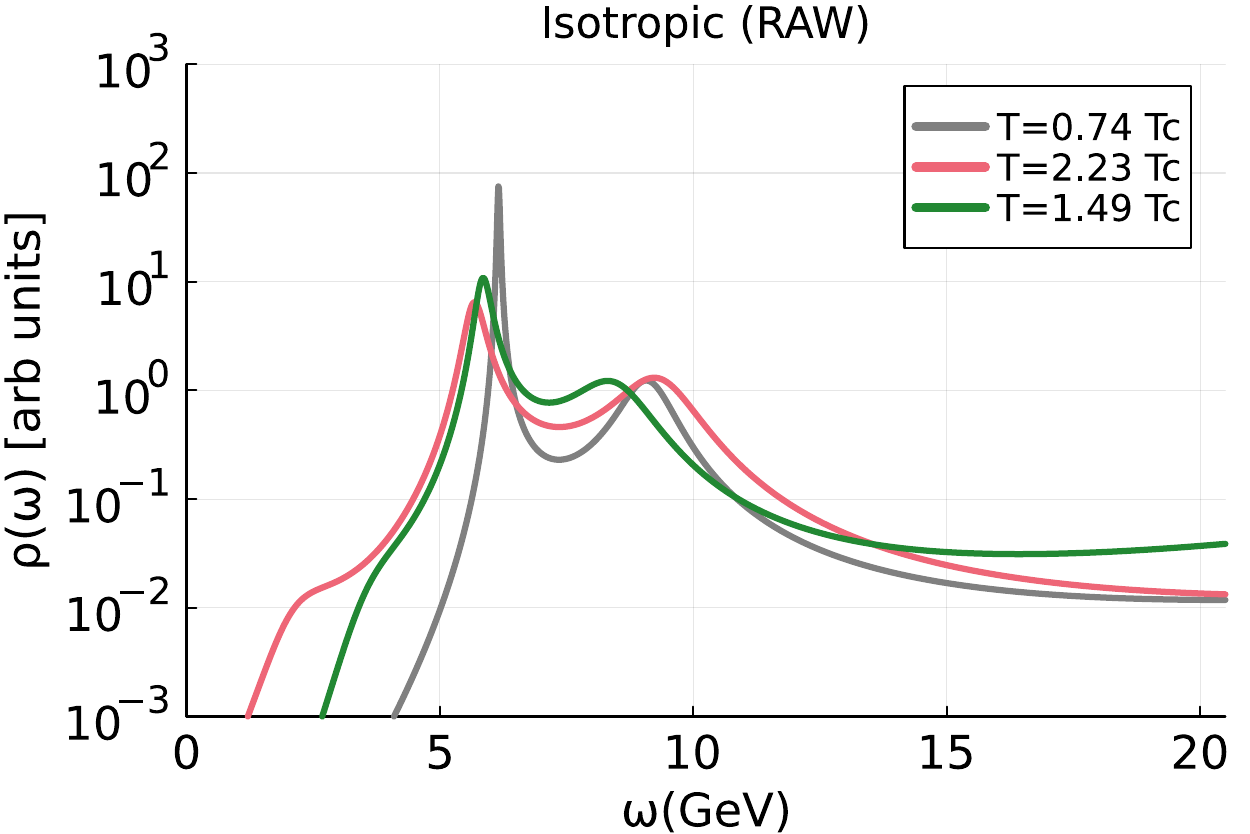}
    \includegraphics[scale=0.4]{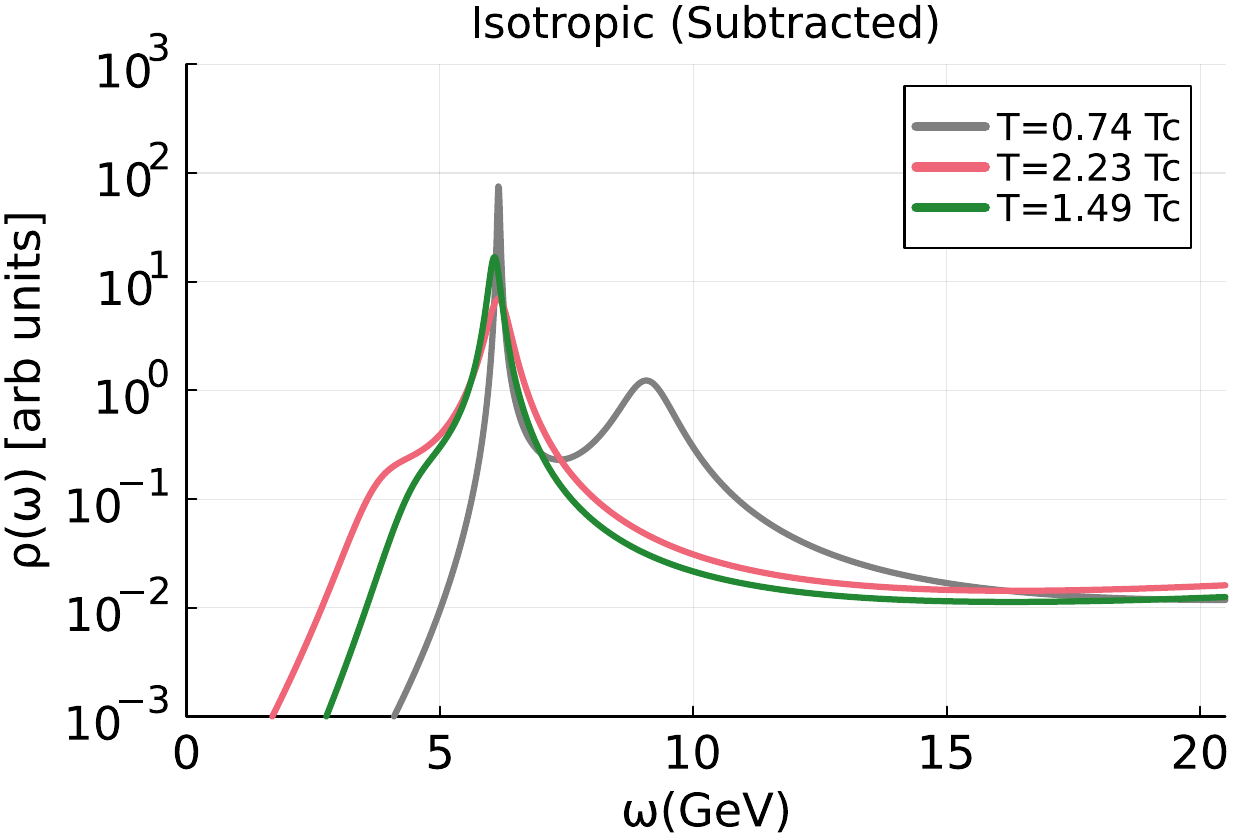}
    \caption{Comparison of BR reconstructed spectral functions at different temperatures $T\in (2.23,1.49,0.74) T_c$ at a separation distance of $r=0.53$fm. The figure on the top shows $\rho(\omega)$ for raw correlators and the figure on the bottom shows $\rho(\omega)$ for subtracted correlator.}
    \label{fig:BRspecs_iso_temp}
\end{figure}
The natural second step now is to compute the spectral functions from UV continuum subtracted data. This does present a challenge since the subtracted correlator will now have statistical errors dominated by the zero temperature data, since those are the ensembles where we have the lowest statistics (\cref{table:isotropic} and \cref{table:anisotropic}). The available statistics are thus reduced by a factor of 5-10 when the analysis will be carried out on the subtracted correlator. Nonetheless, we still perform the subtraction procedure (see \cref{sec:meff_sub}) and proceed with running the BR reconstructions. We proceed in the following way: the mean of the UV contribution is subtracted from each measurement of the correlator. Then, we account for the larger errors in this new subtracted data by multiplying the covariance matrix with the appropriate factor which is determined by comparing relative errors of the zero and finite temperature data. The actual procedure for spectral reconstruction is the same as with the raw correlators.
\begin{figure}[!ht]
    \centering
    \includegraphics[scale=0.4]{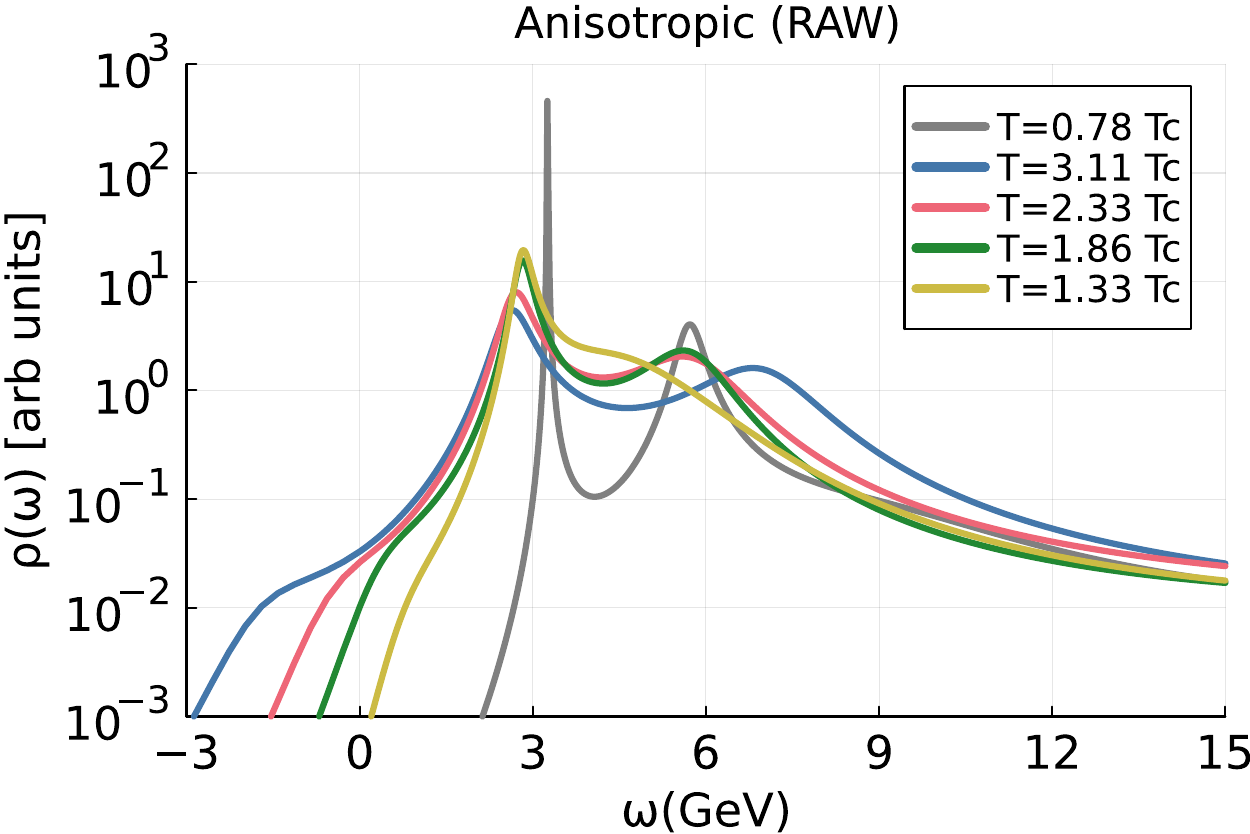}
    \includegraphics[scale=0.4]{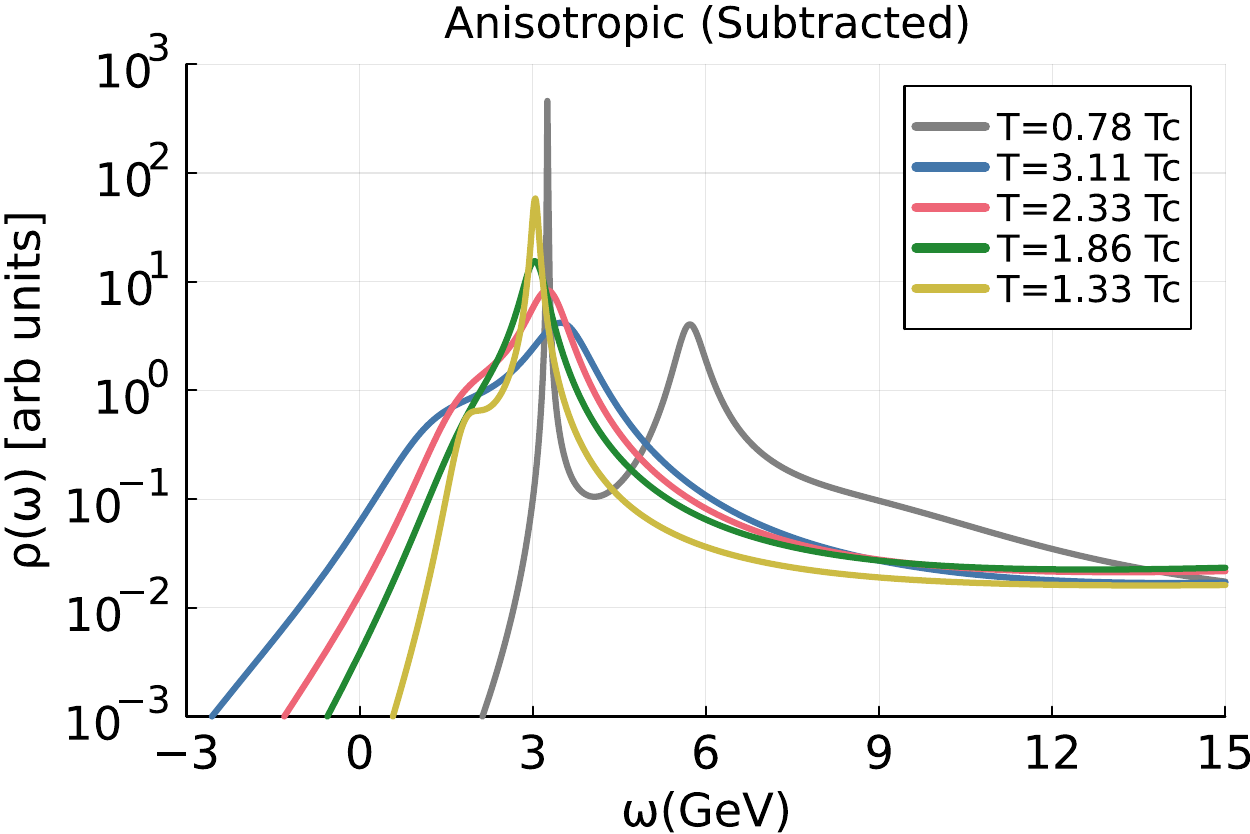}
    \caption{Comparison of BR reconstructed spectral functions at different temperatures at a separation distance of $r=0.702$ fm. The figure on the top shows functions for raw correlators and the figure on the bottom shows the functions for subtracted correlator.}
    \label{fig:BRspecs_aniso_temp}
\end{figure}
We first show a comparison of the spectral functions at different temperatures at a fixed separation distance of $r=0.53$fm in \cref{fig:BRspecs_iso_temp} (isotropic) and $r= 0.702$fm in 
\cref{fig:BRspecs_aniso_temp} (anisotropic).

When compared to reconstructed spectral functions of HTL Wilson Line correlators we see more structure in the spectral function. The spectral functions from the raw data on isotropic lattices show some shoulder structures at small positive frequencies below the main dominant peak, whereas in the case of anisotropic lattices the shoulder structures appear at a negative frequency. In addition, we also find other structures at higher frequencies (around 8-10 GeV for isotropic and 5-6 GeV for anisotropic lattices) which correspond to UV continuum artifacts affecting the small $\tau$ behaviour in the correlator. 
Performing the same analysis on the subtracted correlator still shows a dominant peak and some shoulder structures at small frequencies remain, but the frequencies at which these structures appear are slightly higher and the amplitude is also lower. 
These changes might be due to $\tau$-dependent statistical fluctuations of the $T=0$ correlators at large times, where one would otherwise naively expect just the stable ground state contribution. 
Comparing the raw or subtracted effective masses at large $\tau$ in full QCD \cite{Bala:2021fkm, Bazavov:2023dci} suggests milder changes in the low-frequency region of the spectrum, which however, is known to be strongly operator dependent and susceptible to lattice artifacts.
The structures at higher frequencies that we observe from the raw data do not appear anymore, suggesting that the subtraction procedure has indeed removed the high frequency parts of the spectral function as intended. In addition, the position of the dominant peak in the spectral function also appears to have shifted to slightly higher frequencies. In all cases the width of the spectral peak increases with increasing temperatures, which suggests the presence of an imaginary part which grows with temperature.

\begin{figure}[!ht]
    \centering
    \includegraphics[scale=0.4]{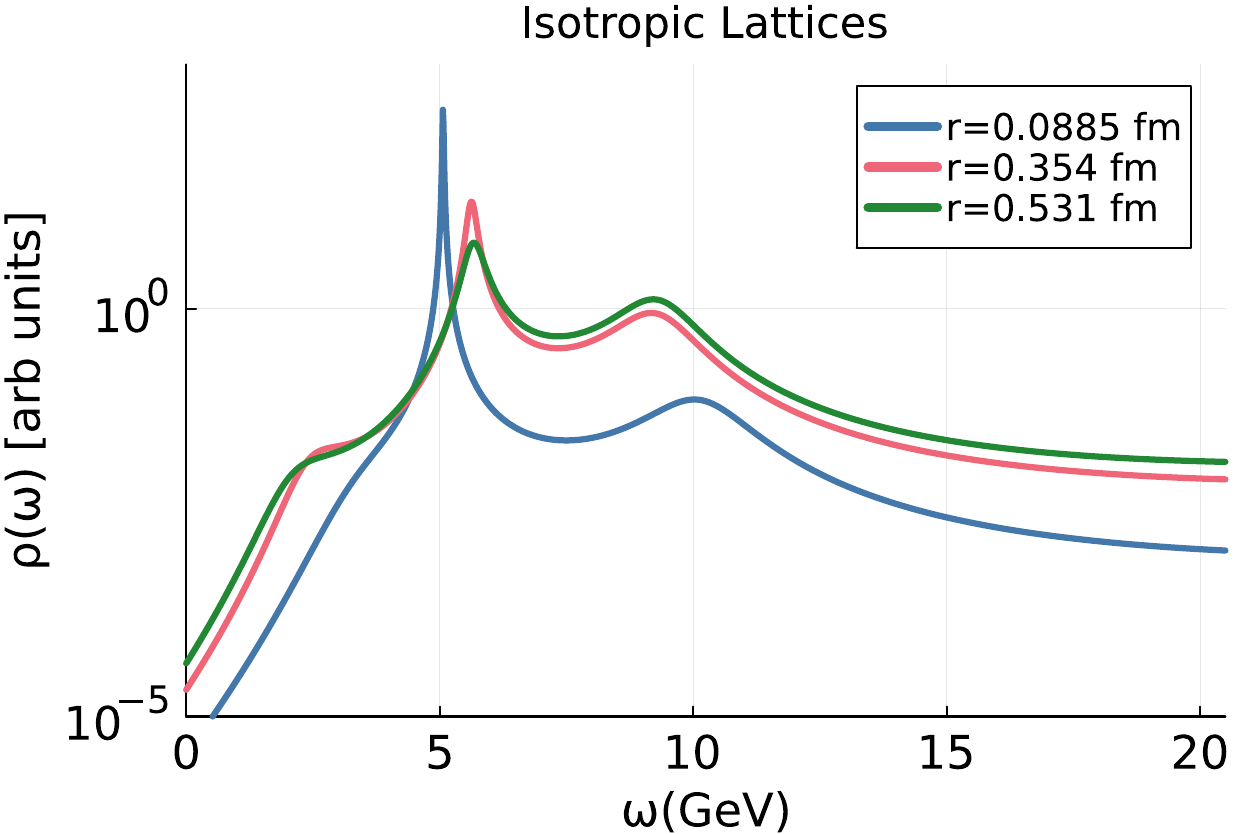}
    \includegraphics[scale=0.4]{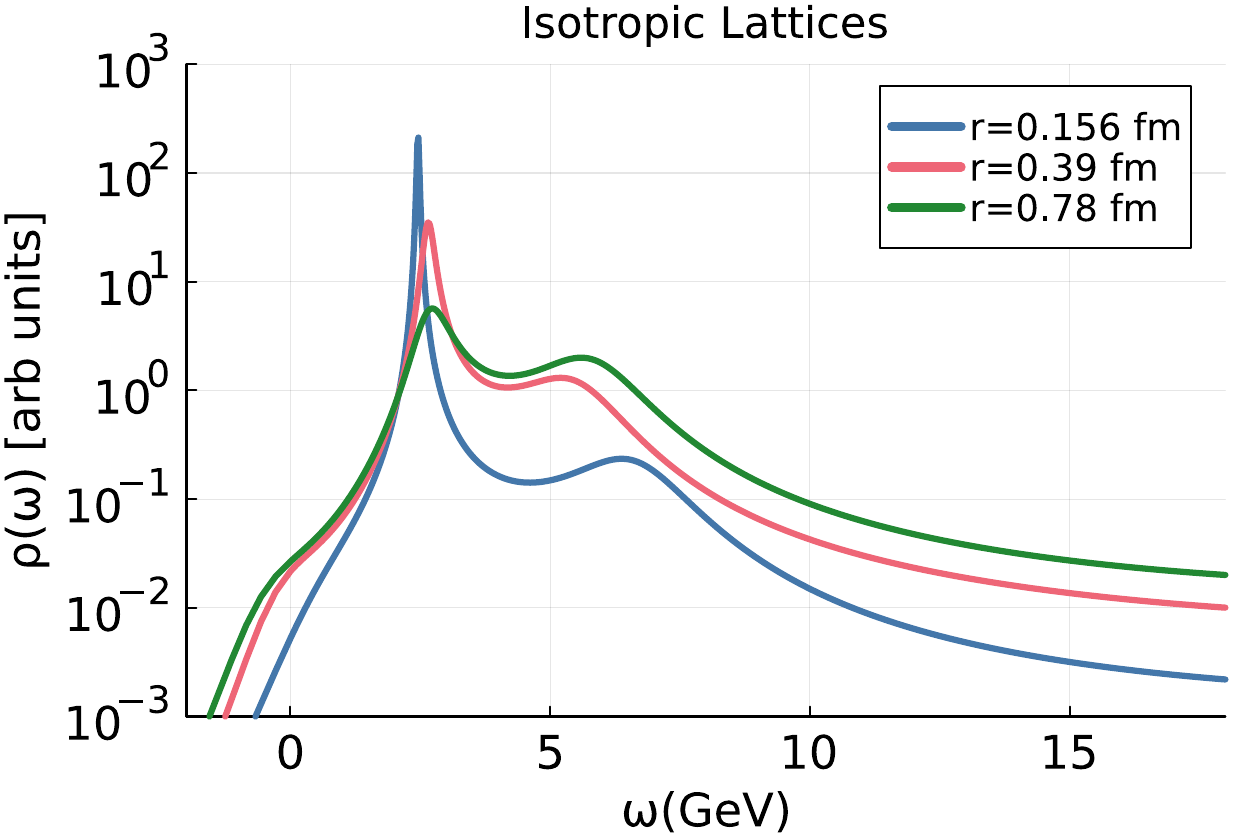}
    \caption{BR reconstructed spectral functions of the (top) isotropic lattices at $T=2.23 T_c$ and (bottom) anisotropic lattices at $T=2.33 T_c$ using the raw correlator data. The three curves each denote spectra at different spatial separation distances.}
    \label{fig:BRspecs_lattice}
\end{figure}

\begin{figure}[!ht]
    \centering
    \includegraphics[scale=0.4]{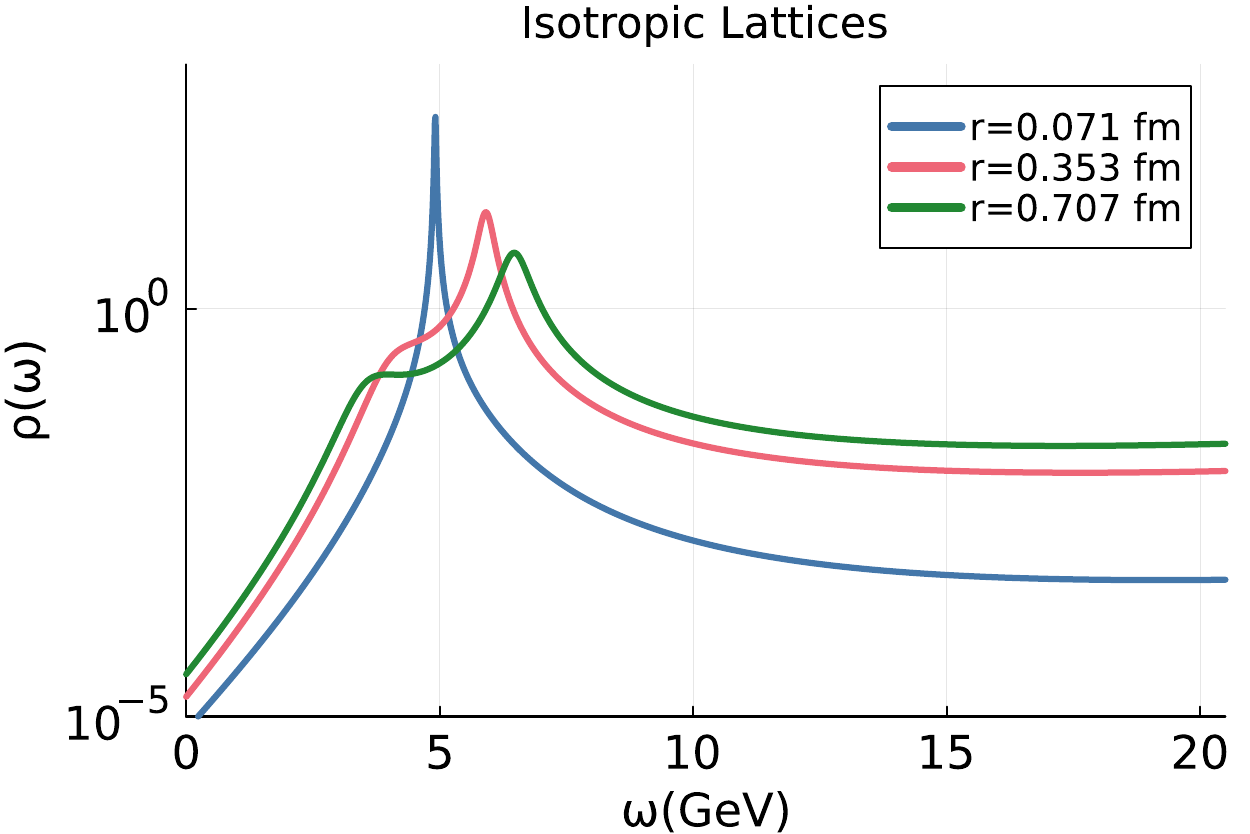}
    \includegraphics[scale=0.4]{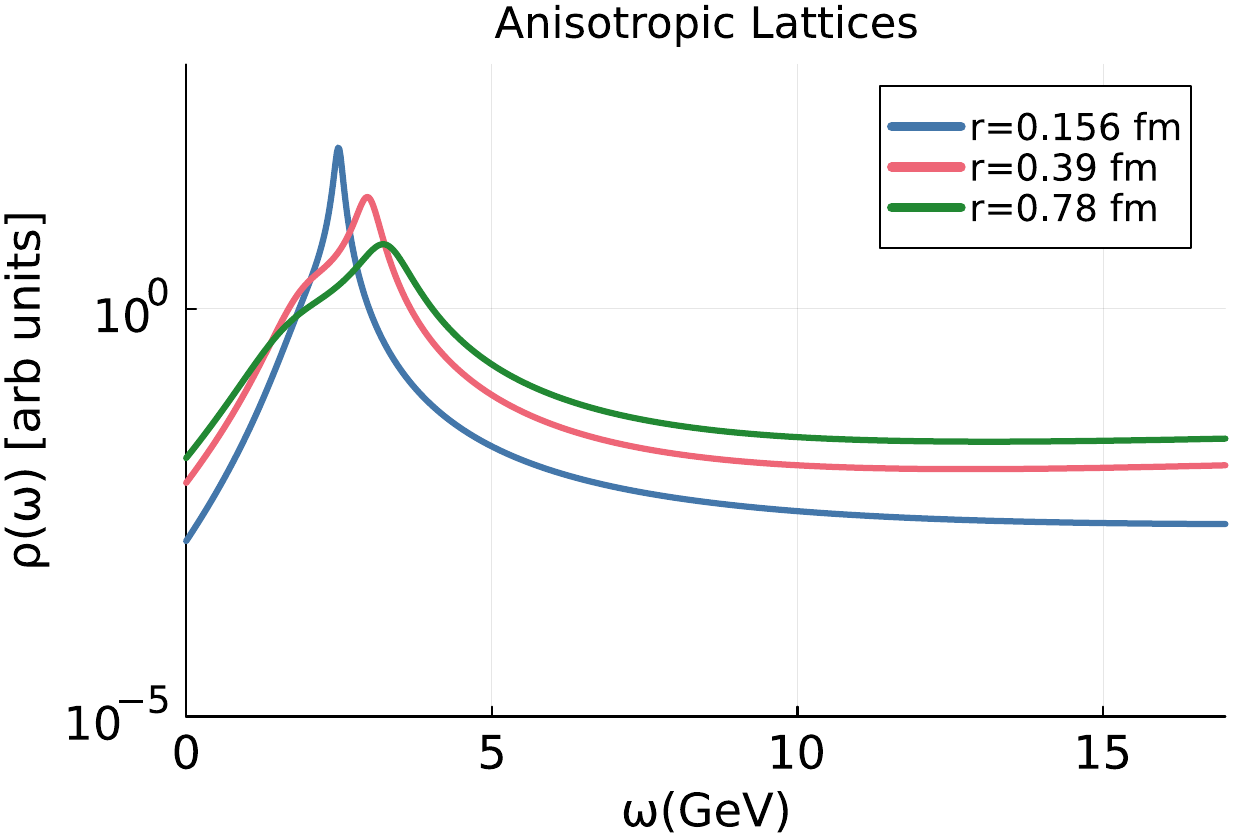}
    \caption{BR reconstructed spectral functions of the isotropic lattices at $T=2.23 T_c$ and anisotropic lattices at $T=2.33 T_c$ using the subtracted correlator data. The three curves each denote spectra at different spatial separation distances.}
    \label{fig:BRspecs_sub}
\end{figure}

Next we compare the spectral functions at a fixed temperature for different separation distances to analyse the change in peak position and width. \Cref{fig:BRspecs_lattice} shows spectral reconstructions using the BR method for raw correlators $T=2.23 T_c$ (isotropic) and $T=2.33 T_c$ (anisotropic) at different separation distances and \cref{fig:BRspecs_sub} shows the same for the subtracted correlators. We observe that the position of the dominant peak for raw correlators stabilises at a fixed position at around $r=0.45$fm (isotropic) and $r=0.39$fm (anisotropic). This suggests the presence of screening in the real part of potential. 
However, for the subtracted correlators no stabilization of the peak position is seen, and hence, no screening.
We hence confirm results obtained in previous studies. The width of the spectral peak also appears to broaden over increasing separation distance suggesting the presence of an increasing imaginary part with separation distance in the potential. We  discuss the potential extraction from spectral functions in \cref{sec:ReV_lat} where we also discuss the systematic and statistical error-budget in our analysis where is presented.


\begin{figure}[!ht]
    \centering
    \includegraphics[scale=0.4]{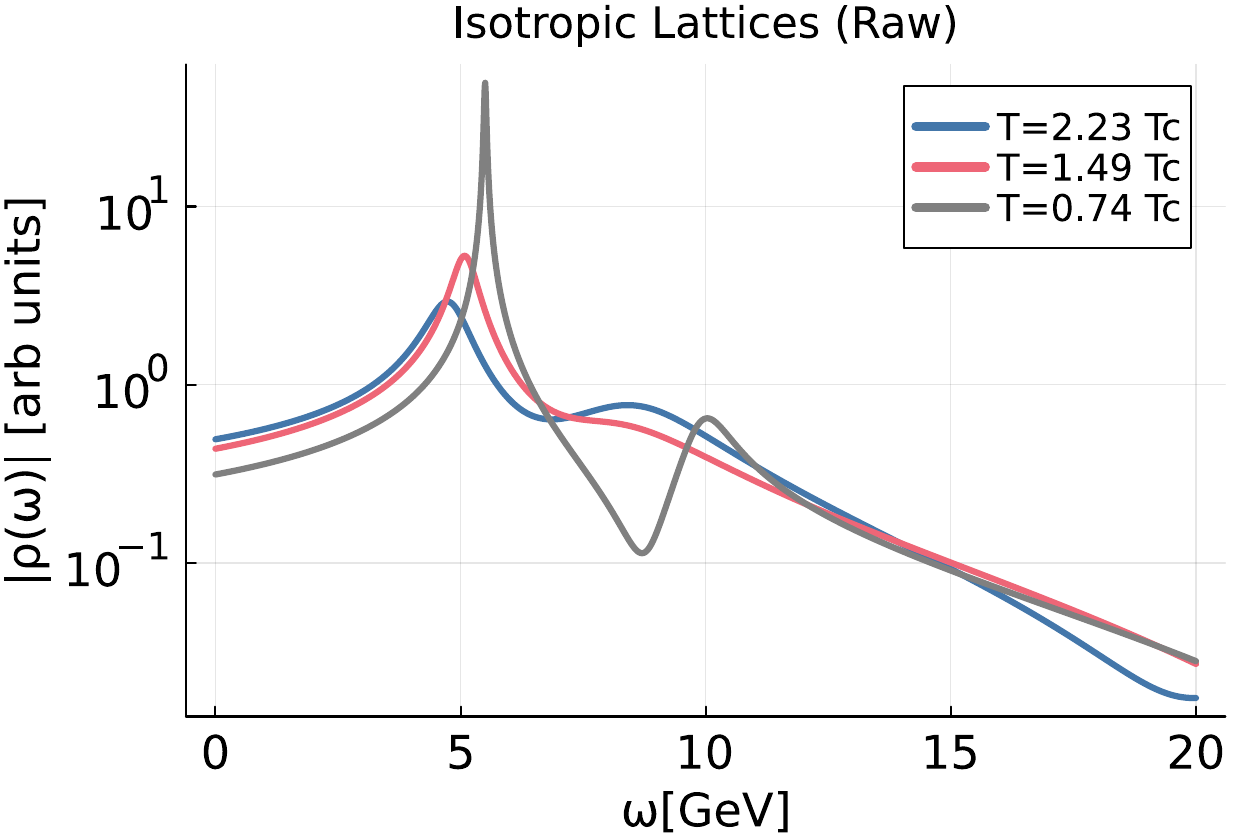}
    \includegraphics[scale=0.4]{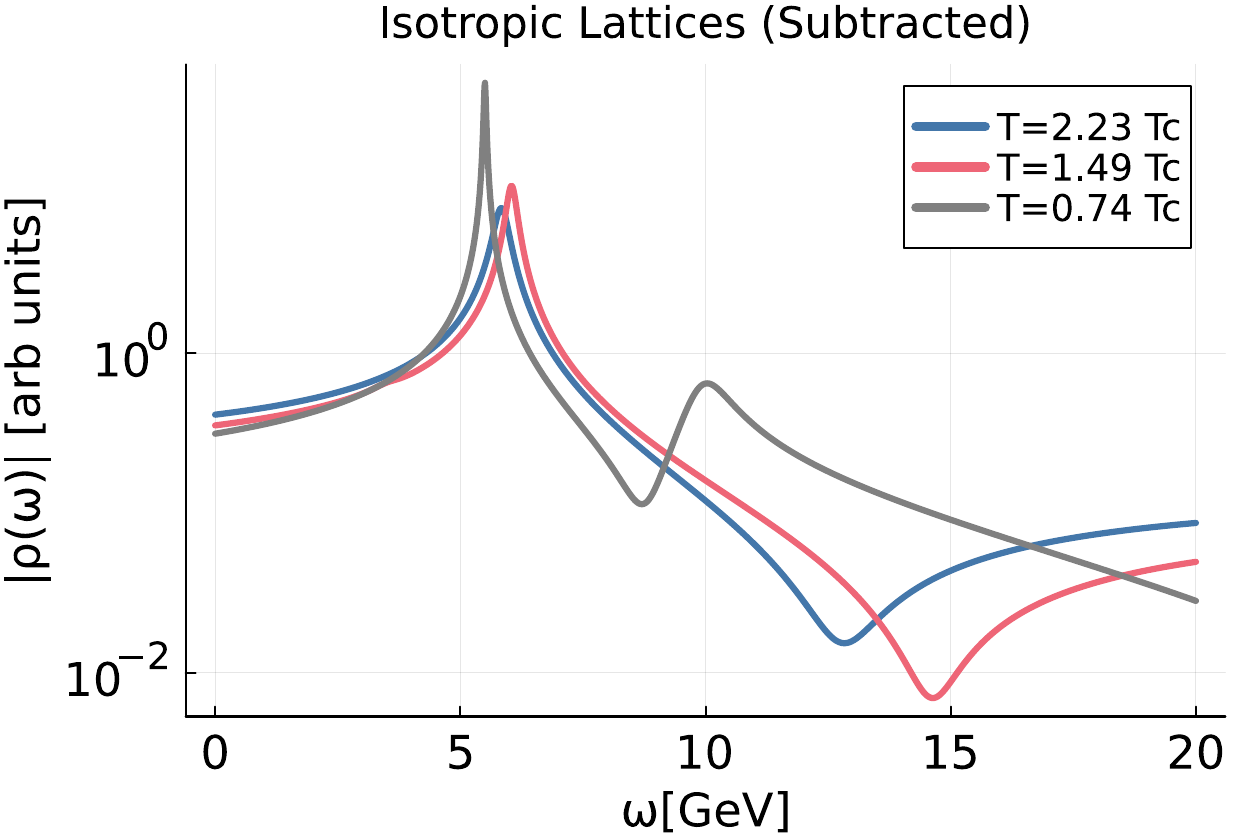}
    \caption{Comparison of Pad\'e reconstructed spectral functions at different temperatures at a separation distance of $r=0.53$ fm. The figure on the top shows functions for raw correlators and the figure on the bottom shows the spectral functions for subtracted correlator. The zero temperature reconstruction is performed on the raw correlator in both the plots.}
    \label{fig:padespecs_iso_temp}
\end{figure}

We proceed to do the same with the Pad\'e interpolation reconstruction. We will display here the spectral functions obtained from the Pad\'e, however, note that we compute the potential from the pole analysis in the next section. As mentioned earlier, the Pad\'e requires very high statistics, which makes it less reliable on the subtracted correlator. The subtraction procedure is carried out in a slightly different way from the BR reconstructions. We divide our data into 16 jackknife bins and first obtain the high-frequency part of the $T=0$ correlator for each bin. This continuum part is then subtracted from the finite temperature data for each jackknife bin. This was only possible for the isotropic lattices where the statistics were high enough, the interpolation fails for the anisotropic case. \Cref{fig:padespecs_iso_temp} shows a comparison of the Pad\'e extracted spectral functions on isotropic lattices for the raw correlators (top) and subtracted correlators (bottom) at a separation distance of $r=0.53$fm. Firstly, the spectral peak of the vacuum ($T=0.74 T_c$) is now somewhat broader than the peak obtained from the BR reconstruction on the same lattice. The spectral functions for raw correlators at all temperatures show a main dominant peak, followed by a second bump at high frequencies at about 7-12 GeV. These bumps correspond to high frequency structures in the UV which dominate the correlator at small $\tau$. These bumps are not observed in the spectral functions of the subtracted correlator. This is an additional indication that the  structures in the UV are indeed removed by the subtraction procedure. The shoulder structures that are observed in the BR also seem to be absent in the Pad\'e reconstructions which could very well be because the Pad\'e is unable to capture them. Furthermore, the subtraction procedure changes the peak position of the spectral function in a non-trivial way. The peak has been shifted to a higher frequency for both the temperatures alike. The reconstructed peak for the subtracted correlator now even sits at a higher position than the vacuum peak obtained from the raw correlators. This suggests that the presence of some non-trivial structures at high frequency can affect the Pad\'e and could lead it to underestimate the peak position or that the subtraction affects the low lying part of the spectral function itself.
This is reminiscent of the changes to the low-frequency shoulder structures seen in the BR reconstructed spectra.

\begin{figure}[!ht]
    \centering
    \includegraphics[scale=0.4]{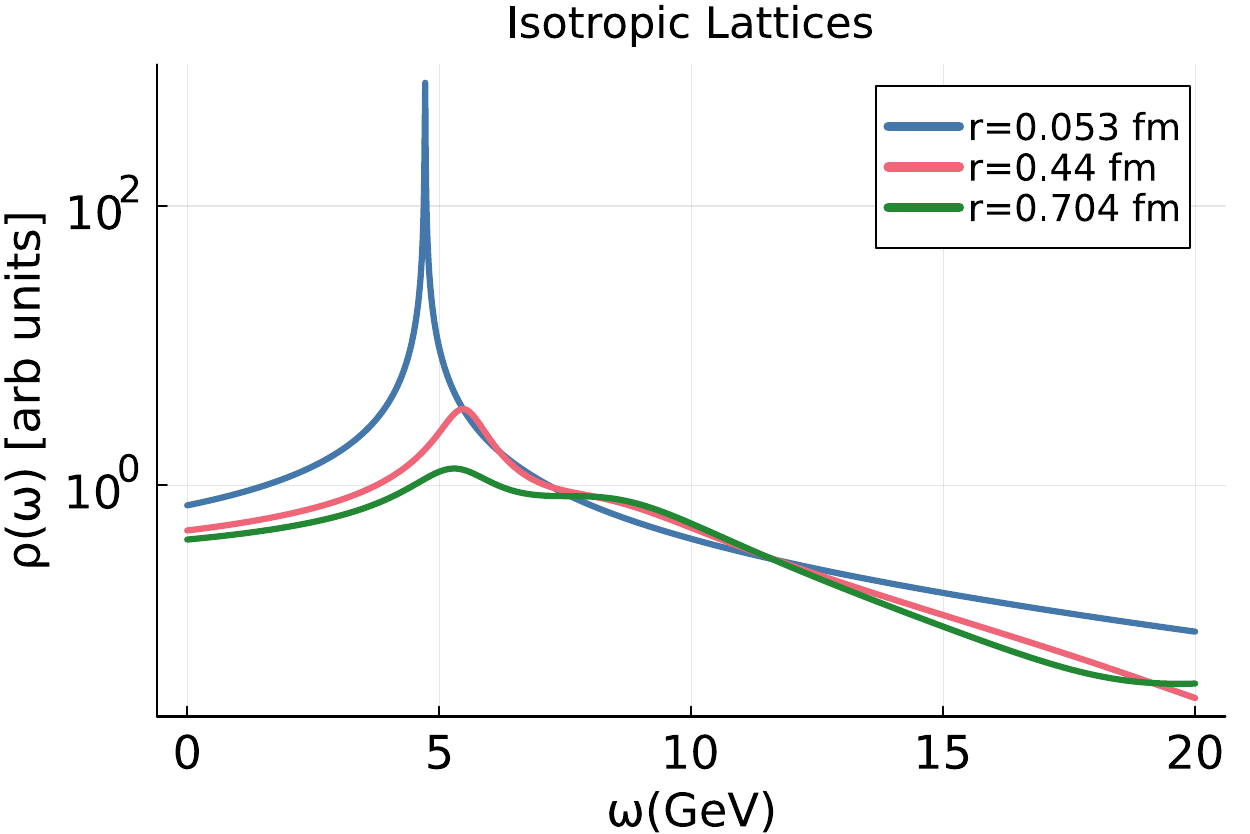}
    \includegraphics[scale=0.4]{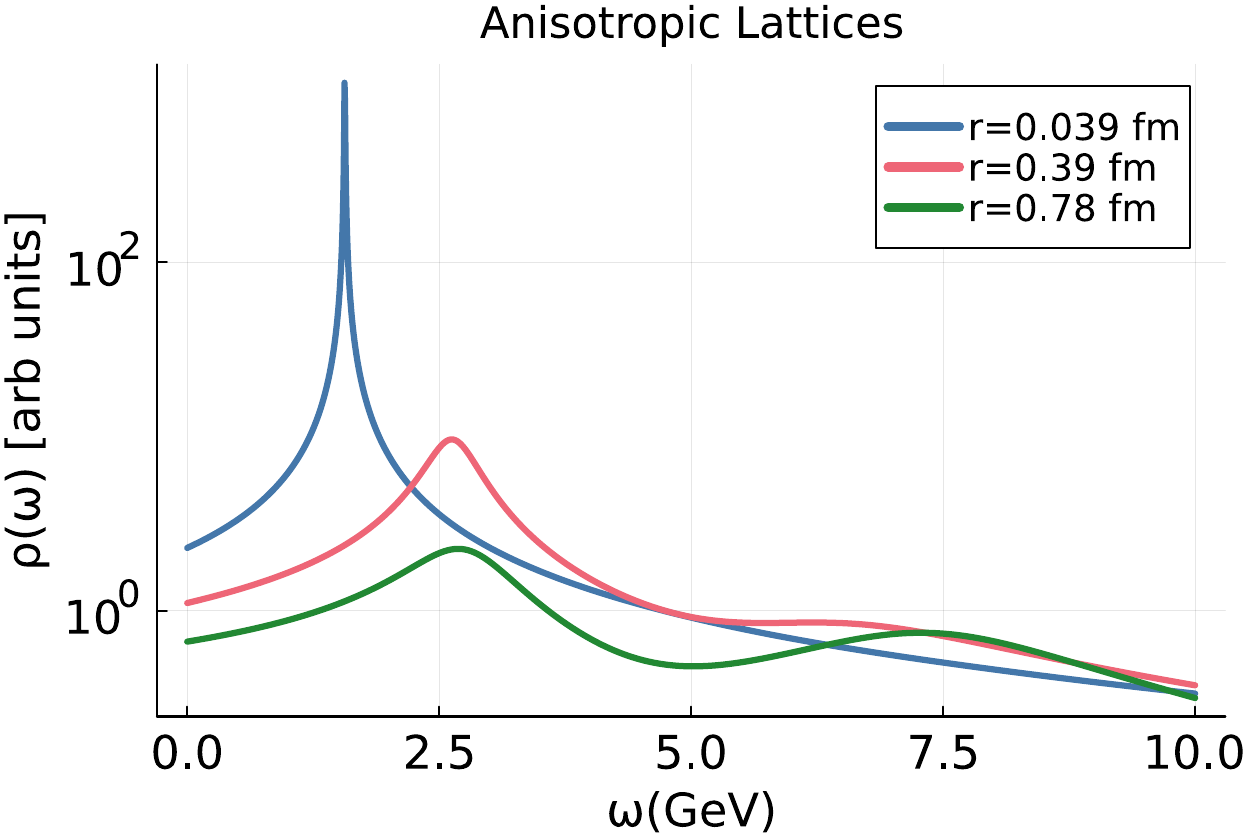}
    \caption{Pad\'e interpolation reconstructed spectral functions of the (top) isotropic lattices at $T=2.23 T_c$ and (bottom) anisotropic lattices at $T=3.11 T_c$ using the raw correlator data. The three curves each denote spectra at different spatial separation distances. }
    \label{fig:Padespecs_lattice}
\end{figure}

\begin{figure}[!ht]
    \centering
    \includegraphics[scale=0.4]{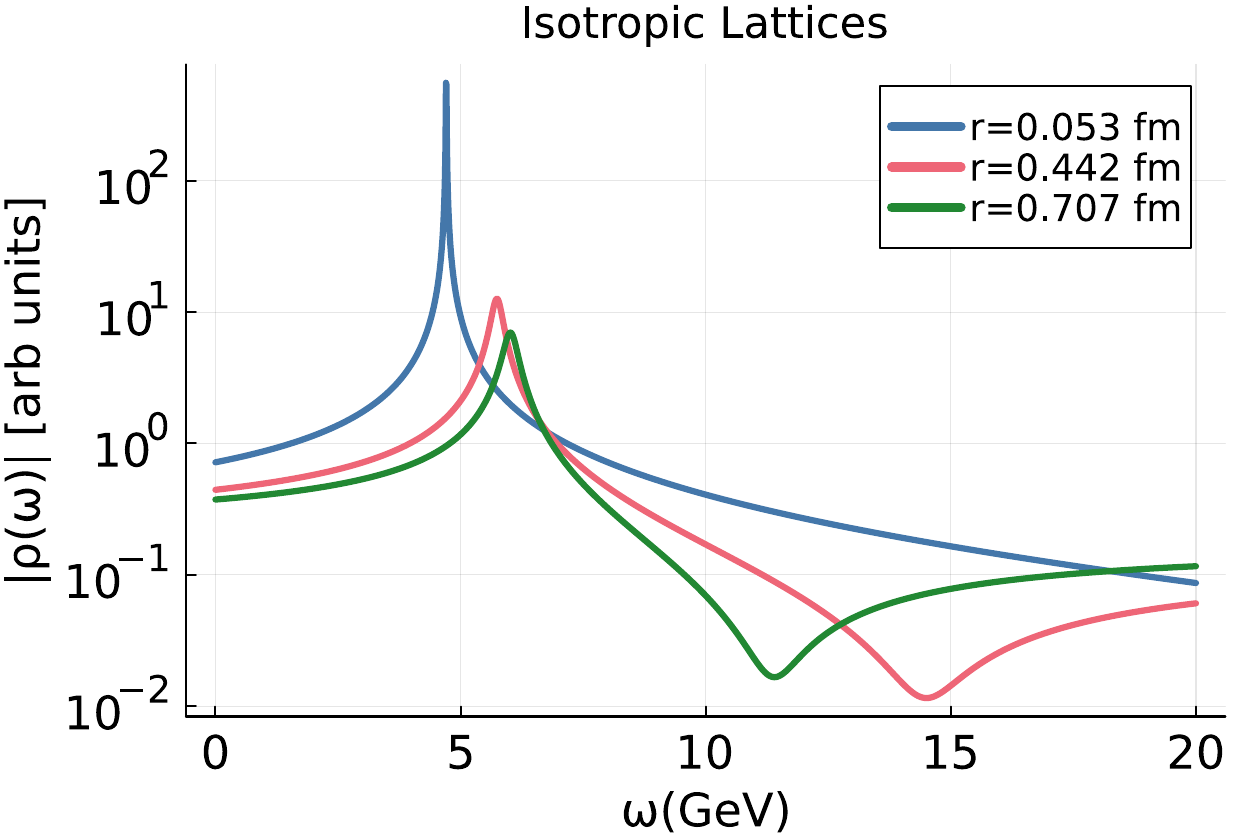}
    \caption{Pad\'e interpolation reconstructed spectral functions of the isotropic lattices at $T=2.23 T_c$ using the subtracted correlator data. The three curves each denote spectra at different spatial separation distances.}
    \label{fig:Padespecs_sub}
\end{figure}

Next, we compare the spectral functions at the same temperatures but at different separation distances. \cref{fig:Padespecs_lattice} shows spectral functions of the raw correlator on isotropic (top) and anisotropic (bottom) lattices and  \cref{fig:Padespecs_sub} shows the same for the subtracted correlator but now only for the isotropic lattices. We find that when using the raw correlator, the position of the dominant peak approaches a constant at a separation distance (around 0.4 fm) suggesting the presence of screening in the real part of the potential. This changes drastically when using the subtracted correlator as the peak position keeps increasing even up to separation distances of 0.7 fm. This behaviour is not compatible with a screened potential and could be interpreted as a vacuum-like rising real part.

We have described and compared spectral functions of the raw and subtracted correlators on both isotropic and anisotropic lattices in this section. Visual analysis suggests that the subtraction procedure removes high frequency structures. At the same time the subtraction affects the reconstructed peaks at small frequency. It also suggests that the raw correlators show signs of screening confirming results from previous studies. However, the subtracted correlators show behaviour compatible with the vacuum. In the next section we will present the real part of the potential extracted using all the methods discussed in this paper along with providing statistical and systematic uncertainty budgets.

\subsection{Real part of the potential}
\label{sec:ReV_lat}
\begin{figure}[!ht]
    \includegraphics[scale=.4]{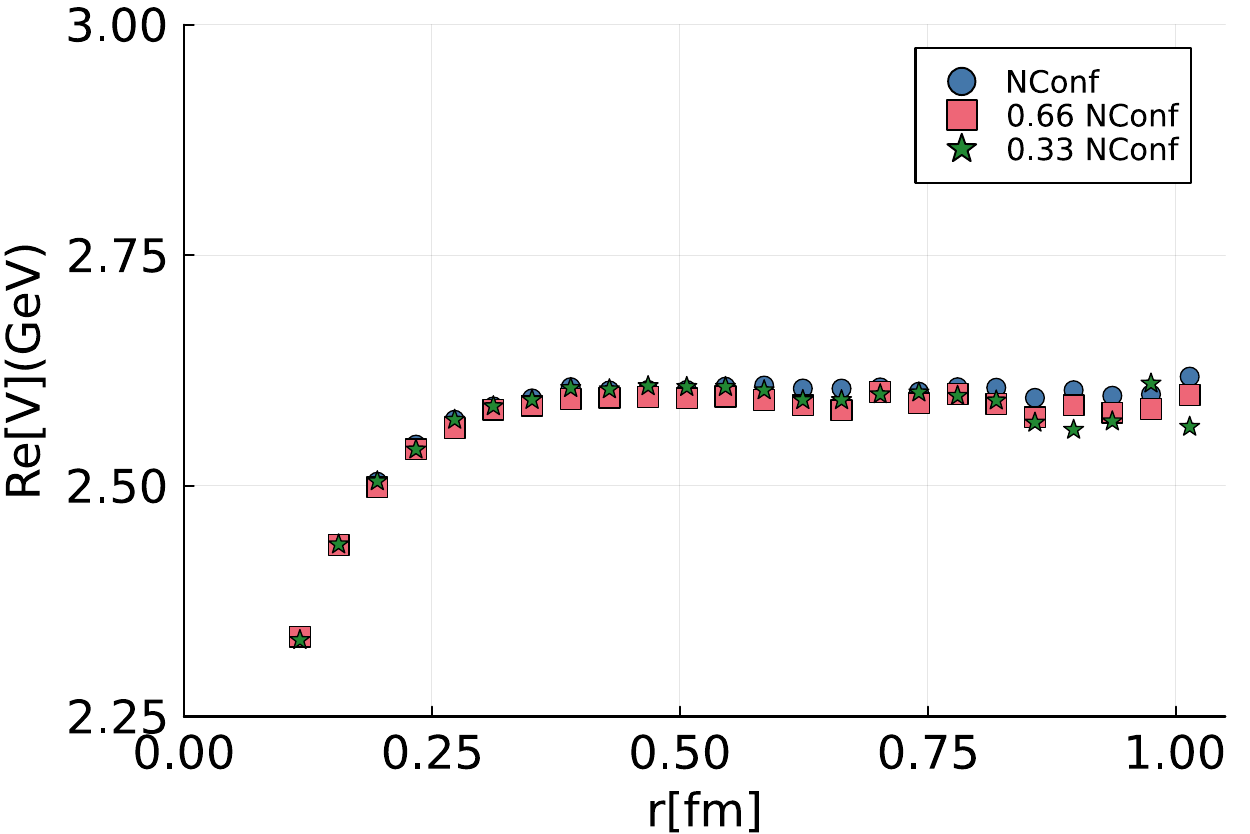}
    \includegraphics[scale=.4]{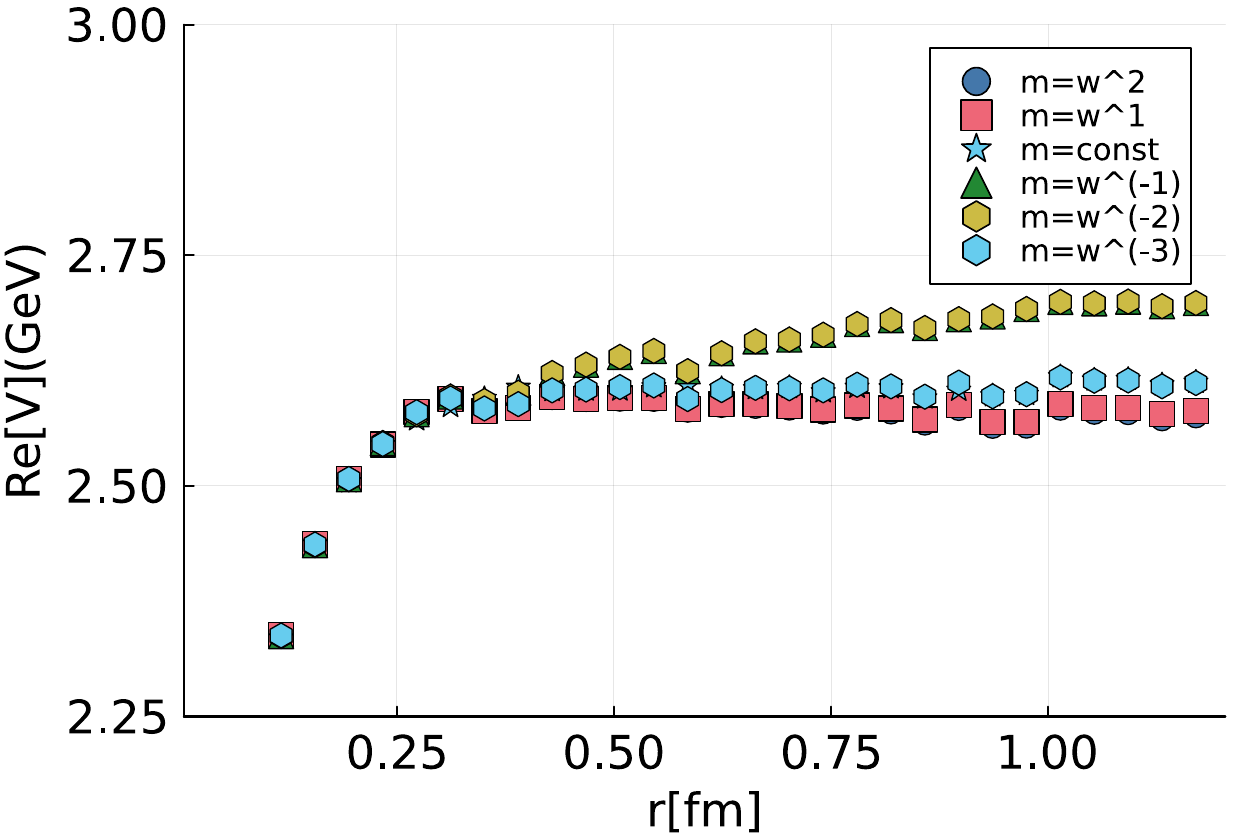}    
    \caption{Assessing the dependence of the Bayesian reconstruction extracted values of ${\rm Re}[V]$ on (top) variation in the statistical error of the input data (bottom) on changes in the default model
    (anisotropic ensemble at $T=3.11 T_c$).}
    \label{fig:Re_V_BR_stats}
\end{figure}

We now discuss the real part of the potential from each of the methods described in the previous sections. We extract the potential from spectral functions by fitting to a functional form according to \cref{eq:PotFitFunc}. In HTL mock data tests we observed that the BR reconstructions can approach a failure mode resulting in ringing behaviour (see \cref{fig:HTLBRdD3spec}) when the errors on the data are reduced without increasing the number of data points. This could result in a sudden jump in the extracted potential above at a certain separation distance (see \cref{fig:HTLBRdD3pot}). To check whether we observe such effects in the lattice data we artificially reduce the statistics on our data to a factor of 2/3 and 1/3 of total measurements. The top part of \cref{fig:Re_V_BR_stats} shows the real part of potential extracted using a constant default model for with varying statistics at $T=3.11 T_c$ ($N_\tau =24$) on anisotropic lattices. Seeing no jump in the extracted real part~suggests the reconstruction is unaffected by ringing artifacts.

\begin{figure}[!ht]
  \includegraphics[scale=.4]{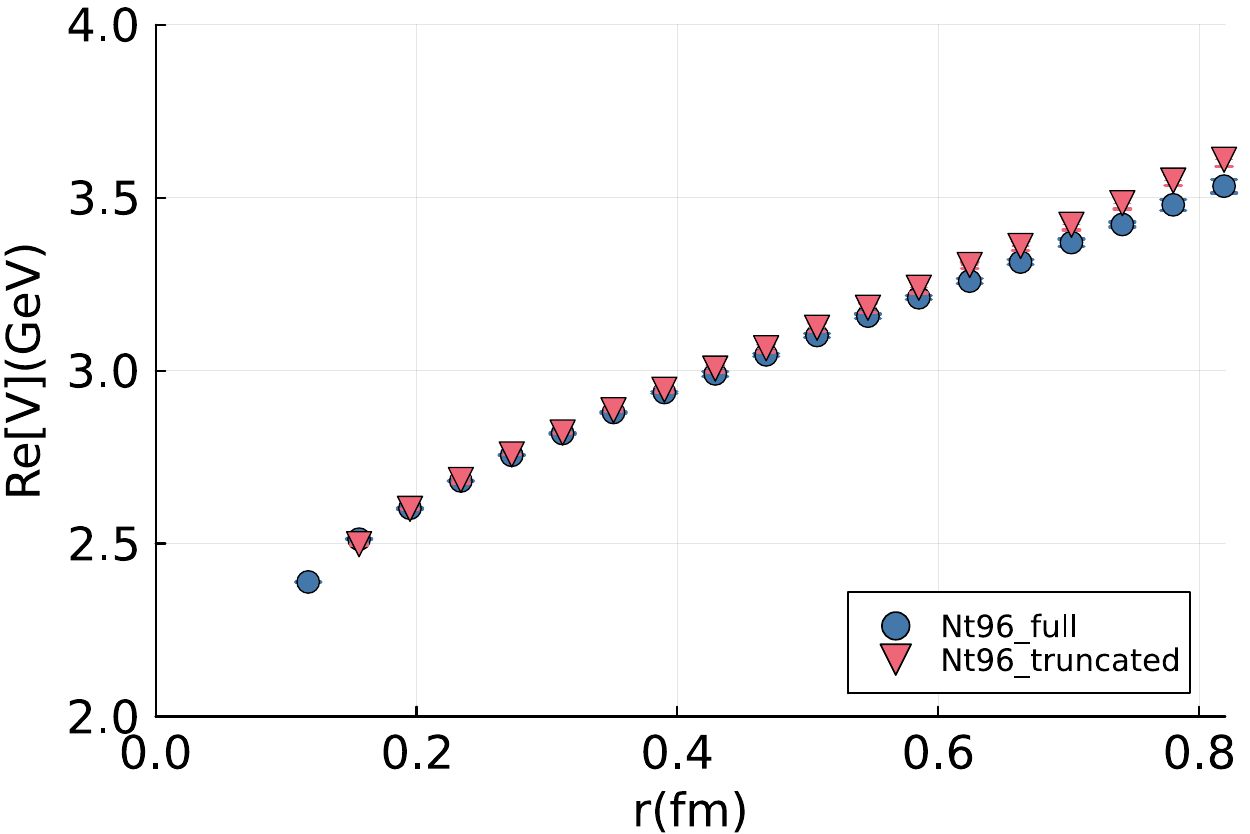}
    \caption{Assessing the dependence of the Bayesian reconstruction at $T=0.78T_C$ on the available extent in imaginary time. Each of these was computed using the full statistics with the same constant default model (isotropic ensemble at $T=0.78 T_c$).}
    \label{fig:Re_V_BR_trunc}
\end{figure}

We then proceed with estimating the systematics of the extracted potential. Generally, the least informative default model is used in Bayesian reconstruction i.e. a constant default model. To test the default model dependence we choose four different default models in addition to the constant default model;  $m(\omega) = \frac{m_0}{(\omega - \omega_{min} + 1)^\kappa}$ where $\kappa = [1,2,-1,-2]$. For the highest temperature ($N_\tau=24$ anisotropic lattices) we also choose an additional value $\kappa=-3$ since we expect the dependence to be the highest at this temperature. The resulting values of the real part for the anisotropic lattices of $T=3.11 T_c$ ($N_\tau=24$) is shown in the bottom panel of \cref{fig:Re_V_BR_stats}. We estimate the systematic error in the extracted potential from the maximal variation among these data points. We point out that the residual slope for some of the default models is much weaker than the slope due to the effective string tension in the confined phases, c.f. \cref{fig:Re_V_BR_trunc}, such that the default model dependence does not hinder distinction between screening or no screening.

In our simulations, the temperature is changed by changing the number of lattice points while keeping the lattice spacing fixed. 
Thus, at higher temperatures we have fewer imaginary time steps and thereby less information. 
Since it is known from precise data in full QCD~\cite{Brambilla:2022het} that the low-lying excited states become negligible at $T=0$ only after $\tau \gtrsim 0.4$ fm, one might be concerned that the $T>0$ analysis could be affected by uncontrolled excited state contamination due to the restricted time range.
To investigate these effects, we carry out the BR reconstruction on the low temperature ($T=0.78 T_c$, $N_\tau=96$) data with artificially truncated Euclidean time range to $\tau/a<24$ as shown in \cref{fig:Re_V_BR_trunc}. 

We observe that truncating the data to $N_\tau=24$ introduces a small upward shift in the potential in line with the expected excited state contamination. This behaviour is similar to what was observed in a previous study on lattice NRQCD S- and P-wave bottomonium states \cite{Kim:2014iga}. This artifact only leads to over-estimation of $\textrm{Re}[V]$ and does not mimic artificial screening.

\begin{figure}
    \includegraphics[scale=.4]{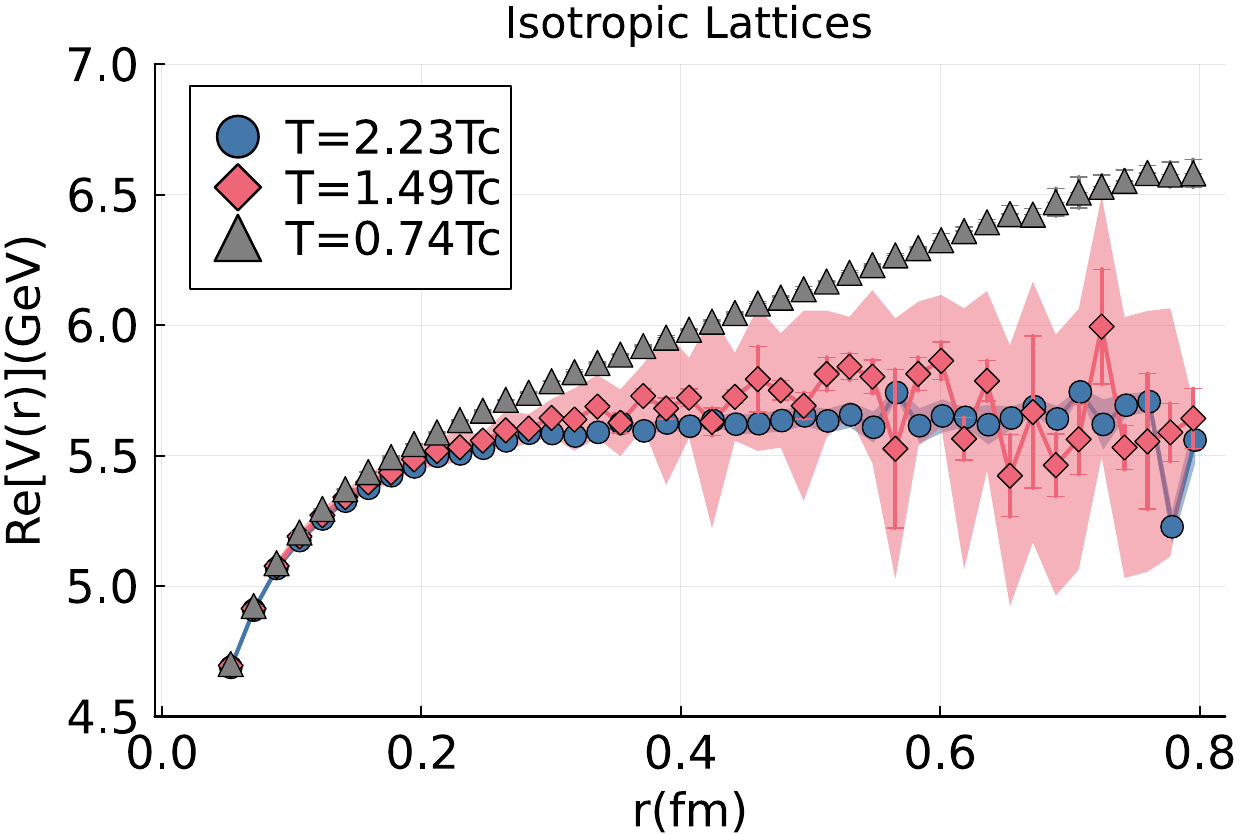}
    \includegraphics[scale=.4]{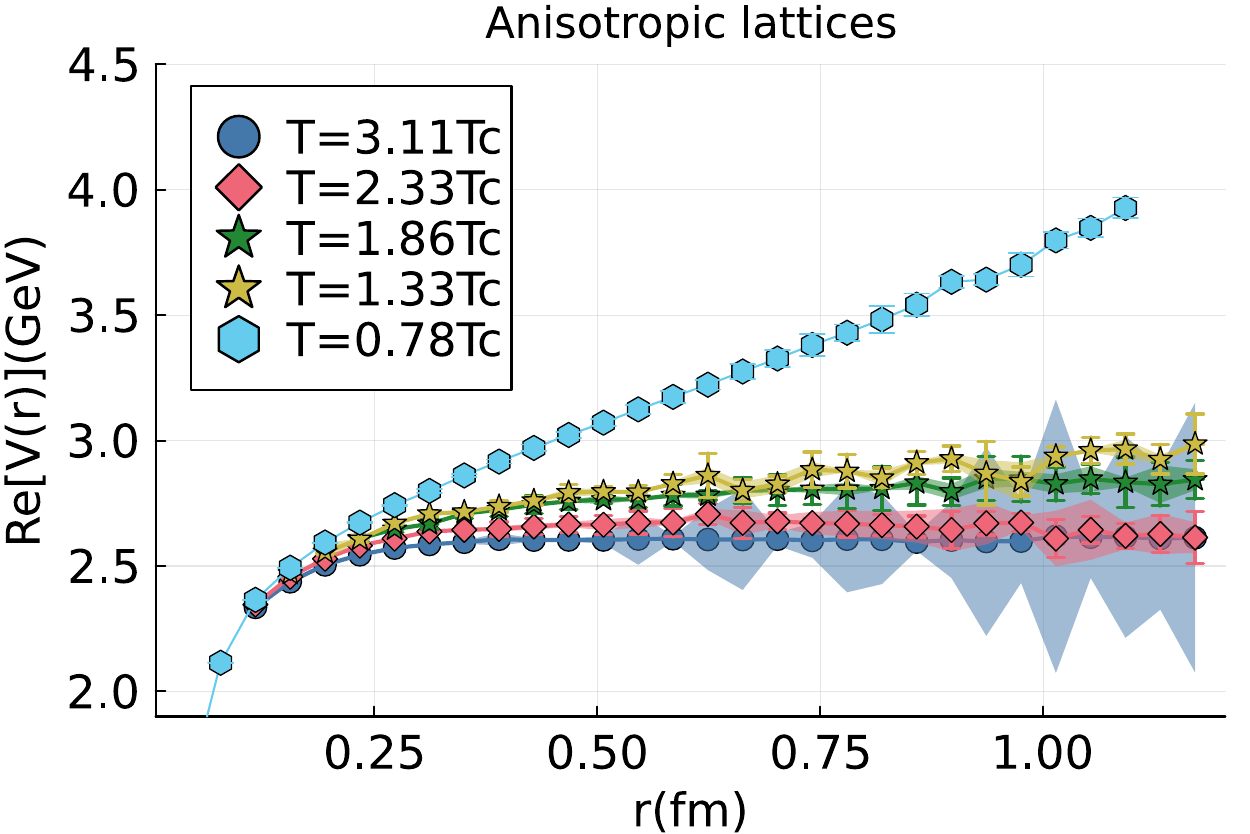}
    \caption{Real part of the potential at different temperatures using the BR method from (top) isotropic and (bottom) anisotropic lattices. The error bands denote systematic errors and the error bars refer to statistical errors. The zero ($T_c$) temperature potentials are calculated by fitting the effective masses to a constant in the regime where a plateau is observed.}
    \label{fig:V_Re_BR}
\end{figure}

\begin{figure}[!ht]
    \includegraphics[scale=.4]{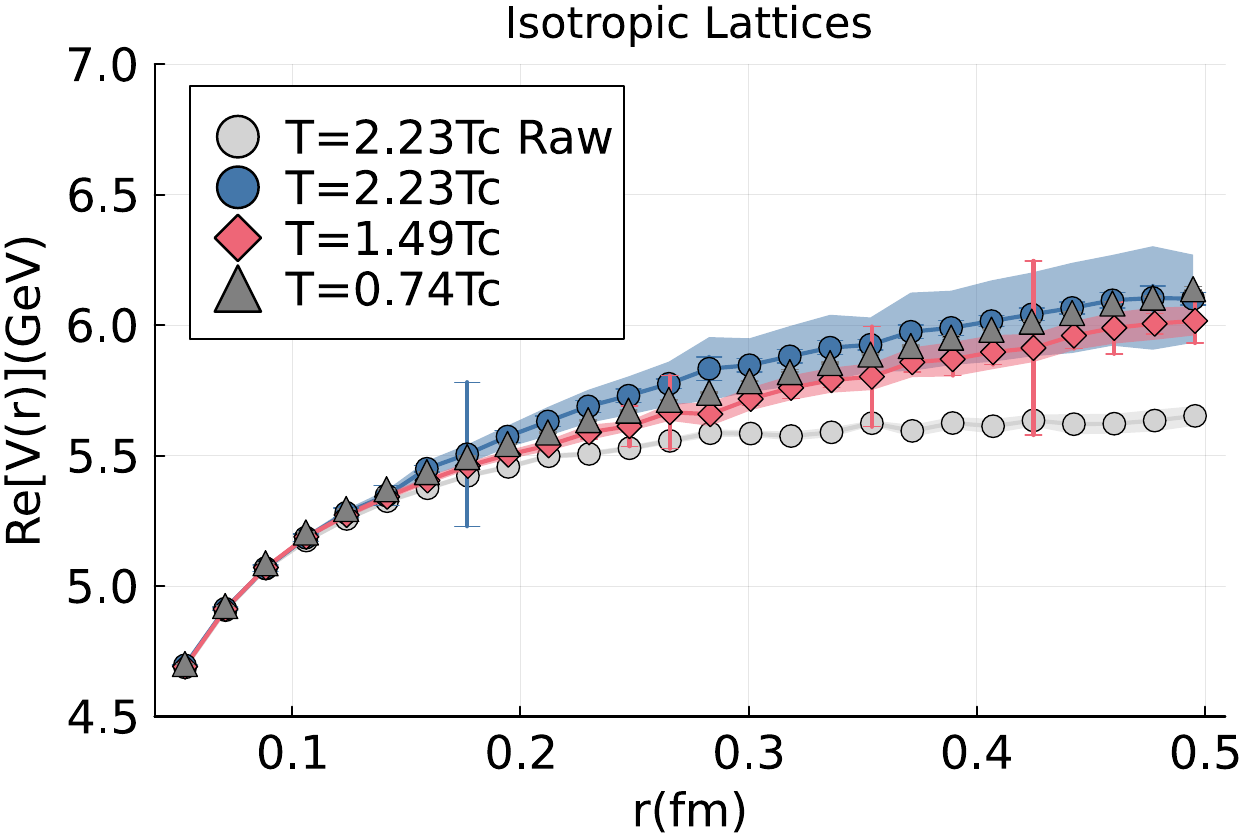}
    \includegraphics[scale=.4]{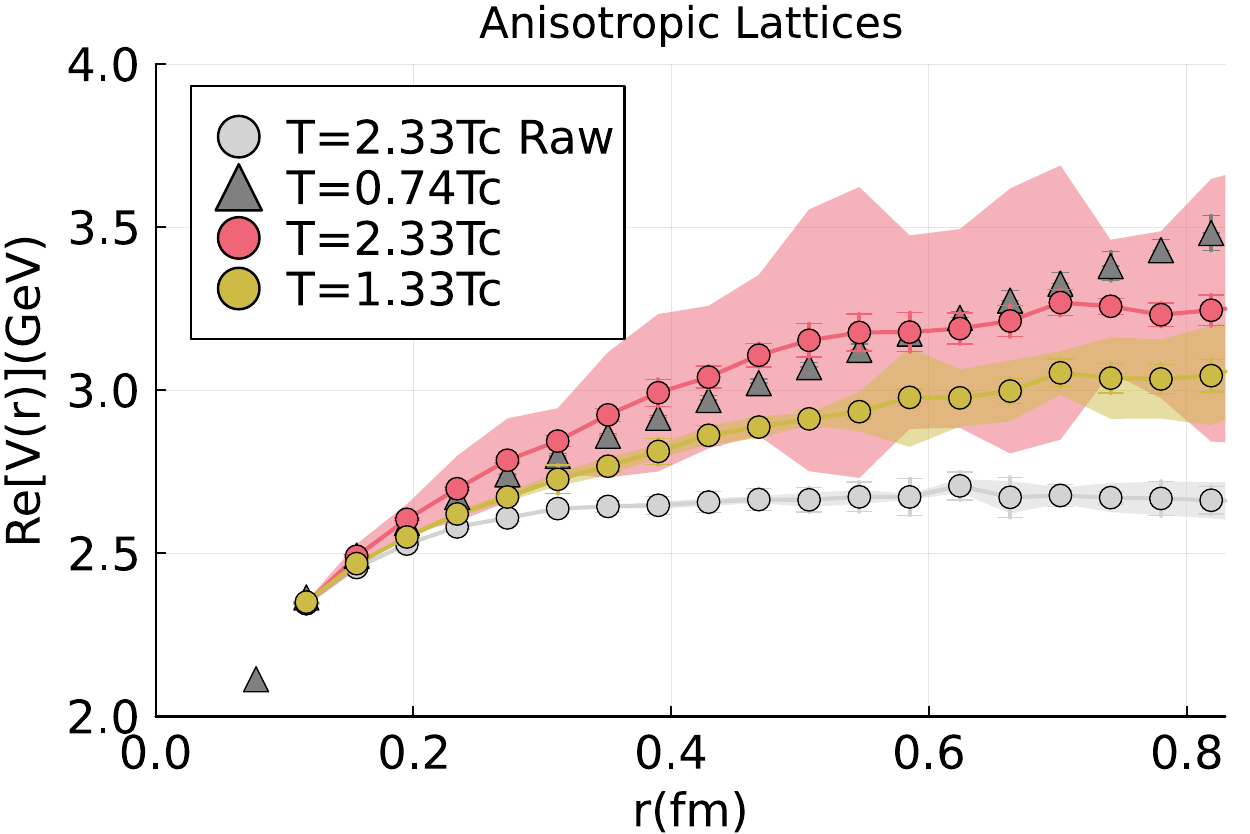}
    \caption{Real part of the potential at different temperatures using the BR method from subtracted (top) isotropic and (bottom) anisotropic lattices. The error bands denote systematic errors and the error bars refer to statistical errors. To understand the effects of subtraction we show (in light grey) the potential obtained from the raw correlator at one temperature.}
    \label{fig:V_Re_BR_sub}
\end{figure}

After combining the statistical and systematic errors the real part of the potential extracted from the raw correlators is shown in \cref{fig:V_Re_BR} and from the subtracted correlator in \cref{fig:V_Re_BR_sub}. The BR analysis on the raw correlator suggests the presence of screening at $T>T_c$ for both the isotropic lattices and anisotropic lattices. However, when applied to the subtracted correlator, we see a substantially different behaviour, the screening has diminished if not disappeared, and the real part is--within the sizable uncertainty--compatible with the zero temperature potential. 
For the isotropic lattices the real part shows almost no temperature dependence within error bars. 
For the anisotropic lattices, there is some inconclusive temperature dependenceincompatible with the screened behaviour found from the raw correlators. This is an intriguing finding and further investigation is needed to find the sources of the discrepancy of the real part of the potential between subtracted and unsubtracted correlator.

\begin{figure}[!ht]
    \includegraphics[scale=.4]{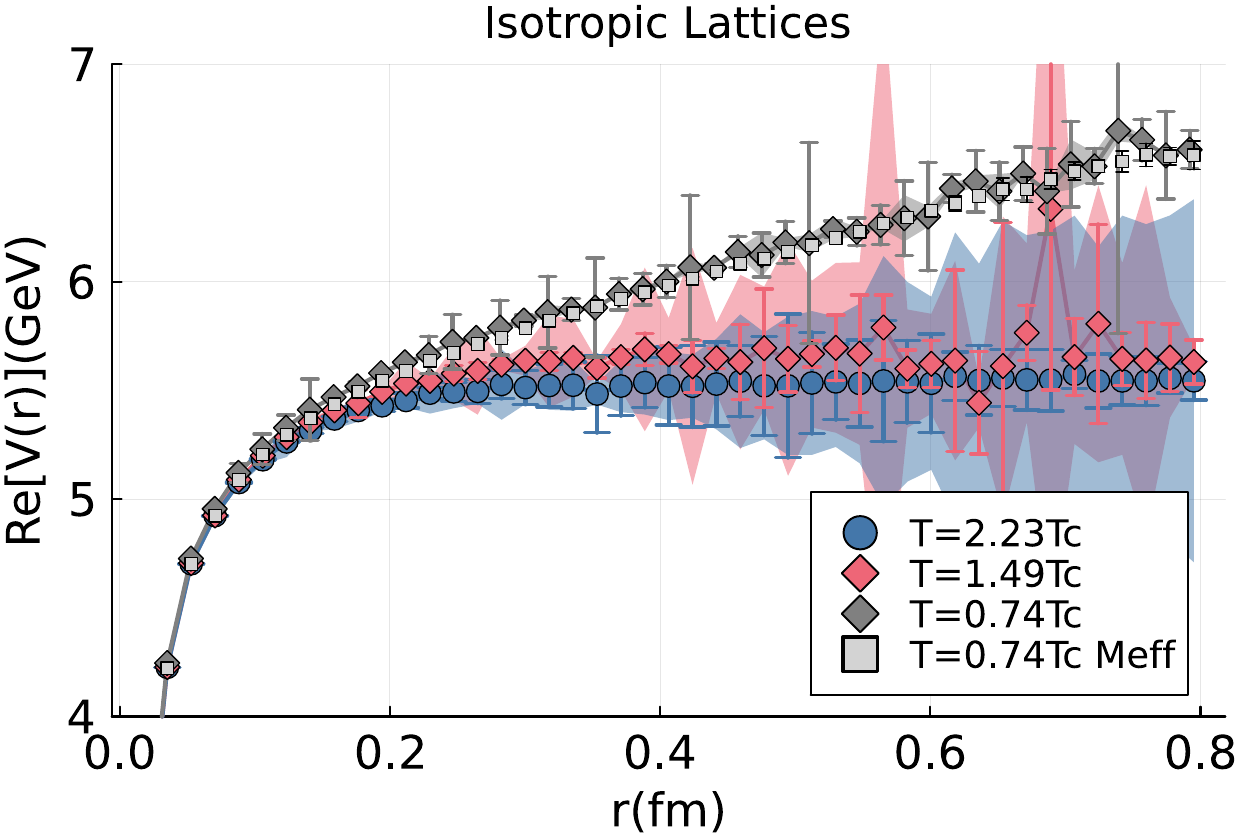}
    \includegraphics[scale=.4]{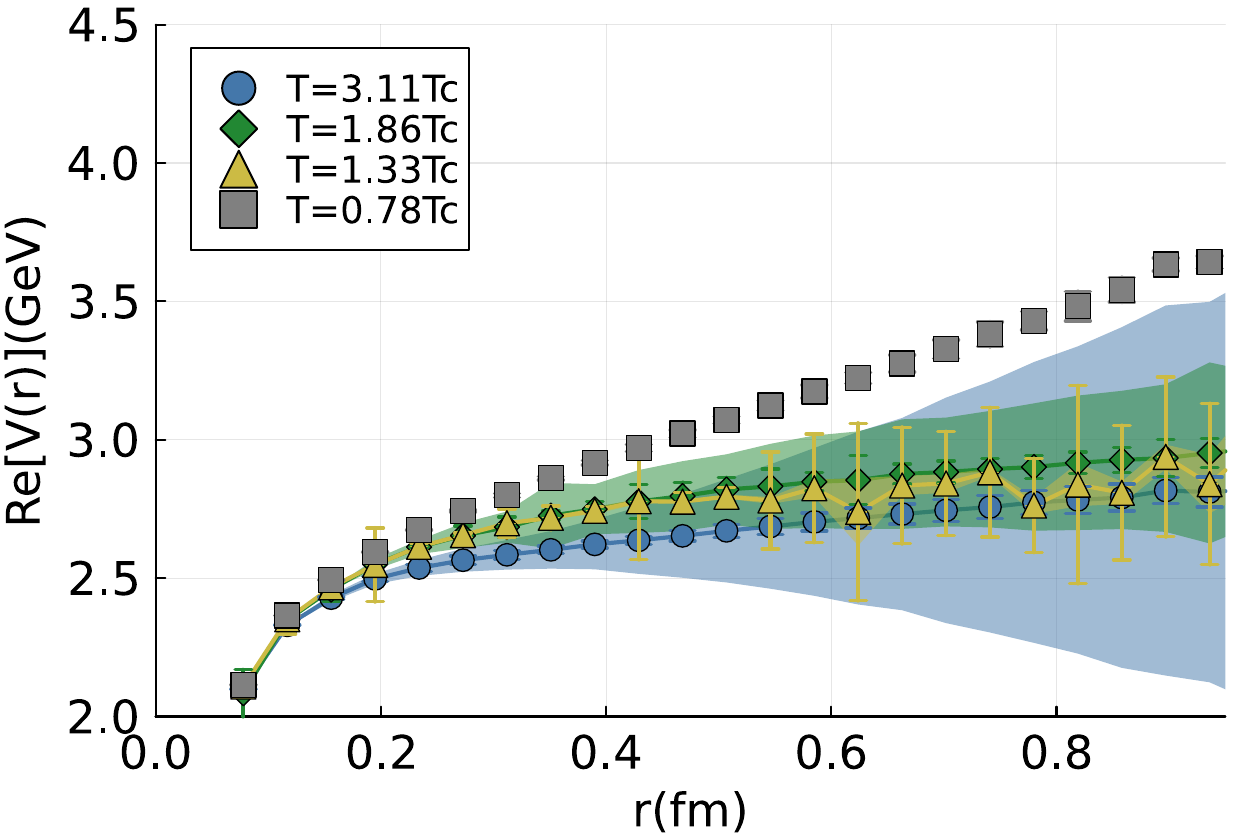}
    \caption{Real part of the potential at different temperatures using the Pad\'e method from (top) isotropic and (bottom) anisotropic lattices. The error bands denote systematic errors and the error bars refer to statistical errors.}
    \label{fig:V_Re_Pade}
\end{figure}

\begin{figure}[!ht]
    \includegraphics[scale=.4]{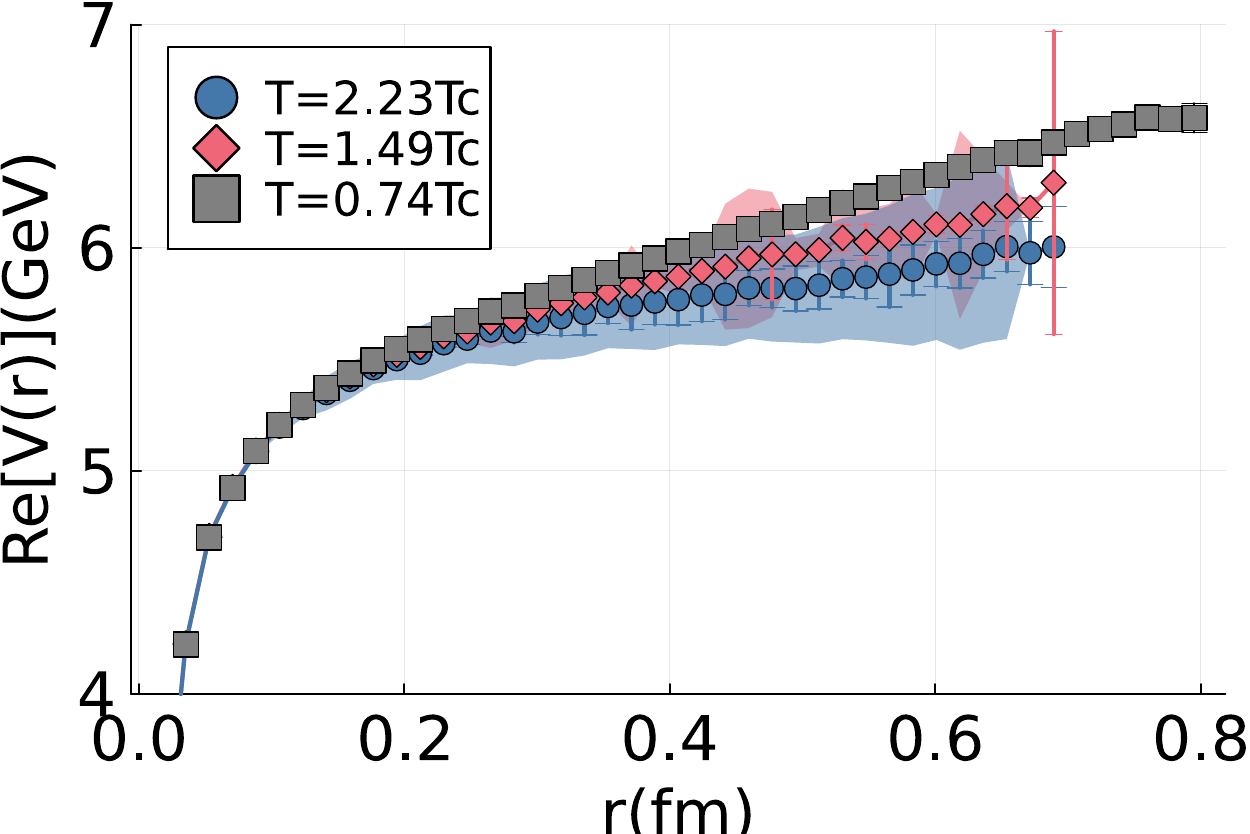}
    \caption{Real part of the potential at different temperatures using the Pad\'e method from  isotropic lattices with the subtracted correlator.  The error bands denote systematic errors and the error bars refer to statistical errors.}
    \label{fig:V_Re_Pade_sub}
\end{figure}

Given that we see such different behaviour of the real part of the potential in the Bayesian reconstruction we proceed with the extraction of the potential using the Pad\'e interpolation to confirm whether it, too, shows similar results. We calculate the potential by inspecting the dominant pole of the Pad\'e interpolation function in \cref{Eq:ContFrac} as described in \cref{sec:pade}. The error budget is estimated in two ways. The first one is the statistical errors. These are obtained from a jackknife procedure with 16 jackknife bins. Since the Pad\'e does not have any explicit regularization we have to be conservative in estimating systematic errors. We proceed in the following way. Let us choose the minimum number of points at which a good interpolation\footnote{By good interpolation we mean an interpolation function which passes through all the data points which are not used as in explicit input for the interpolation function along with having no kinks or discontinuities.} with a smooth function is seen. Then we add one and two more points from the negative frequency side. This gives us three different realisations of the potential and thus a range of uncertainty. In our tests we have observed that the systematic uncertainty dominates the statistical uncertainty at higher temperatures and gradually starts reducing when decreasing temperatures. This is primarily due to the fact that more points in the $\tau$ direction become available, so adding additional points in the interpolation has less of an effect. The real part of the extracted potential using the Pad\'e on raw correlators is shown in \cref{fig:V_Re_Pade} and using the subtracted correlators is shown in \cref{fig:V_Re_Pade_sub}.

We observe that even though the systematic errors are large, the real part of the Pad\'e interpolated potential flattens off at large distances for the raw correlators. The extracted potential on the raw correlators for both isotropic and anisotropic lattices is in agreement with the BR reconstructed real part, i.e. it is compatible with screening. The subtracted correlator analysis can only be performed reliably on the isotropic lattices due to the lack of statistics in the zero temperature lattices in the anisotropic case. The real part extracted in this way shows more of a temperature dependence when compared to the BR reconstructions, but, still does not flatten out at large distances as one would expect from a screened potential. The Pad\'e analysis confirms the behaviour observed with the Bayesian reconstruction.

The next method we discuss is the potential extraction from the subtracted correlator using the Gaussian fit Ansatz. In \cref{fig:Meff_sub} we observed that after the zero temperature subtraction, the effective masses showed linear behaviour at intermediate $\tau$ followed by a downward bending. The agreement with the linear model extends to small $\tau$ values with decreasing temperatures. 
The downward bend at large $\tau$ can be parameterized in terms of a sum of delta functions at frequencies much lower than the peak: as in QCD \cite{Larsen:2019bwy, Bala:2021fkm, Bazavov:2023dci} we find a single delta function sufficient given the data.
We thus parameterise the UV subtracted correlator as: 
\begin{equation}
    C_{sub}(r,t) \approx A \exp({- \Omega \tau + \frac{1}{2} \Gamma^2 \tau ^2 + O(\tau^3)}) + A_{cut} \exp(-{\omega_{cut} \tau})
    \label{eq:gaussfit}
\end{equation}

The spectral function can thus be represented as, suggested in \cite{Bala:2021fkm}
\begin{equation}
    \rho(\omega, T) = A(T) \exp({-\frac{|\omega - \Omega(T)|^2 }{2 \Gamma(T)^2}}) + A^{cut}(T) \delta(\omega - \omega^{cut}(T))
\end{equation}

\begin{figure}[!ht]
    \includegraphics[scale=.4]{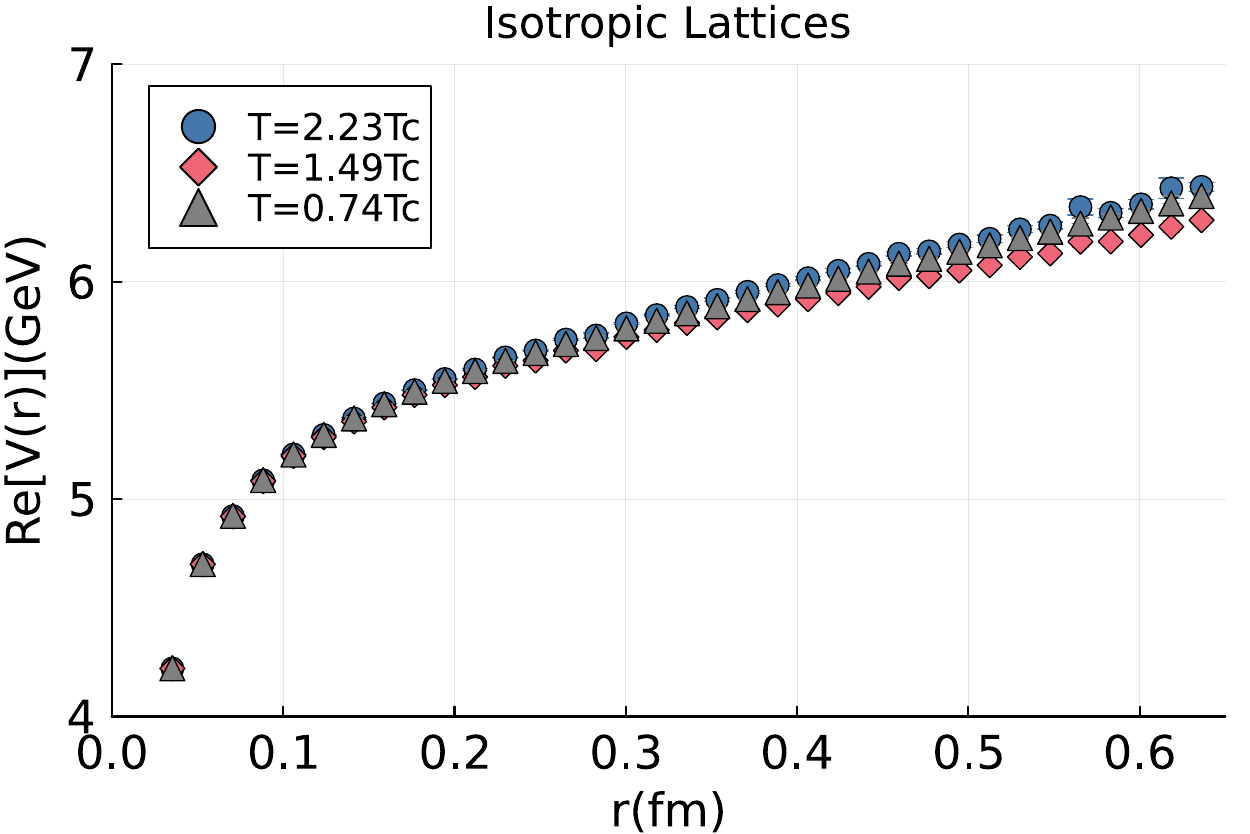}
    \includegraphics[scale=.4]{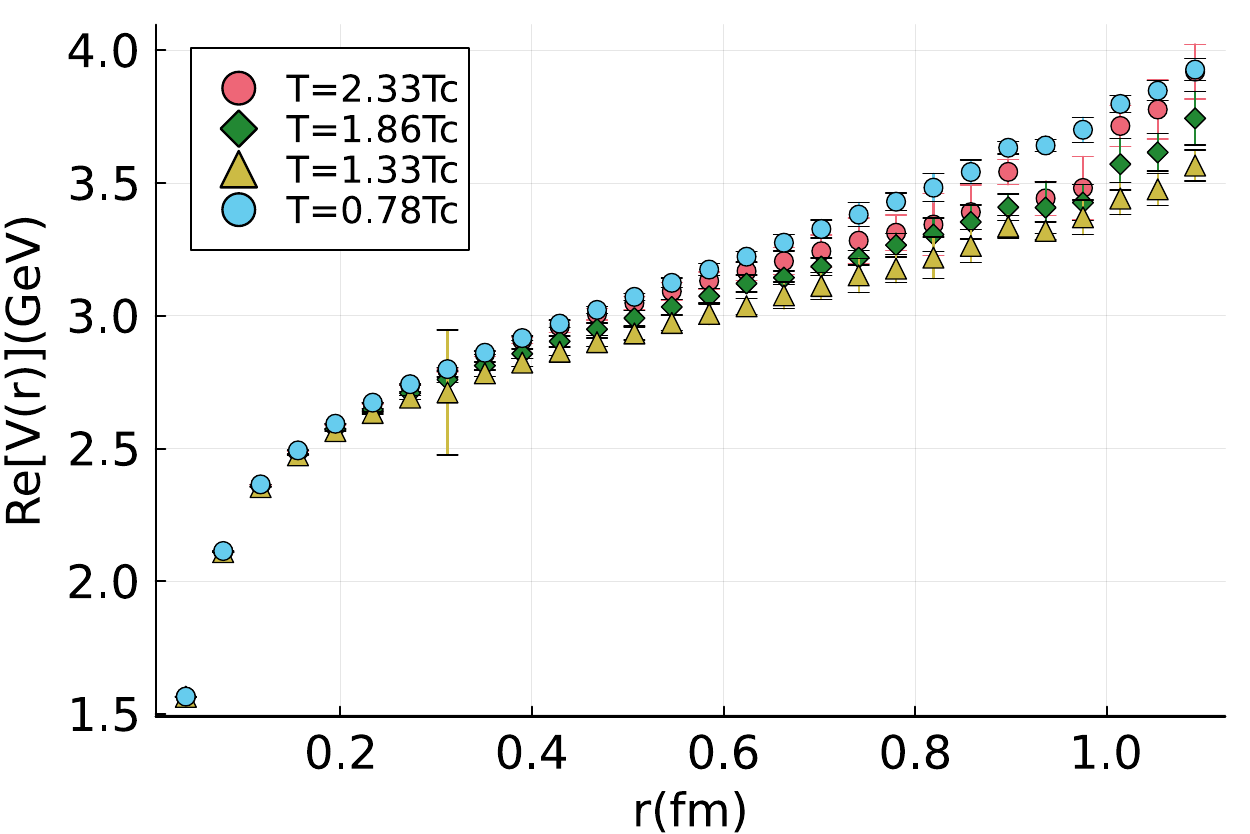}
    \caption{The figure shows the real part of the potential using Gaussian Fits on isotropic (top) and anisotropic (bottom) lattices. The potential below $T_c$ is computed by fitting the effective mass with a constant in the regime where a clear plateau is observed.}
    \label{fig:Gaussian_pot}
\end{figure}

The real part of the potential as obtained from the Gaussian fits is shown in \cref{fig:Gaussian_pot}. The error bars represent statistical uncertainty and are obtained by the jackknife procedure with 16 bins. We observe that for the isotropic lattices the behaviour of $\textrm{Re}[V]$ is compatible with that at zero temperature without a clear trend of temperature dependence. For the anisotropic lattices, there is more temperature dependence present, but, without a clear trend the behaviour is still compatible with an unscreened potential. We discuss the goodness of these fits in \cref{sec:quality}.

\begin{figure}[!ht]
    \includegraphics[scale=.4]{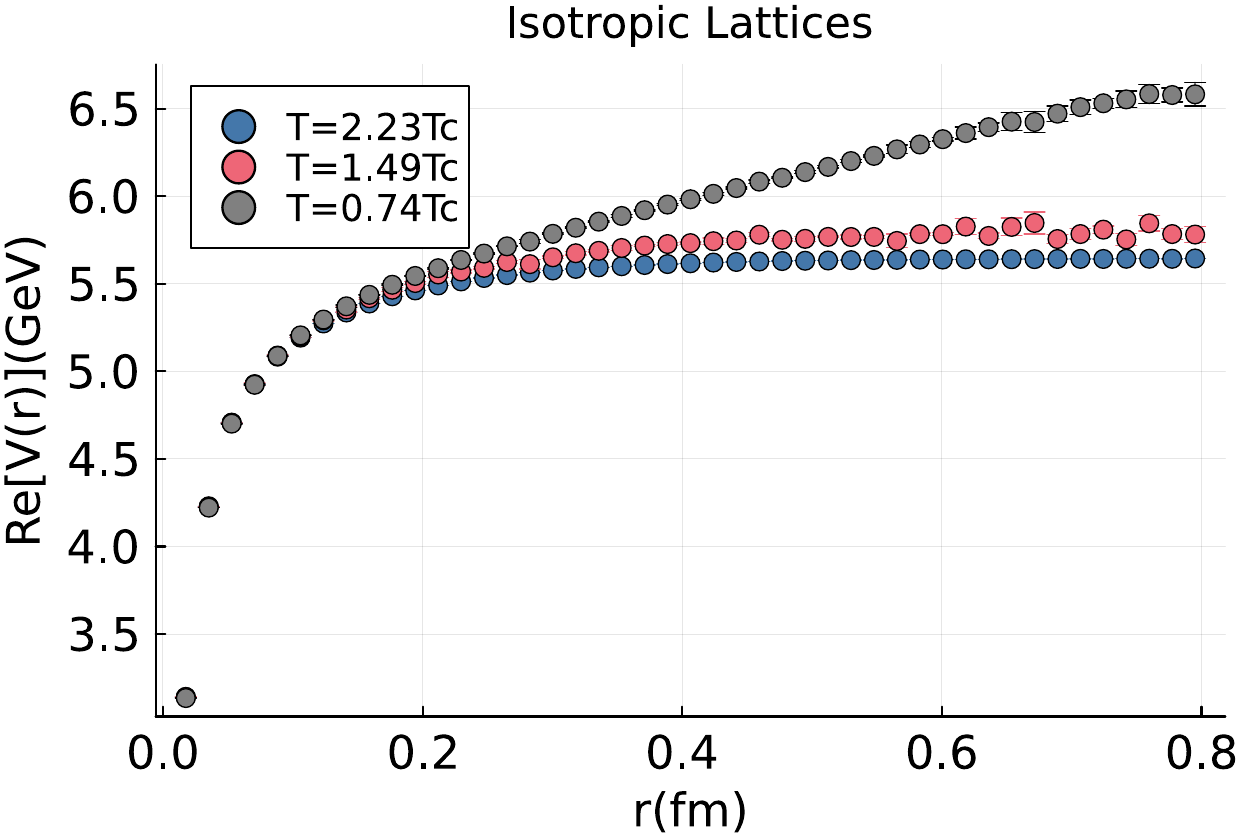}
    \includegraphics[scale=.4]{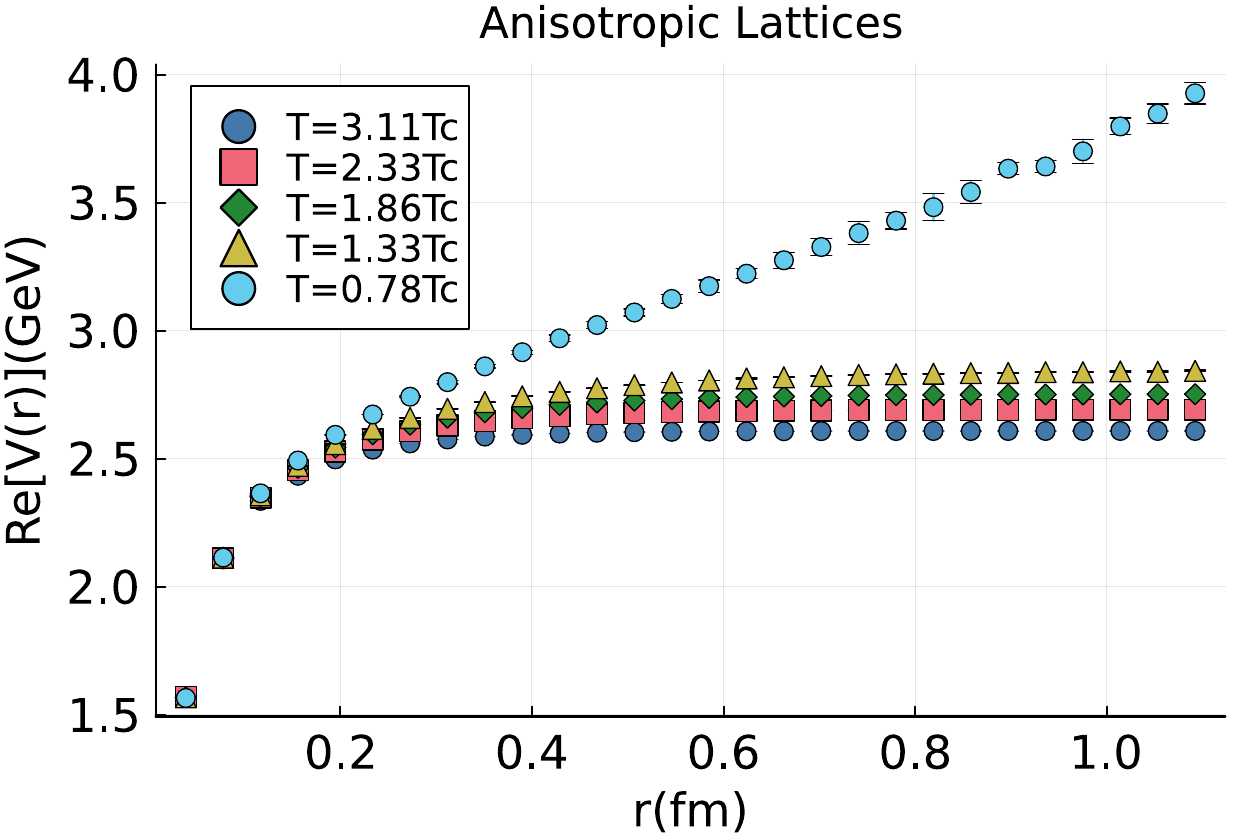}
    \caption{Real part of the potential at different temperatures using the HTL inspired method from (top) isotropic and (bottom) anisotropic lattices. The error error bars refer to statistical errors.}
    \label{fig:VRe_baladatta}
\end{figure}

The last method we present in our discussion is the HTL-inspired fits described in \cref{sec:bala_datta}. In the previous study on (\texorpdfstring{\boldmath{$2+1$}}{2+1})-flavour HISQ lattices we had observed that the results of $\textrm{Re}[V]$ from this method were different from results obtained from Pad\'e and the Gaussian fits. Here, we wish to establish whether that is still the case. When two higher order terms $c_1$ and $c_2$ in \cref{eq:BDfit} are included in the fits in the data fits the functional form up to a significantly large $\tau $ region away from $\tau=\beta/2$ (see \cref{fig:HTL_fit}), and the potential is estimated from the parameters of this fit. The goodness of these fits is shown in \cref{sec:quality}.  We show the extracted potential using the HTL-inspired fits in \cref{fig:VRe_baladatta}. The error bars represent statistical errors obtained from a jackknife analysis with 16 bins. The figure shows the real part as a function of separation distance using the raw correlator data, but we find that the results remain unchanged when using the subtracted data. We observe a screened potential from all of these fits.

\subsection{Imaginary part of the potential}

The mock data analyses carried out in \cref{sec:methodsmock} indicated that with the currently available data quality of $N_\tau \sim 20$ and $\Delta D/D=10^{-2}$, only a robust reconstruction of the real-part is possible. Both the BR and the Pad\'e method applied to the HTL Wilson line data underestimated the imaginary part by around a factor of $1/2$. In light of these limitations the goal of this section is quite modest: we investigate whether there exists a non-zero imaginary part of the in-medium static potential. A careful estimate of the total uncertainty budget will be provided.

Our analysis indicates that despite the sizable uncertainties, a non-zero imaginary part is present for temperatures in the deconfined phase $T>T_c$. At the same time we confirm that in the confined phase $T<T_c$ both methods show an imaginary part that is compatible with zero.

\begin{figure}[!ht]
    \includegraphics[scale=.4]{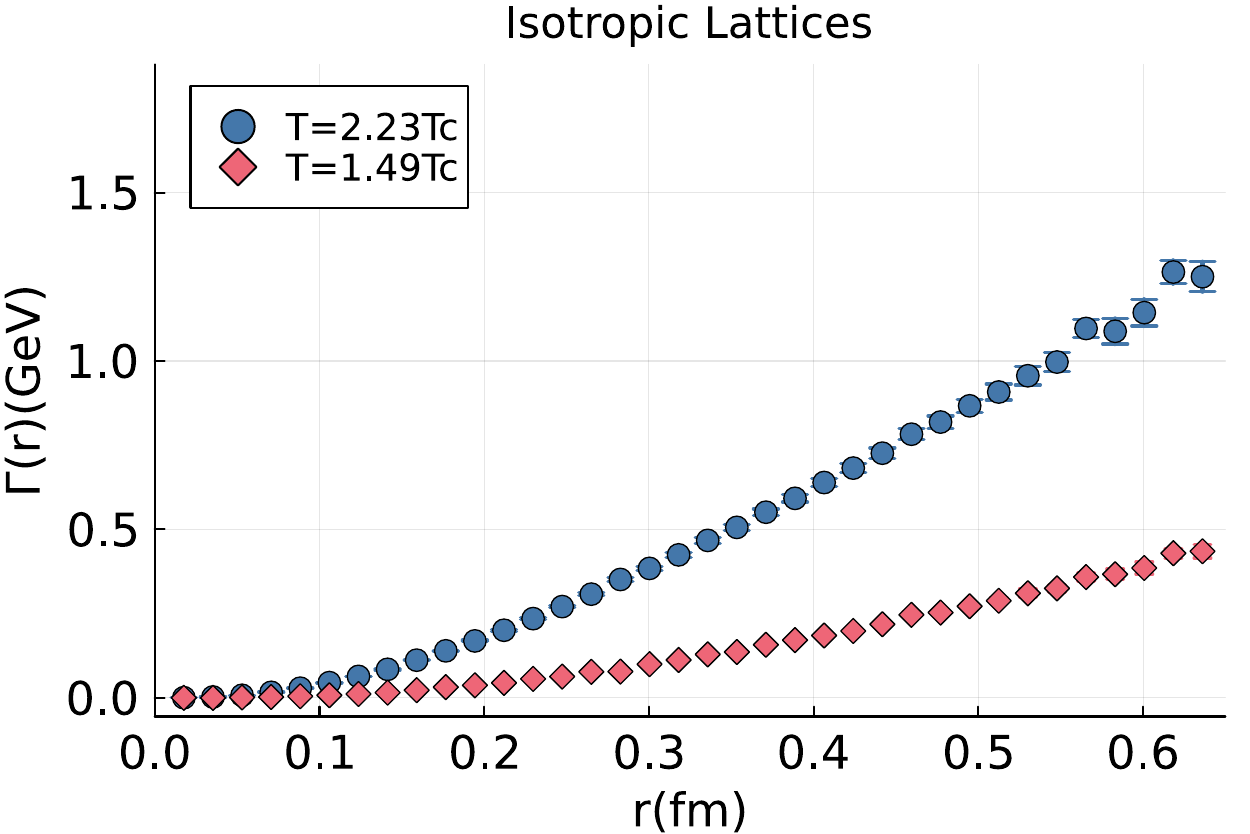}
    \includegraphics[scale=.4]{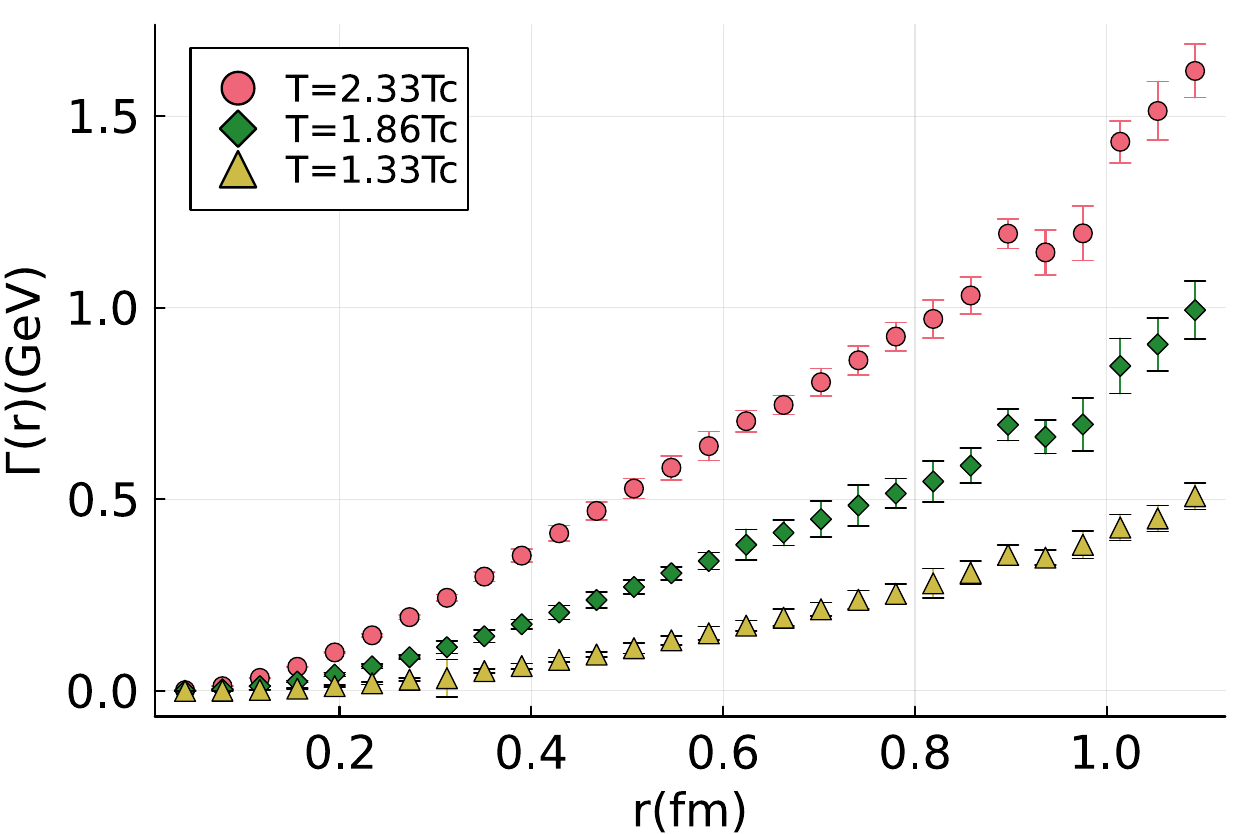}
    \caption{The figure shows the extracted Gaussian width parameter $\Gamma$ (related to the imaginary part of the potential) on isotropic (top) and anisotropic (bottom) lattices. The error bars refer to statistical errors computed from the jackknife analysis.}
    \label{fig:Gaussian_potim}
    \end{figure}

We start first by revisiting the effective masses in \cref{fig:qnchdmeff}. For both isotropic and anisotropic lattices they exhibit a plateau at intermediate $\tau$ values for $T<T_c$ which is compatible with a delta-function like spectral function and thus a zero imaginary part. For the higher temperature lattices ($T>T_c$) the plateau is absent and instead one finds a finite slope at intermediate $\tau$. As discussed before, this linear behaviour extends over an even longer imaginary time duration in the subtracted correlator as shown in \cref{fig:Meff_sub}. The presence of this slope may be interpreted as the finite width of a Gaussian or cut-off Lorentzian spectral peak encoding the imaginary part of the potential. The corresponding Gaussian model fits were discussed in \cref{sec:ReV_lat}, and we show the width parameter as a proxy for width in \cref{fig:Gaussian_potim}. As the authors of Refs. \cite{Bazavov:2023dci} have pointed out, while the width parameter is model dependent, while the associated second cumulant is not.

\begin{figure}[!ht]
    \includegraphics[scale=.4]{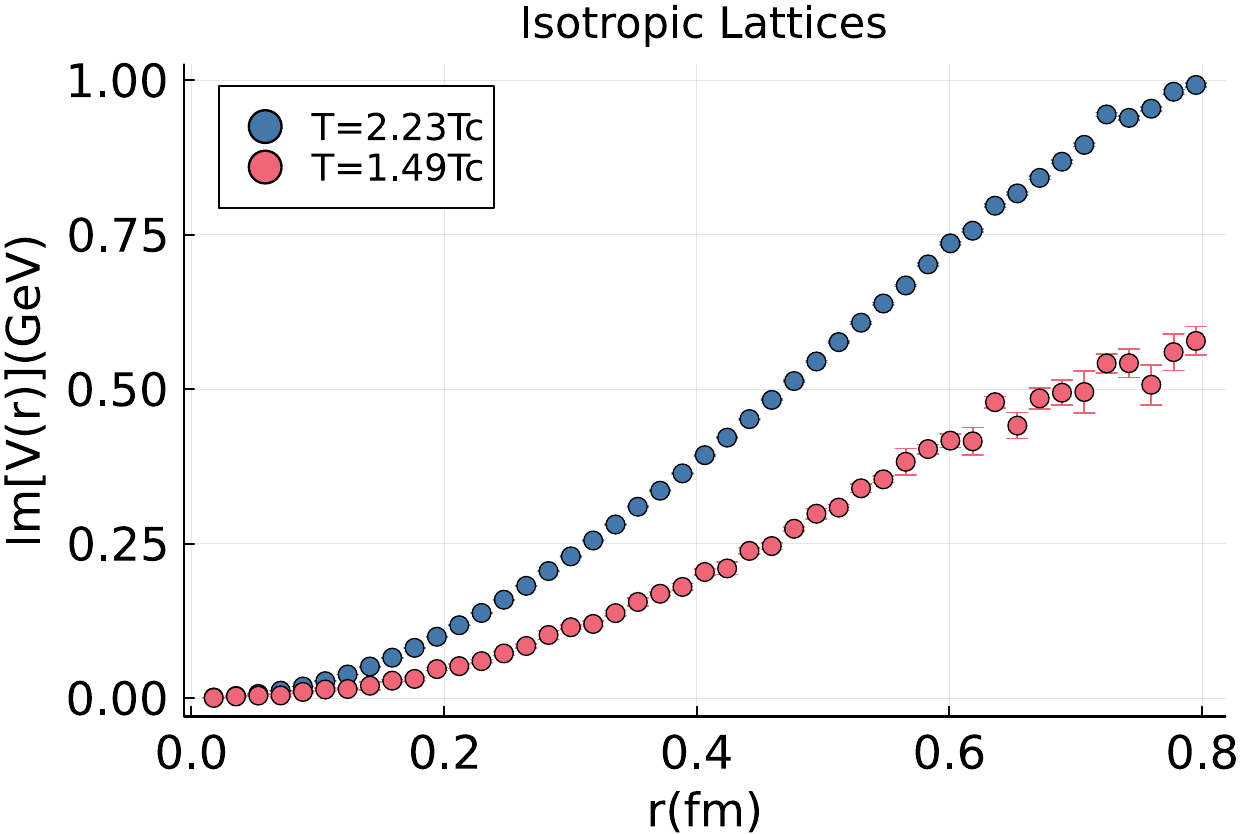}
    \includegraphics[scale=.4]{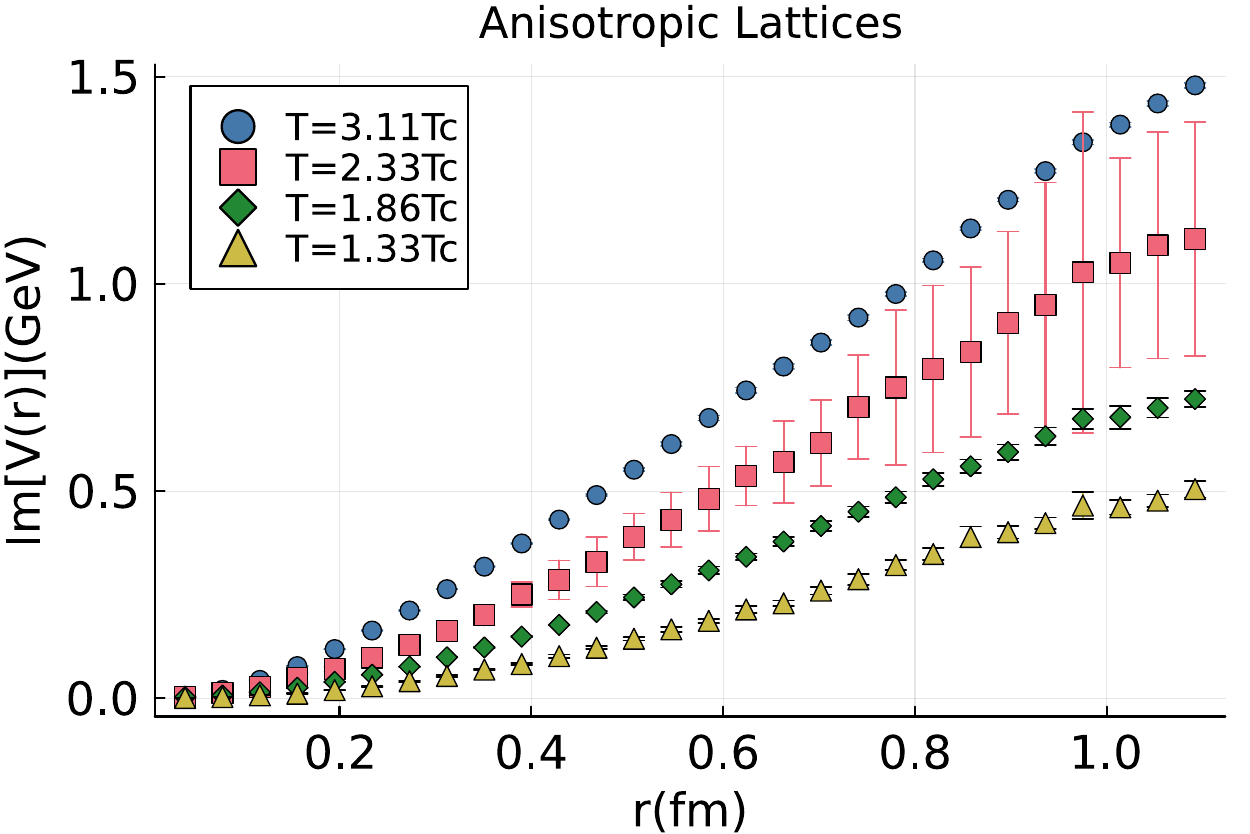}
    \caption{Imaginary part of the potential at different temperatures using the HTL inspired method from (top) isotropic and (bottom) anisotropic lattices. The error bars refer to statistical errors computed from the jackknife analysis}.
    \label{fig:VIm_baladatta}
\end{figure}

Besides the Gaussian model Ansatz, the HTL-inspired Ansatz with two extra terms, too, is able to fit the data within the errors except for the first and the last few points (number of points depend on the temperature and separation distances) as shown in  \cref{fig:HTL_fit_good_isotropic,fig:HTL_fit_good_anisotropic}), with the error per degree of freedom being slightly worse.
The slope in the effective masses is attributed to a finite imaginary part, the extracted values of which are given in \cref{eq:BDfit} in \cref{fig:VIm_baladatta}.

In order to determine the presence or absence of an imaginary part in a model independent fashion using the BR or Pad\'e method, we need to obtain a reliable uncertainty estimate of the extracted spectral widths.  

\begin{figure}[!ht]
  \includegraphics[scale=.4]{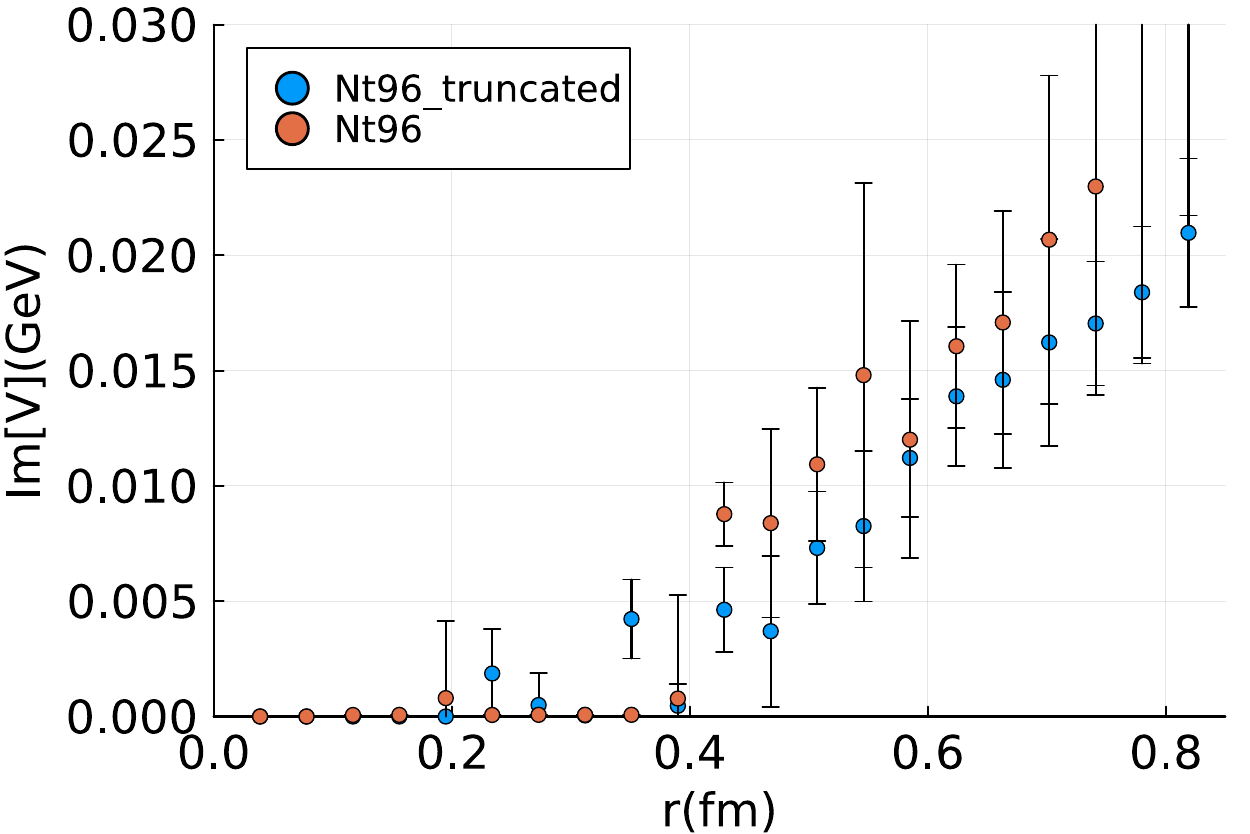}
    \caption{Assessing the dependence of the Bayesian reconstruction at $T=0.78T_C$ on the available extent in imaginary time. Each of these was computed using the full statistics with the same constant default model.}
    \label{fig:Im_V_BR_trunc}
\end{figure}

One known artifact is related to the fact that the temperature is changed by changing the physical length of the Euclidean time domain. Because it is known that this change can affect the resolution of Bayesian reconstruction methods, we need to make sure that it does not artificially introduces a finite width. To this end we carry out the BR reconstruction on the low temperature ($T=0.78 T_c$, $N_\tau=96$) correlator data, after artificially truncating the Euclidean time to only $24$ points, the same extent as available at the highest temperature. We then compare the outcome of this reconstruction with the results from reconstruction on the fully available data-points in \cref{fig:Im_V_BR_trunc}. 
In both cases we find a statistically non-vanishing imaginary part for $r > 0.4$ fm even in the confined phase, but more than one order of magnitude smaller than in the deconfined phase.
We observe that the mean of the extracted imaginary part from the truncated data has a tendency to sit slightly below the one from the full data. However, when error bars are considered the two are virtually indistinguishable.

\begin{figure}[!ht]
    \includegraphics[scale=.4]{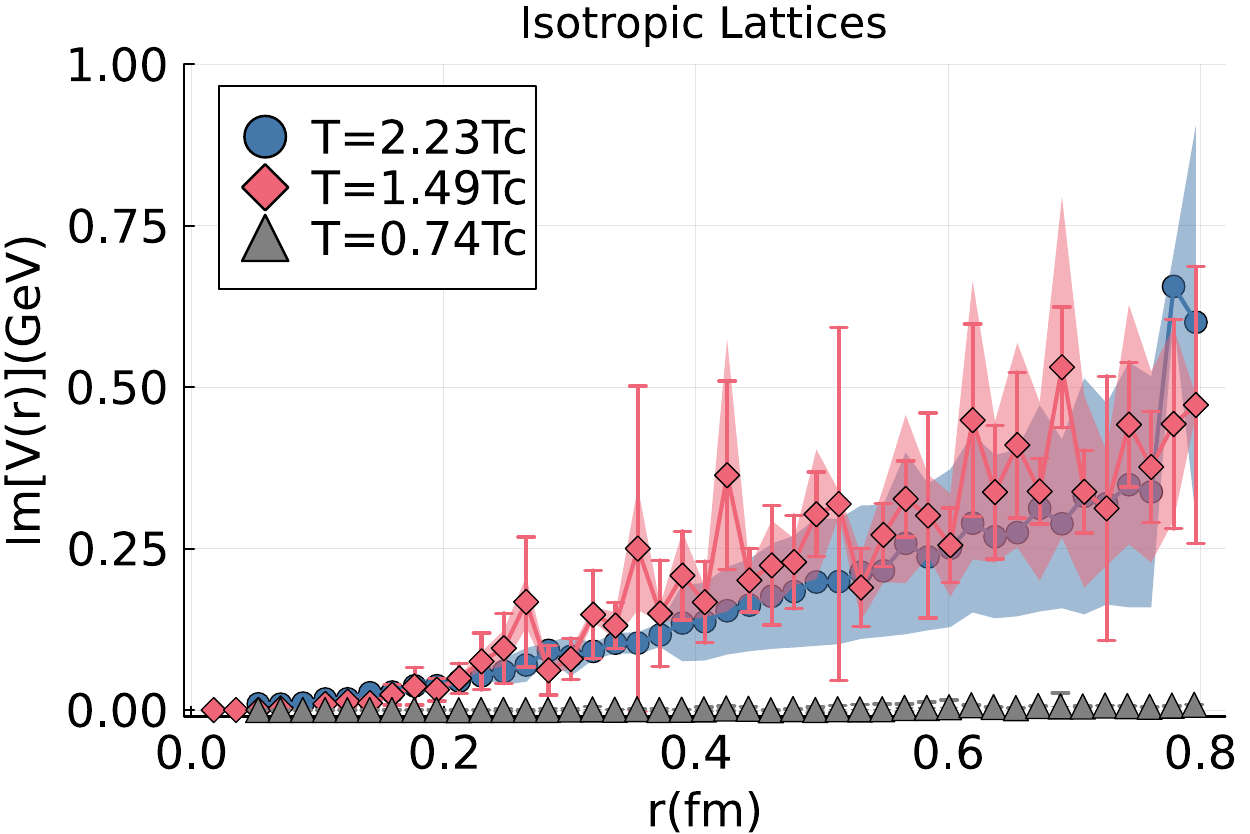}
    \includegraphics[scale=0.4]{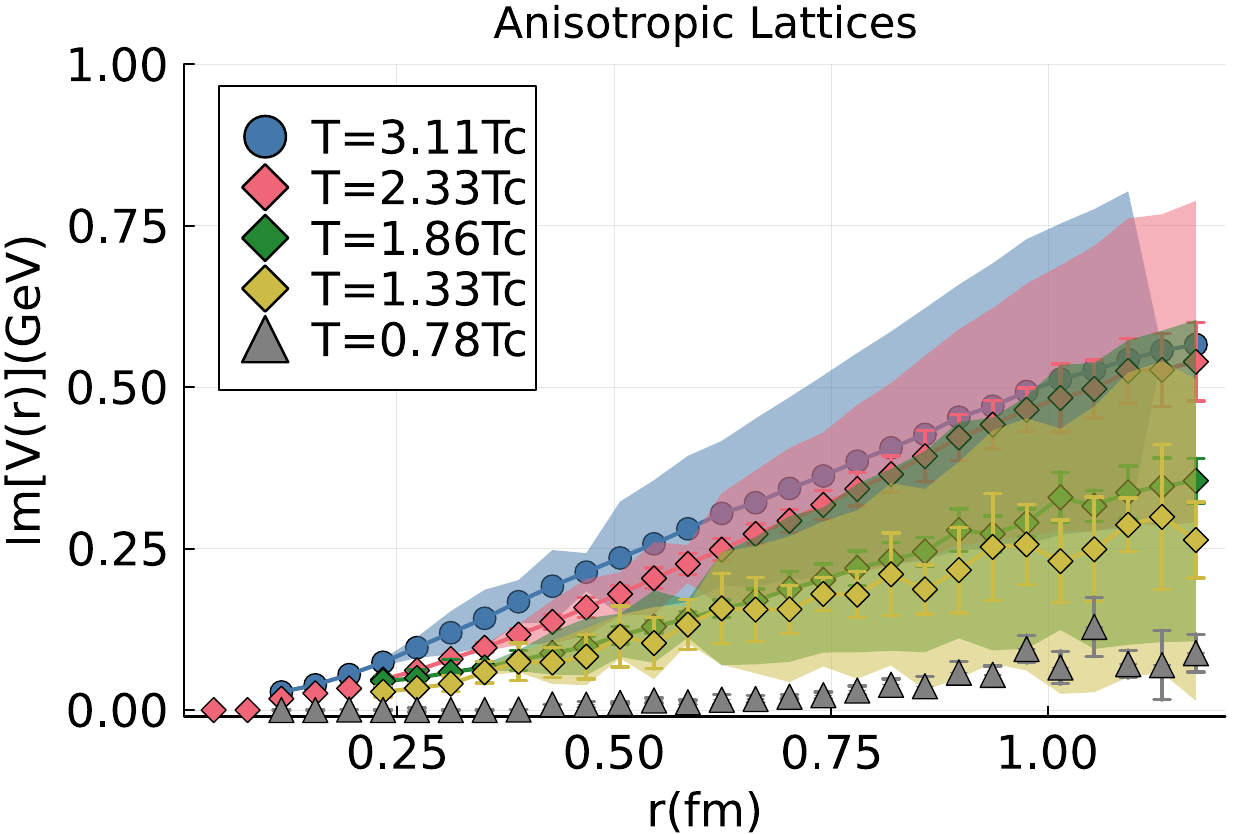}
    \caption{Imaginary part of the potential at different temperatures using the BR method from subtracted (top) isotropic and raw (bottom) anisotropic lattices. The error bands denote systematic errors and the error bars refer to statistical errors. The imaginary part is consistent with zero till around 0.6 fm.}
    \label{fig:V_Im_BR_sub}
\end{figure}

The BR extracted imaginary part is shown in \cref{fig:V_Im_BR_sub} using the raw correlator. For the isotropic case, and the anisotropic case wewsee that the high temperature lattices show a finite imaginary part for both, while the low temperature lattices show a behaviour that is compatible with zero up to $r > 0.7$ fm). The error-bars are computed in the same way as in the real part (see \cref{sec:ReV_lat}).

\begin{figure}[!ht]
    \includegraphics[scale=.4]{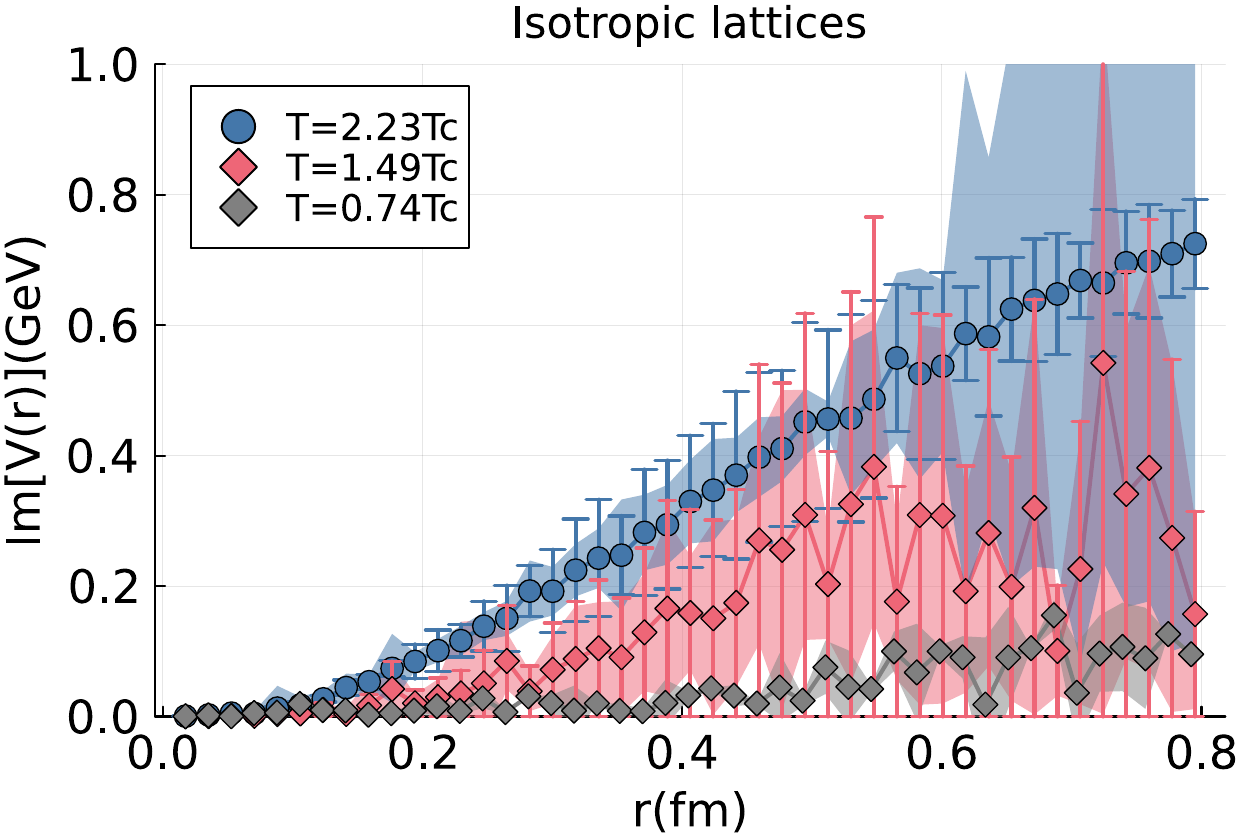}
    \includegraphics[scale=.4]{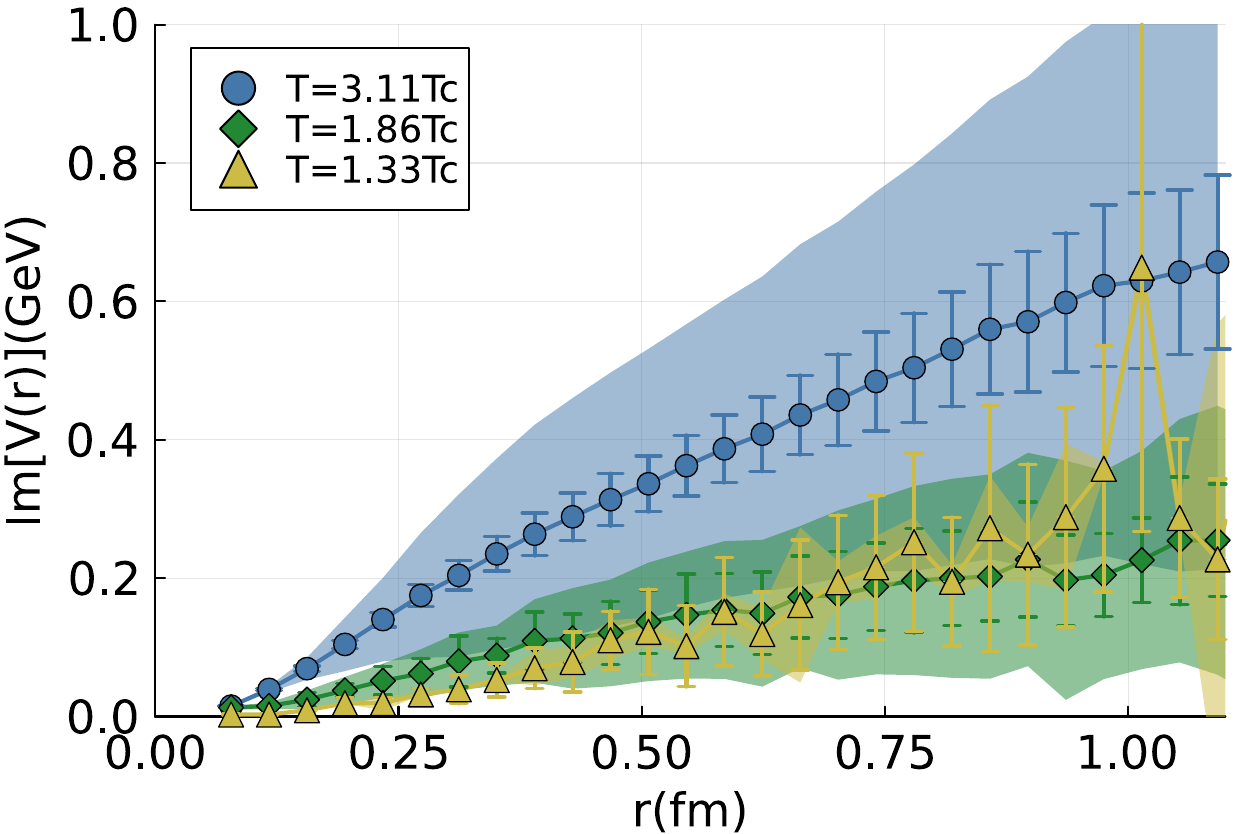}
    \caption{Imaginary part of the potential at different temperatures using the Pad\'e method from (top) isotropic and (bottom) anisotropic lattices. The error bands denote systematic errors and the error bars refer to statistical errors. We observe that for both isotropic and anisotropic the imaginary part is zero up to 0.7 fm after which we see a small increase.}
    \label{fig:VIm:Pade}
\end{figure}

Lastly, in \cref{fig:VIm:Pade} we show the extracted imaginary part from the Pad\'e interpolation from the raw correlator. The imaginary part is nothing but the imaginary part of the dominant pole of the Matsubara correlator whose real part had been used to extract the real part of the potential. The statistical and systematic errors have been computed using the same procedure as for the real part. For the isotropic lattices we clearly see that there is a non-zero imaginary part for $T=2.23 T_c$ even when the large uncertainties are considered. The uncertainties are too large for the case with $T=1.49 T_c$ to make such a claim, but nonetheless the obtained mean is non-zero. For the low temperature case $T=0.74 T_c$ the extracted imaginary part is compatible with zero up to at least $r > 0.6$ fm. For the anisotropic lattices, we see that there is a positive imaginary part for all $T>T_c$. Our attempt to extract real and imaginary part of the dominant pole fails for the low temperature ($T=0.78 T_c$, $N_\tau=96$) case due to lack of statistics for those ensembles.

\section{Summary and Conclusions}
\label{sec:conclusions}

We have re-investigated the static quark-antiquark potential in a gluonic medium at finite temperature using high resolution isotropic and anisotropic quenched QCD lattices. We deploy four different independent methods, all of which have different underlying assumptions in analyzing the spectral structure of Euclidean correlators. We first started with analysing the correlation functions and their effective masses. 
The small $\tau$ behaviour of the latter showed non-trivial temperature dependence (on anisotropic lattices) unlike what the authors of \cite{Bala:2021fkm} reported. Yet the case for zero temperature subtraction can still be made keeping in mind the possibility of under-subtracting and over-subtracting which could affect the reconstruction adversely. 

After analysing the correlation functions themselves we investigated the real part of the potential using each of the methods described in the paper on both the subtracted and raw correlator. We find that the results from the raw correlator using the BR, Pad\'e and HTL-inspired fits agreed with previous studies on quenched lattices and showed the presence of screening in the real part of the potential for both isotropic and anisotropic lattices. Since the Gaussian fit Ansatz is only applicable to the subtracted correlator we do not deploy it on the raw correlator. In contrast when the same analysis (except HTL-inspired fit, which is hardly impacted by data outside of $\beta/4 < \tau < 3\beta/4$) is performed on the subtracted correlators we observe that the real part of the potential no longer shows screening. These different outcomes due to the subtraction are rather puzzling and further analysis is needed to identify the source of these differences.

The intention behind the subtraction procedure is to remove structures in the spectral function that are irrelevant for the potential physics by allowing the lowest lying peak to dominate more of the available data it attempts to reduce the severity of the ill-posed inverse problem. Since the Bayesian reconstruction relies on properties such as smoothness in the spectral function, presence of some non-smooth structures induced e.g. by the lattice cutoff in the UV can distort the spectral reconstruction and one may speculate that they could cause it to systematically underestimate the peak position. In such a hypothetical scenario it is possible that the subtracted correlator does not suffer from these discontinuities and thus allows for a more accurate determination of the actual spectral function. On the other hand, even though the motivation for subtraction is well motivated in full QCD and on isotropic lattices in quenched QCD, the agreement of effective masses at small $\tau$ is not as good on anisotropic lattices in quenched QCD. 

The BR is able to reconstruct the Wilson loop peak in our mock data analysis with a spectrum that is rising at its high-frequency end, see \cref{fig:wspecs_HTL}. However, the raw Wilson line correlator on the lattice has instead a second, bumpy structure at higher frequencies, see Figures \ref{fig:BRspecs_iso_temp} or \ref{fig:BRspecs_aniso_temp}. This is apparently much more difficult for Bayesian inference.
Such a second bump also shows up in the Pad\'e analysis of the raw Wilson line correlator on the lattice. 
Therefore, we suggest focusing the attention on such high-frequency bumps.
  
Given the quality of data with $N_\tau \sim 24$ and $\Delta D/D =10^{-2}$, our mock data tests underestimated the imaginary part of the potential. For this reason we decided to focus on merely determining the presence or absence of an imaginary part of the potential at different temperatures. We have shown that all our methods indicate the presence of a non-zero imaginary part in the deconfined phase that increases with temperature and separation distance.

\section*{Acknowledgements}
G.~P.~ and A.~R. gladly acknowledge support by the Research Council of Norway under the FRIPRO Young Research Talent grant 286883. The study has benefited from computing resources provided by  
UNINETT Sigma2 - the National Infrastructure for High Performance Computing and Data Storage in Norway under project NN9578K-QCDrtX "Real-time dynamics of nuclear matter under extreme conditions" and the UNIX-system at University of Stavanger. This research used awards of computer time provided by the PRACE awards on Marconi100 at CINECA, Italy.
J.H.W.’s research is funded by the Deutsche Forschungsgemeinschaft (DFG, German Research Foundation)---Projektnummer 417533893/GRK2575 ``Rethinking Quantum Field Theory''.

\begin{backmatter}

\section*{Competing interests}
  The authors declare that they have no competing interests.

\bibliographystyle{stavanger-mathphys}


\bibliography{references}


\end{backmatter}

\appendix
\section{Quality of fits}
\label{sec:quality}
\begin{figure}
    \centering
    \includegraphics[scale=0.4]{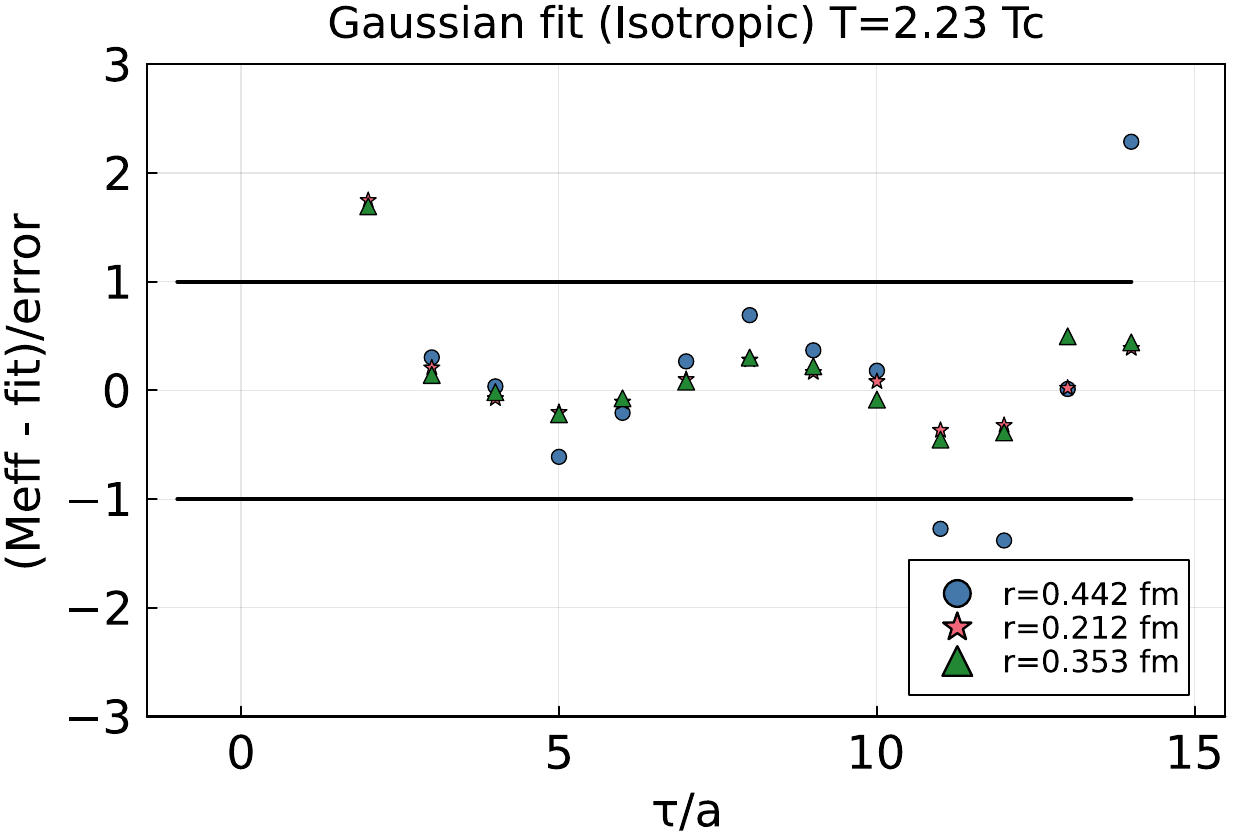}
    \includegraphics[scale=0.4]{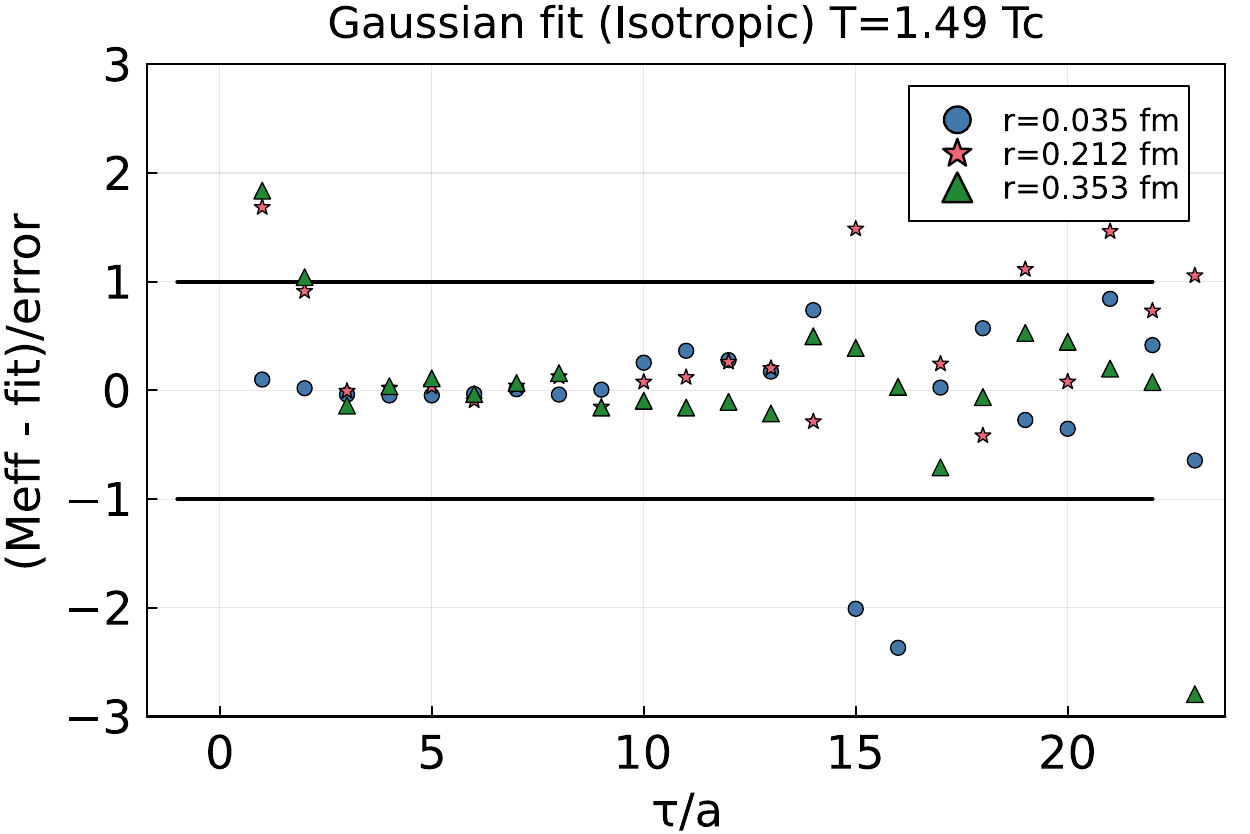}
    \caption{Goodness of Gaussian fit on subtracted isotropic lattices at three different separation distances.}
    \label{fig:gauss_fit_good_isotropic}
\end{figure}

\begin{figure}
    \centering
    \includegraphics[scale=0.4]{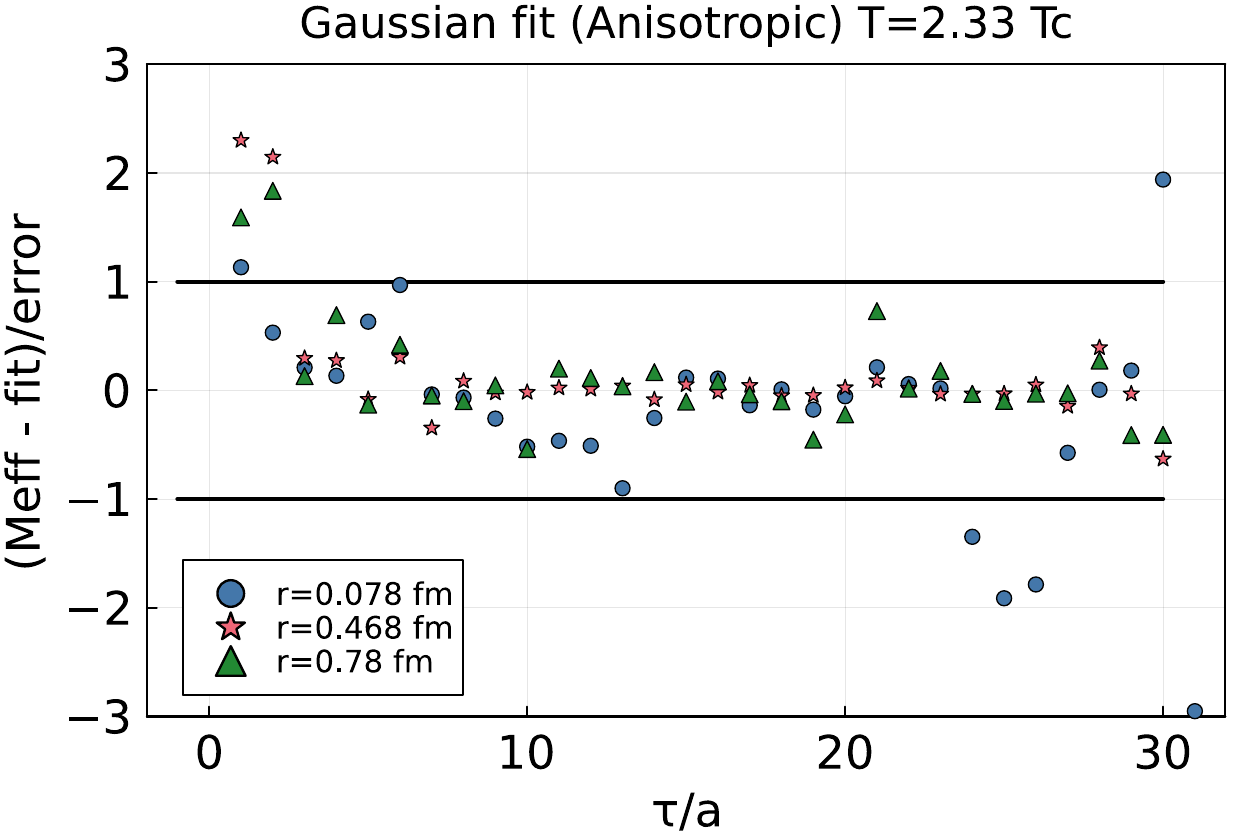}
    \includegraphics[scale=0.4]{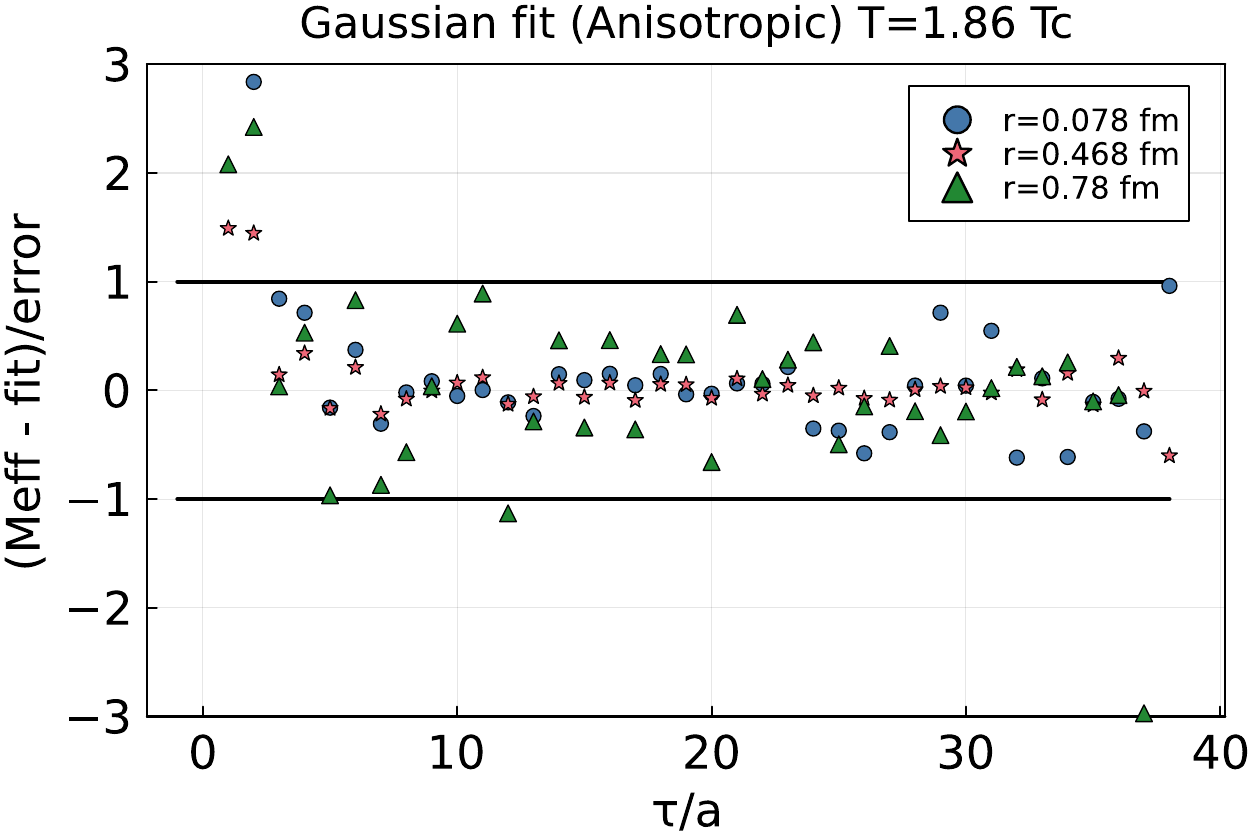}
    \includegraphics[scale=0.4]{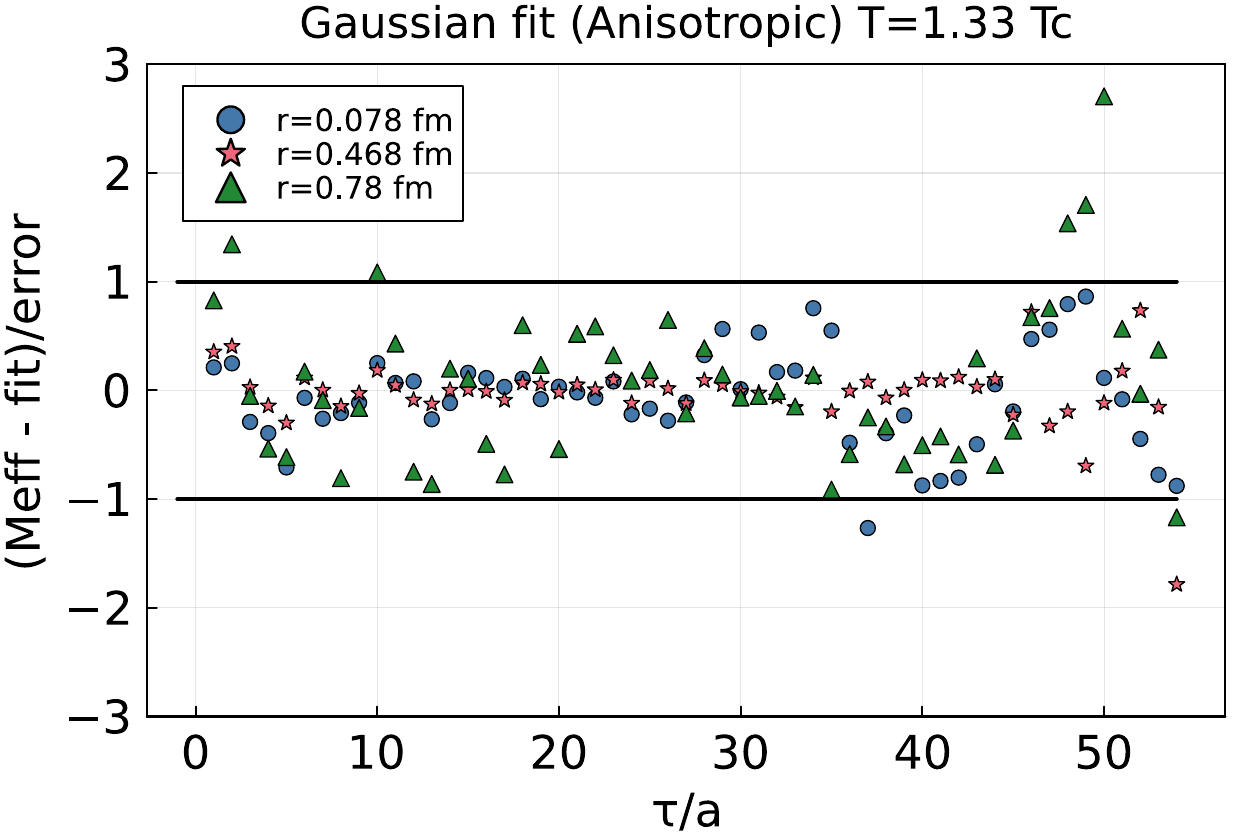}
    \caption{Goodness of Gaussian fit on subtracted anisotropic lattices at three different separation distances.}
    \label{fig:gauss_fit_good_anisotropic}
\end{figure}

In this appendix we discuss the goodness of fits on the euclidean correlator. \Cref{fig:gauss_fit_good_isotropic} and \cref{fig:gauss_fit_good_anisotropic} show the goodness of the Gaussian fits on isotropic and anisotropic lattices respectively. See \cref{eq:gaussfit} for the fit form. The Gaussian fits the subtracted correlator throughout the $\tau$ range upto a relatively large separation distance except for the first few points and the last couple of points in some cases.

\begin{figure}
    \centering
    \includegraphics[scale=0.4]{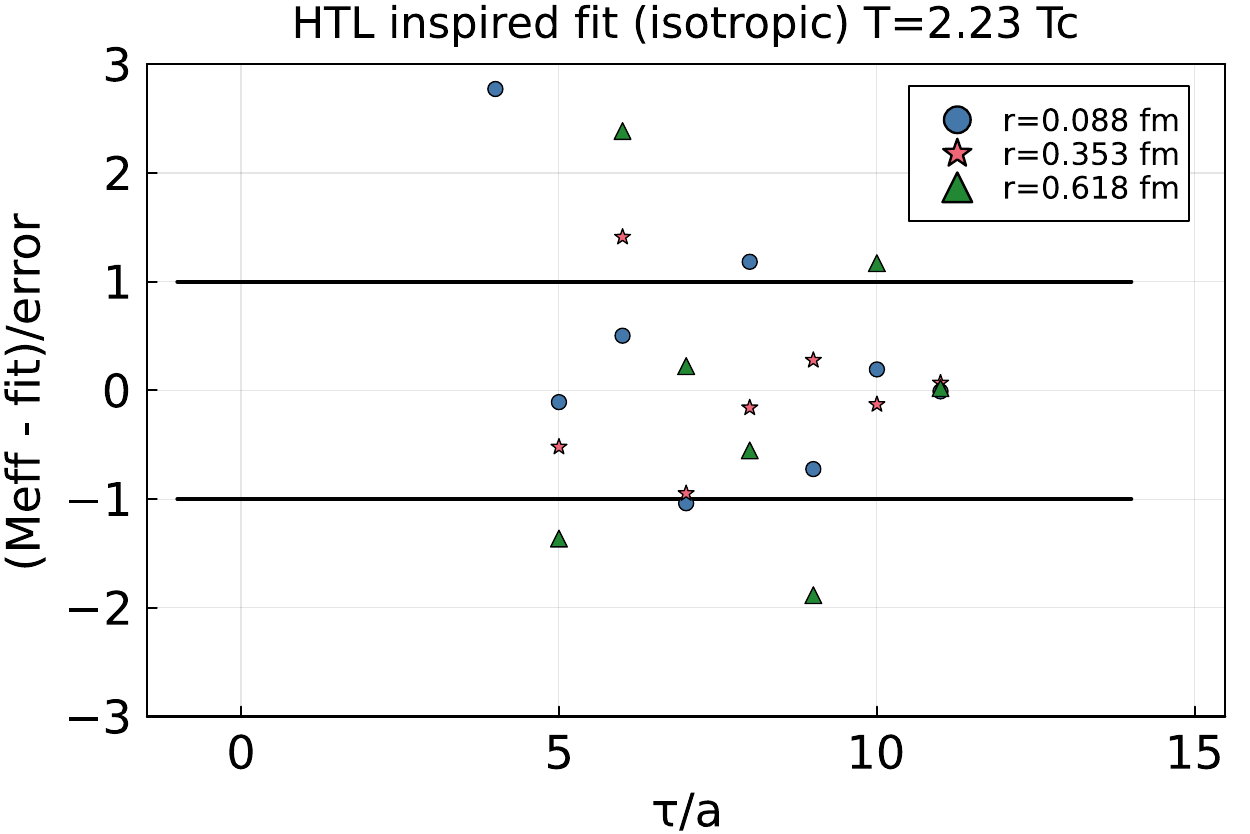}
    \includegraphics[scale=0.4]{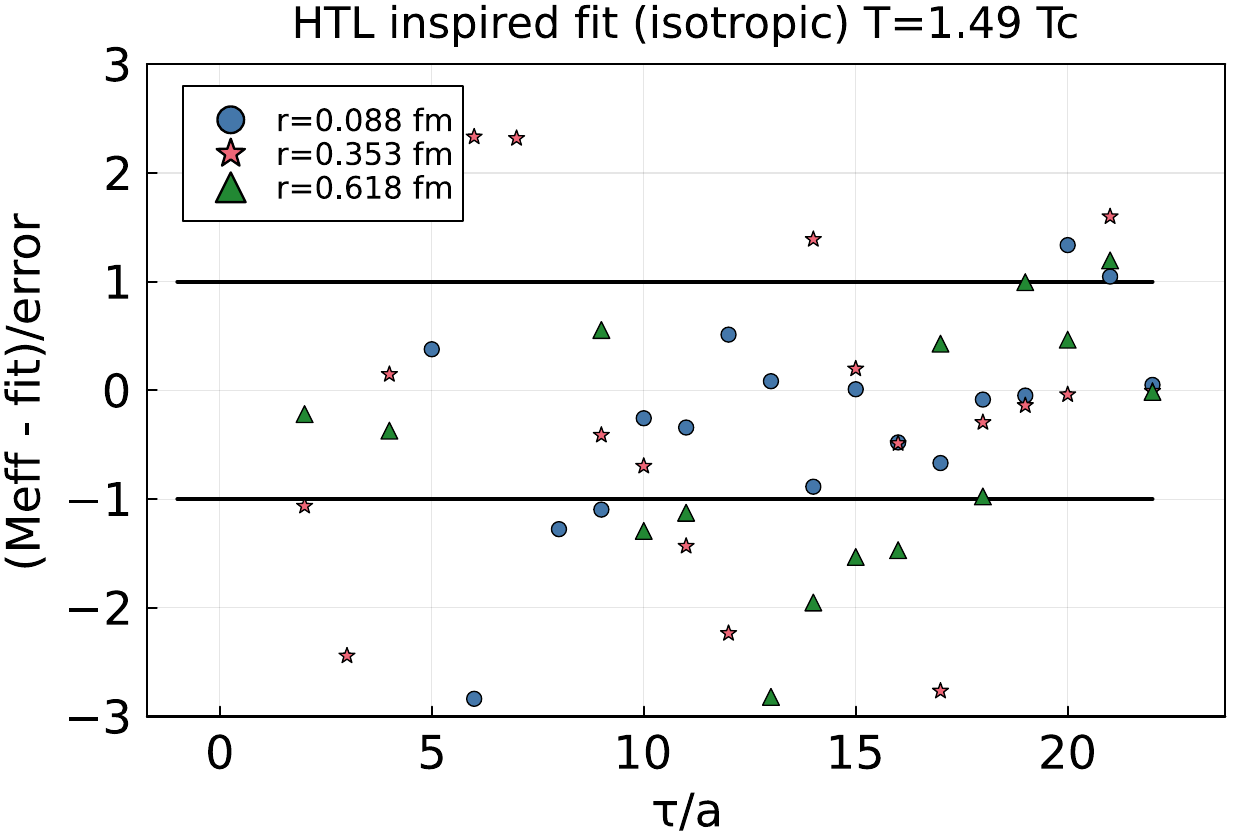}
    \caption{Goodness of HTL-inspired fit on isotropic lattices at three different separation distances.}
    \label{fig:HTL_fit_good_isotropic}
\end{figure}

\begin{figure}
    \centering
    \includegraphics[scale=0.37]{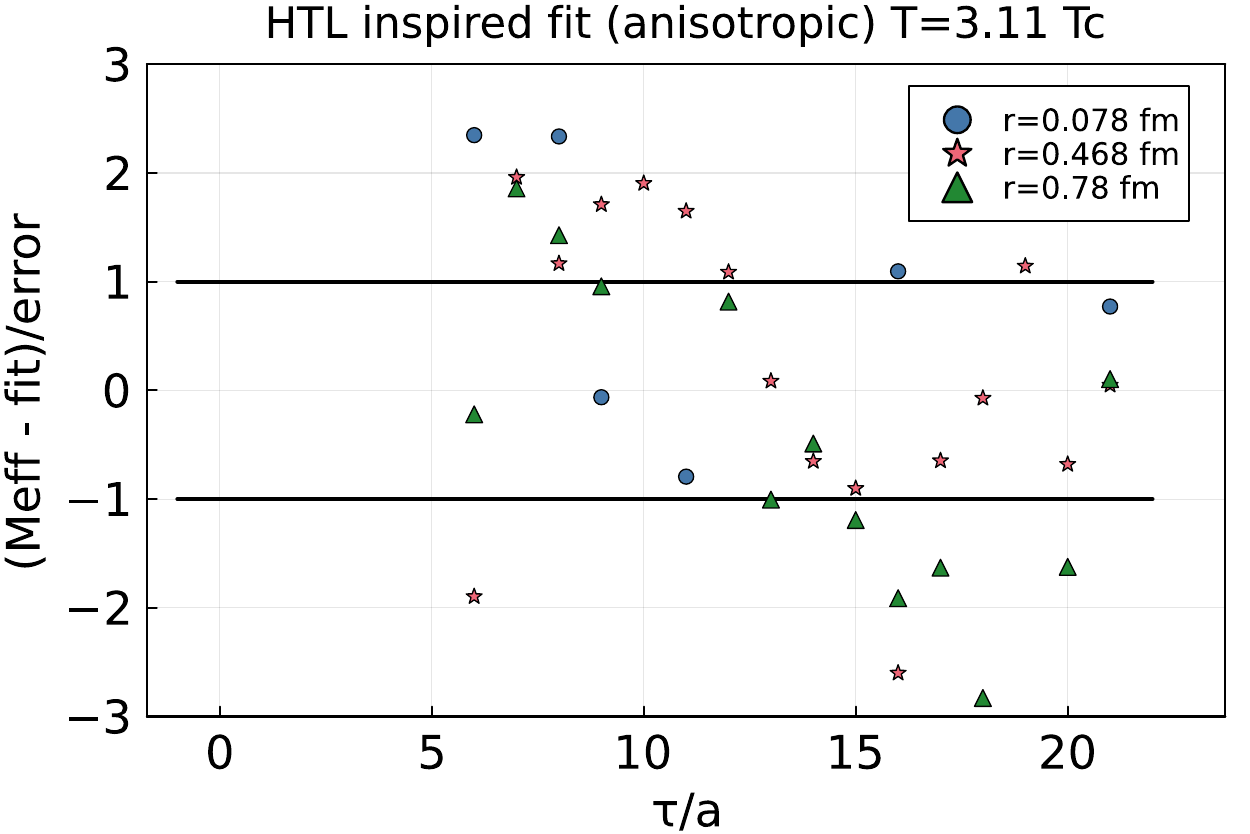}
    \includegraphics[scale=0.37]{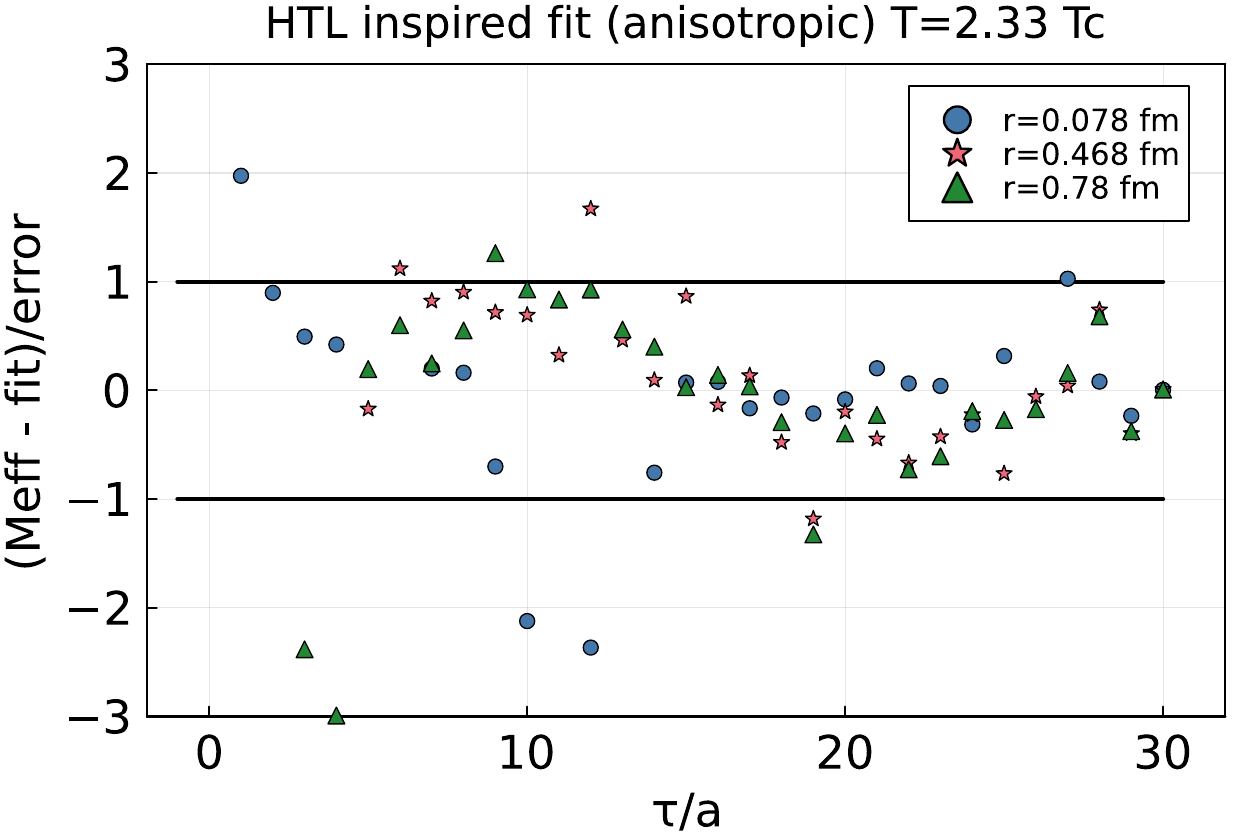}
    \includegraphics[scale=0.37]{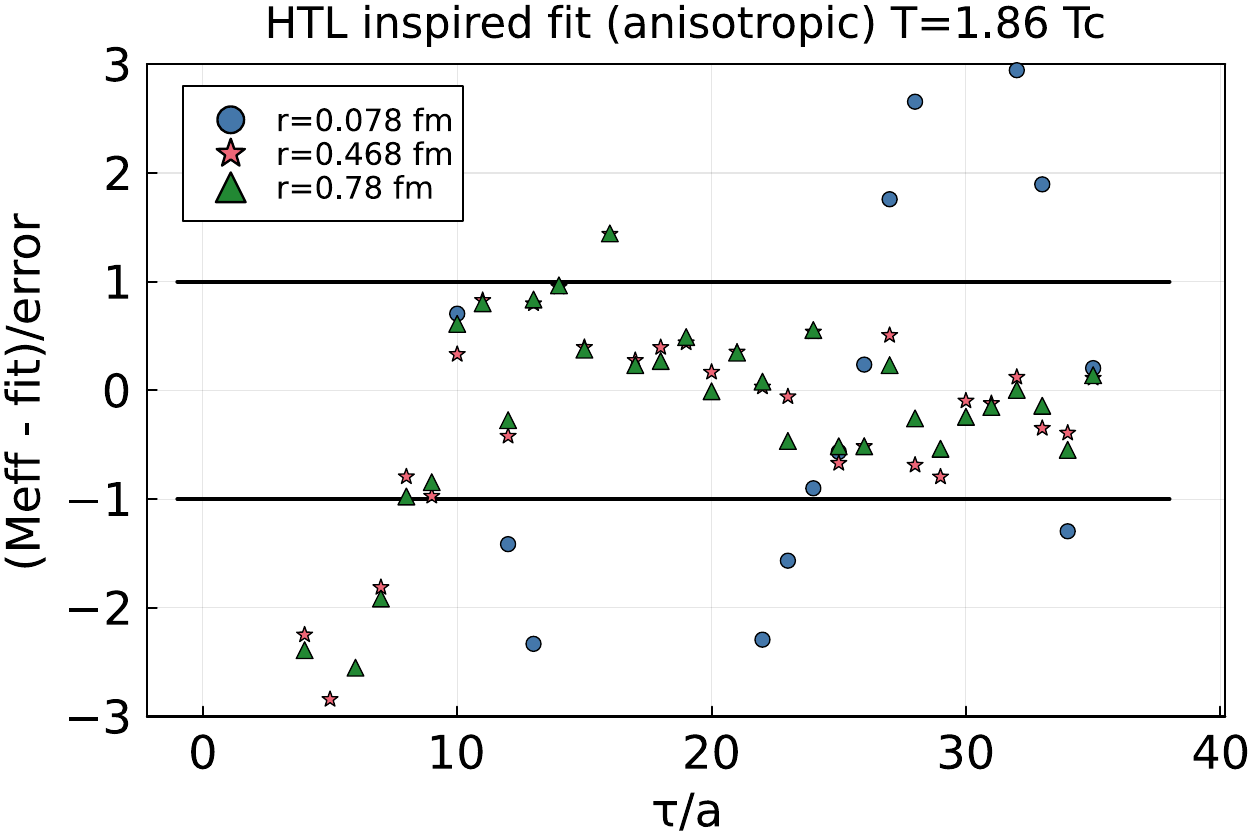}
    \includegraphics[scale=0.37]{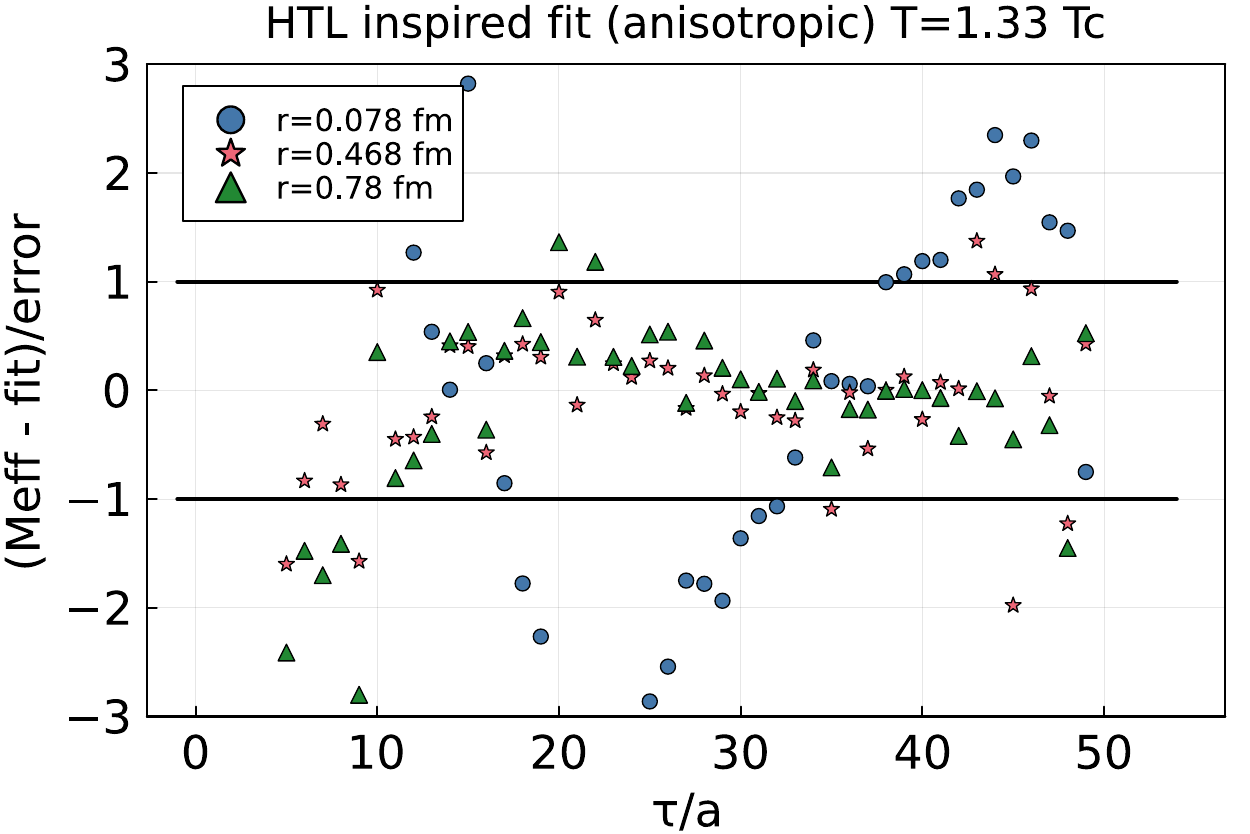}
    \caption{Goodness of HTL-inspired fit on anisotropic lattices at three different separation distances.}
    \label{fig:HTL_fit_good_anisotropic}
\end{figure}

\Cref{fig:HTL_fit_good_isotropic} and \cref{fig:HTL_fit_good_anisotropic} show the goodness of the HTL-inspired fits on isotropic and anisotropic lattices respectively. See \cref{eq:BDfit} for the fit form. We observe that the form fits the data quite well through a large $\tau$ range outside of $\tau=\beta/2$. The fitting range is increased with decreasing temperature.
\end{document}